\documentclass[useAMS,usenatbib]{mn2e}

\usepackage{times}

\pdfoutput=1
\usepackage{amssymb}
\usepackage[fleqn]{amsmath}
\usepackage{graphicx}
\usepackage{url}

\numberwithin{equation}{section}

\title[Main-sequence dark stars at the Galactic centre]{Dark stars at the Galactic centre -- the main sequence}
\author[P.C. Scott, M. Fairbairn and J. Edsj\"o]{Pat Scott$^{1}$\thanks{E-mail: pat@physto.se, malc@cern.ch, edsjo@physto.se}, Malcolm Fairbairn$^{2,3}$\footnotemark[1] and Joakim Edsj\"o$^{1}$\footnotemark[1]\\
$^{1}$Cosmology, Particle Astrophysics and String Theory, Department of Physics, Stockholm University \& Oskar Klein Centre for \\
Cosmoparticle Physics, AlbaNova University Centre, SE-106 91 Stockholm, Sweden\\
$^{2}$Theory Division, CERN, CH-1211, Geneva 23, Switzerland\\
$^{3}$Physics, Kings College London, Strand, London WC2R 2LS, UK}
\begin{document}

\date{Accepted 2008 November 19.  Submitted 2008 November 17; in original form 2008 October 5.}

\volume{394}
\pagerange{82--104}
\pubyear{2009}

\maketitle

\label{firstpage}

\begin{abstract}

In regions of very high dark matter density such as the Galactic centre, the capture and annihilation of WIMP dark matter by stars has the potential to significantly alter their evolution.  We describe the dark stellar evolution code \textsf{DarkStars}, and present a series of detailed grids of WIMP-influenced stellar models for main sequence stars.  We describe the changes in stellar structure and main sequence evolution which occur as a function of the rate of energy injection by WIMPs, for masses of 0.3--2.0\,M$_\odot$ and metallicities $Z$ = 0.0003--0.02.  We show what rates of energy injection can be obtained using realistic orbital parameters for stars at the Galactic centre, including detailed consideration of the velocity and density profiles of dark matter.  Capture and annihilation rates are strongly boosted when stars follow elliptical rather than circular orbits.  If there is a spike of dark matter induced by the supermassive black hole at the Galactic centre, single solar-mass stars following orbits with periods as long as 50 years and eccentricities as low as 0.9 could be significantly affected.  Binary systems with similar periods about the Galactic centre could be affected on even less eccentric orbits.  The most striking observational effect of this scenario would be the existence of a binary consisting of a low-mass protostar and a higher-mass evolved star.  The observation of low-mass stars and/or binaries on such orbits would either provide a detection of WIMP dark matter, or place stringent limits on the combination of the WIMP mass, spin-dependent nuclear-scattering cross-section, halo density and velocity distribution near the Galactic centre.  In some cases, the derived limits on the WIMP mass and spin-dependent nuclear-scattering cross-section would be of comparable sensitivity to current direct-detection experiments.

\end{abstract}


\section{Introduction}

\label{intro}

Observations continue to support the existence of non-baryonic dark matter \citep[DM;][]{bergstrom00,bertone05,bullet,wmap5} with a cosmological abundance around five times that of baryonic matter but of unknown composition.  Weakly-interacting massive particles (WIMPs) are a popular and convenient class of dark matter candidates because their weak-scale masses and couplings naturally give rise to an appropriate thermal relic abundance in the early universe.

Typical WIMPs, such as the lightest neutralino in supersymmetry \citep{Jungman96}, posses non-zero nuclear-scattering and self-annihilation cross-sections.  The nuclear-scattering cross-section makes it possible for WIMPs to collide elastically with nuclei in massive bodies such as stars, obtain velocities less than the local escape velocity and become gravitationally bound \citep{Press85, Griest87, Gould87b, Gould87a}.  This population of WIMPs will continue to scatter off nuclei in the star, sinking down to the core and eventually annihilating with other captured WIMPs.

If enough WIMPs are captured, the structure of the host body may be altered by the energy produced in WIMP annihilations, or by energy transport caused by the WIMP-nucleus scattering events themselves.  This was first realised over 30 years ago in the context of heavy neutrinos \citep{Steigman78}.  The potential for transport effects to modify the structure of stellar cores was initially developed by \citet{Spergel85} and \citet{Faulkner85}.  Implications of annihilation for stellar evolution were first explored by \citet{SalatiSilk89} and \citet{BouquetSalati89a}.  A series of subsequent studies \citep{Gilliland86, Renzini87, Spergel88, Faulkner88, Bouquet89, BouquetSalati89b, Deluca89, Salati90, Dearborn90b, Dearborn90, GH90, CDalsgaard92, Faulkner93} focused upon possible impacts of energy transport by `cosmion' WIMPs designed to solve the solar neutrino problem.  With the advent of neutrino oscillations this problem has of course disappeared.  Furthermore, the existence of much more stringent limits upon the WIMP-nucleon scattering cross-sections \citep[e.g.][]{Desai04, Angle08a, CDMS08, Behnke08} and an improved understanding of the distribution of dark matter on galactic scales \citep[e.g.][]{merritt05,vialactea} mean that the likelihood of seeing changes induced purely by WIMP energy transport seems somewhat diminished.  Indeed, later efforts to constrain WIMP physics via helioseismology have proven fruitless \citep{Bottino02}.

Recent times have seen a resurgent interest in the impacts of WIMPs upon stars, now focussing almost exclusively upon the influence of annihilation. \citet{Moskalenko07} and \citet{Bertone07} showed that it could be possible to see white dwarfs heated by WIMP annihilation, at the Galactic centre and in globular clusters respectively.  \citet*{Spolyar08} and \citet*{Natarajan08b} showed that WIMP annihilation might be able to partially inhibit the formation of PopIII stars, resulting in giant, cool, primordial stars supported entirely by annihilation energy.  In previous letters the current authors  presented the first numerical simulations of the structure and evolution of WIMP-burning main sequence stars, employing and comparing both a simple static structure code and a preliminary version of the evolutionary code we present here \citep*{Fairbairn08, Scott08a}.  We found that WIMP annihilation in stellar cores diminishes nuclear burning and causes them to re-ascend the Hayashi track, in agreement with the analytical estimates of \citet{SalatiSilk89}.

\citet{Iocco08a} and \citet{Freese08a} performed simplified capture calculations on models of `naturally-formed' PopIII stars, showing that even if the stars were to form normally, they might later accrete sufficient dark matter to alter their appearance.  The dark matter densities considered in these studies and in that of \citet{Spolyar08} were confirmed as reasonable by \citet{Freese08b}, using a more detailed treatment of the collapse of the primordial dark matter--gas halo.  Both groups went on to consider different stages of the pre-main sequence evolution of WIMP-influenced PopIII stars: \citet{Freese08c} employed polytropic models in an attempt to understand the evolution of the stars postulated by \citet{Spolyar08}, and \citet{Iocco08b} followed the evolution from the tip of the Hayashi track using a full stellar evolution code.  Both found stalling phases, but of different durations and at different stages of the stars' formative evolution.  \citet*{Yoon08} and \citet{Taoso08} have now presented simulations of main sequence PopIII stars assumed to have formed normally, but then allowed to evolve with the effects of WIMP capture and annihilation.  The last three studies show extended main sequence lifetimes and stalling on the Hayashi track, in agreement with our earlier conclusions at non-zero metallicity and the results we present here.

Those working on PopIII stars have referred to WIMP-burning stars as `dark stars', whilst we and others working at non-zero metallicities have typically used the terms `WIMP burners' or `dark matter burners'.  In the interests of cohesiveness and simplicity, we will simply adopt the former term.  We do acknowledge that the term `dark star' is something of a misnomer, since stars burning dark matter are not strictly dark.  As we shall see in the following pages, except for cases where the ambient dark matter density is extremely high, their luminosities are at least reduced relative to normal stars.

In this paper, we present a detailed analysis of the effects of dark matter capture, annihilation and energy transport upon the structure and evolution of main sequence stars, specifically those which might exist at the Galactic centre.  In Sect.~\ref{models} we give the full description of the \textsf{DarkStars} code and its input physics alluded to in \citet{Fairbairn08} and \citet{Scott08a}.  Sect.~\ref{ms} presents the properties of main sequence dark stars, based upon a grid of stellar models covering a range of masses and metallicities.  We take up the questions of the distribution of dark matter close to the Galactic centre in Sect.~\ref{halo}, and the properties of stellar orbits there in Sect.~\ref{potential}.  In Sect.~\ref{gc}, we present results from further grids of evolutionary models computed with realistic treatments of the environment and orbits expected near the Galactic centre.  We also discuss existing and potential observations in Sect.~\ref{gc}, then give some final remarks on the prospect of detecting or constraining the nature of dark matter through such observations in Sect.~\ref{conclusions}.

\section{Theory and modelling}
\label{models}

\subsection{Capture, annihilation and energy injection}

The total population of WIMPs $N(t)$ present in a star is given \citep{Jungman96} by the equation
\begin{equation}
\label{WIMPpop}
    \frac{\mathrm{d}N(t)}{\mathrm{d}t} = C(t) - 2A(t)-E(t),    
\end{equation}
where $C(t)$ is the rate at which WIMPs are captured, $A(t)$ is the rate at which annihilations occur and $E(t)$ is the evaporation rate. The factor of 2 in the annihilation term arises because each annihilation destroys two Majorana WIMPs. In many cases of interest evaporation is negligible, but we will return to this point later.

Many approximations to the full expression for $C(t)$ derived by \citet{Gould87b} have appeared in the literature, with widely varying accuracies.  Here we attempt to present the full theory in a compact and usable form.  We will also build upon the following in Sect.~\ref{halo} when we consider alternative halo models.  For a star capturing WIMPs from an infinitely distant halo, the capture rate is 
\begin{equation}
\label{cap}
    C(t) = 4\pi\int^{R_\star}_0 r^2\int^\infty_0 \frac{f(u)}{u}w\Omega_v^-(w)\,\mathrm{d}u\,\mathrm{d}r,
\end{equation}
where $r$ is the local height in the star, $u$ is the incoming WIMP velocity before it is influenced by the star's gravitational field and $f(u)$ is the WIMP velocity distribution in the halo.  The local escape velocity at a height $r$ is $v(r,t)$, and $w = w(u,r,t)\equiv\sqrt{u^2+v(r,t)^2}$ is the velocity an incoming WIMP obtains by the time it reaches a height $r$.  $\Omega_v^-(w)$ is the rate at which a WIMP with velocity $w$ scatters to a velocity less than $v$, and thereby becomes captured.  This formula does not apply to capture from an already-bound population of WIMPs, such as occurs in an adiabatically-contracting DM-gas cloud.

For a scattering nucleus of mass $m_\mathrm{nuc}$ and a WIMP mass $m_\chi$, kinematics dictate that the only collisions able to scatter a WIMP to velocities less than $v$ are those where the fraction $\Delta$ of the WIMP energy lost in the collision obeys
\begin{equation}
    \frac{u^2}{w^2}\le\Delta\le\frac{\mu}{\mu_+^2}, 
\end{equation}
with
\begin{equation}
    \mu\equiv\frac{m_\chi}{m_\mathrm{nuc}}, \quad\mu_\pm\equiv\frac{\mu\pm 1}{2}.
\end{equation}
This is clearly only possible for values of $u$ that obey 
\begin{equation}
    \frac{u^2}{w^2}\le\frac{\mu}{\mu_+^2},
\end{equation}
which is equivalent to
\begin{equation}
    u^2\le\frac{\mu v^2}{\mu_-^2}.
\end{equation}
The partial capture rate is then given by
\begin{equation}
    \label{omega}
    \begin{split}
    \Omega_v^-(w) & = \sum_i \Omega_{v,i}^-(w) \\
    & = \sum_i w\sigma_i n_i(r,t) \,\frac{\mu_i}{\mu_{+,i}^2}\,\theta\big(\frac{\mu_i v^2}{\mu_{-,i}^2}-u^2\big) \\
    & \quad\times\int^{\mu_i/\mu_{+,i}^2}_{u^2/w^2}\,|F_i(\Delta)|^2\,\mathrm{d}\Delta. 
    \end{split}
\end{equation}
Here $i$ denotes the $i$th nuclear species, $n_i$ is its local number density in the star and $F_i(\Delta)$ is the $i$th nuclear form factor.  $\theta$ is the Heaviside step function.  The total cross-section $\sigma_i$ for scattering of WIMPs on the $i$th nucleus can be approximated as \citep{Jungman96,darksusy}
\begin{equation}
    \sigma_i = \beta^2\Big[\sigma_\mathrm{SI} A_i^2 + \sigma_\mathrm{SD} \frac{4(J_i+1)}{3J_i}|\langle S_{\mathrm{p},i}\rangle+\langle S_{\mathrm{n},i}\rangle|^2\Big],
\end{equation}
where
\begin{equation}
    \beta = \frac{m_\mathrm{nuc}(m_\chi + m_\mathrm{p})}{m_\mathrm{p}(m_\chi + m_\mathrm{nuc})}
\end{equation}
is the ratio of the reduced masses of the WIMP-nucleus and WIMP-proton systems.  Here $\sigma_\mathrm{SI}$ and $\sigma_\mathrm{SD}$ are the hydrogen-normalised spin-independent and spin-dependent nuclear-scattering cross-sections respectively, $A_i$ is the atomic number of the nucleus, $J_i$ is its spin and $\langle S_{\mathrm{p},i}\rangle$ and $\langle S_{\mathrm{n},i}\rangle$ are the expectation values of the spins of its proton and neutron systems, respectively.

Assuming an exponential form factor
\begin{equation}
\label{expff}
    |F(\Delta)|^2 = \exp\Big(-\frac{m_\chi w^2\Delta}{2E_0}\Big)
\end{equation}
for heavy elements and a delta function for hydrogen, the integral in Eq.~\ref{omega} can be performed analytically.  Here $E_0$ is the coherence energy arising from the characteristic nuclear radius.  When $m_\mathrm{nuc}$ is expressed in GeV/c$^2$, it can be approximated as
\begin{equation}
\label{e0}
    E_0 \approx \frac{5.8407 \times 10^{-2}}{m_\mathrm{nuc}(0.91m_\mathrm{nuc}^{1/3}+0.3)^2}\,\mathrm{GeV}.
\end{equation}
Making the further assumption that WIMP velocities in the halo follow an isothermal distribution with dispersion $\bar{v}$, the velocity distribution in the rest frame of the dark matter halo is
\begin{equation}
\label{galacticframe}
    f_0(u) = \frac{4}{\sqrt{\pi}} \Big(\frac{3}{2}\Big)^{3/2} \frac{\rho_\chi}{m_\chi} \frac{u^2}{\bar{v}^3} \exp\Big(-\frac{3u^2}{2\bar{v}^2}\Big),
\end{equation}
with $\rho_\chi$ the ambient WIMP density.  In the frame of a star moving with velocity $v_\star$ through the halo, this becomes
\begin{equation}
\label{stellarframe}
    f_\star(u) = f_0(u) \exp\Big(-\frac{3v_\star^2}{2\bar{v}^2}\Big) \frac{\sinh(3uv_\star/\bar{v}^2)}{3uv_\star/\bar{v}^2}.
\end{equation}
Using Eqs.~\ref{omega}, \ref{expff} and \ref{stellarframe} it becomes possible to perform the velocity integral in Eq.~\ref{cap} analytically.  One converts the step function in Eq.~\ref{omega} to a finite upper limit of integration $u_{\mathrm{max},i}(r,t) = v(r,t)\sqrt{u}/\mu_{-,i}$ and obtains
\begin{equation}
\label{standardcap}
    C(t) = 4\pi\int^{R_\star}_0 r^2 \sum_i\,\big[W_i(u_{\mathrm{max},i}(r,t)) - W_i(0)\big]\,\mathrm{d}r.
\end{equation}
Here
\begin{gather}
    \begin{split}
    \label{heavystandardcap}
    W_i(u) & \equiv \int\frac{f_\star(u)}{u}w\Omega_{v,i}^-(w)\,\mathrm{d}u \\
    & = \frac{\sigma_i n_i(r,t) \rho_\chi \mu_{+,i}^2E_{0,i}}{m_\chi^2 \bar{v} v_\star \mu_i} \sqrt{\frac{3}{2}} 
      \Bigg\{ (B+G)^{-\frac{1}{2}} \Upsilon(G) \\
    & \quad\times \mathrm{e}^{-BGv_\star^2/(B+G)} - (B+H)^{-\frac{1}{2}}\Upsilon(H) \\ 
    & \quad\times \mathrm{e}^{-\frac{H}{B+H}\big[B(v_\star^2+v(r,t)^2)+Hv(r,t)^2\big]}\Bigg\},
    \end{split}\\
    G \equiv \frac{m_\chi}{2E_0}, \quad H \equiv G\frac{\mu}{\mu_+^2}, \quad B \equiv \frac{3}{2\bar{v}^2}, \\
    \Upsilon(X) \equiv \Upsilon^-(X)-\Upsilon^+(X), \\
    \Upsilon^\pm(X) \equiv \mathrm{erf}\bigg\{\frac{Xu+B(u\pm v_\star)}{\sqrt{B+X}}\bigg\}
\end{gather}
for heavier elements, and
\begin{gather}
    \begin{split}
    \label{Hstandardcap}
    W_\mathrm{H}(u) & = \frac{\sigma_\mathrm{H} n_\mathrm{H}(r,t) \rho_\chi}{m_\chi \bar{v} v_\star} \sqrt{\frac{3}{2\pi}} 
      \Bigg\{\frac{v(r,t)^2\Xi}{2\sqrt{B}} + \frac{\mu_{-,\mathrm{H}}^2}{4\mu_\mathrm{H} B^{3/2}} \\
    & \quad\times \Big(2\sqrt{B}\big[(v_\star - u)\mathrm{e}^{-B(u+v_\star)^2} + (v_\star + u) \\
    & \quad\times \mathrm{e}^{-B(u + v_\star)^2 + 4B v_\star u}\big] - \big[1 + 2B v_\star^2\big]\Xi\Big)\Bigg\}, 
    \end{split}\\
    \Xi \equiv \sqrt\pi \bigg\{\mathrm{erf}\big[\sqrt{B}(u-v_\star)\big] - \mathrm{erf}\big[\sqrt{B}(u+v_\star)\big]\bigg\}
\end{gather}
for the special case of hydrogen.

The annihilation rate $A(t)$ is simply the integral of the local annihilation rate per unit volume $a(r,t)$
\begin{equation}
\label{ann}
    A(t) = 4\pi\int_0^{R_\star}r^2a(r,t)\,\mathrm{d}r,
\end{equation}
which is given by
\begin{equation}
\label{localann}
    a(r,t) = \frac12\langle\sigma_\mathrm{a}v\rangle_0n_\chi(r,t)^2.
\end{equation}
Here $n_\chi(r,t)$ is the local WIMP number density in the star and $\langle\sigma_\mathrm{a}v\rangle_0$ is the non-relativistic limit of the velocity-averaged annihilation cross-section.  The energy injected by WIMP annihilations $\epsilon_\mathrm{ann}$ per unit mass of nuclear matter is
\begin{equation}
\label{epsann}
    \epsilon_\mathrm{ann}(r,t) = \frac{2a(r,t)m_\chi \mathrm{c}^2}{\rho_\star(r,t)} - \nu_\mathrm{loss}(r,t),
\end{equation}
where $\rho_\star(r,t)$ is the local stellar density and $\nu_\mathrm{loss}(r,t)$ accounts for the fraction of the WIMPs' rest-mass energy which escapes in the form of neutrinos.  We assume that the WIMPs annihilate only into standard model particles which, apart from the neutrinos, very quickly deposit their energy in the surrounding gas.  Together with the energy injection/removal rate $\epsilon_\mathrm{trans}$ due to conductive transport by WIMPs, this gives the total local WIMP energy term
\begin{equation}
\label{epsWIMP}
    \epsilon_\mathrm{WIMP}(r,t) = \epsilon_\mathrm{ann}(r,t) + \epsilon_\mathrm{trans}(r,t).
\end{equation}
This acts as an additional source term in the standard stellar luminosity equation at every height in the star.

\subsection{Conductive energy transport and distribution}
\label{condanddistro}

The simplest way to describe the density of WIMPs in a star is to assume that they have thermalised with the stellar matter.  We assume that thermalisation occurs instantaneously, as it would be too cumbersome to allow for a non-thermalised evolution of the WIMP distribution at the same time as evolving a star (no formalism beyond explicit Monte Carlo simulations exists at this stage for doing such a thing).  We expect this approximation to break down when capture rates change rapidly, such as during the short-term evolution of stars on elliptical orbits at the Galactic centre (Sec.~\ref{elliptics}).  Our primary interest is in the longer-term behaviour of such stars, which should not be strongly effected by the thermalisation process.

Thermalisation can be local, such that the WIMP energies reflect the local temperature at every height in the star, or global, such that they reflect only a single overall characteristic temperature $T_\mathrm{W}$.  In the first case, WIMPs are in local thermodynamic equilibrium (LTE) with the stellar matter, whilst in the second case they are isothermally distributed.  The isothermal assumption gives the Boltzmann distribution for particles in a gravitational well
\begin{equation}
\label{rhoiso}
    \begin{split}
    n_{\chi,\mathrm{iso}}(r,t) & = N(t)\frac{g_j\mathrm{e}^{-E_j/\mathrm{k}T}}{\sum_{j'} g_{j'}\mathrm{e}^{-E_{j'}/\mathrm{k}T}} \\
    & = N(t)\frac{\mathrm{e}^{-m_\chi\phi(r,t)/\mathrm{k}T_\mathrm{W}(t)}}
      {\int^{R_\star}_0 4\pi r'^2 \mathrm{e}^{-m_\chi\phi(r',t)/\mathrm{k}T_\mathrm{W}(t)}\,\mathrm{d}r'},
    \end{split}
\end{equation}
where $\phi(r,t)$ is the local value of the gravitational potential, $g_j$ is the statistical weight of the $j$th energy level (1 in this case) and $E_j=m_\chi\phi(r,t)$ is its gravitational potential energy.  Because WIMPs cluster so strongly in the centre of a star, previous analyses typically assumed that the area in which they reside can be approximated by a sphere of uniform density $\rho_\mathrm{c}(t)$, the central density.  This sets $T_\mathrm{W}(t)$ to the central temperature $T_\mathrm{c}(t)$, and gives
\begin{equation}
    \phi(r,t) \approx \frac{2\pi}{3}\mathrm{G}\rho_c(t)r^2. 
\end{equation}
This takes Eq.~\ref{rhoiso} to
\begin{equation}
\label{approxrhoiso}
    n_\chi(r,t) \approx N(t)\frac{\mathrm{e}^{-\frac{r^2}{r_\chi^2}}}{\pi^{3/2}r_\chi^3},
\end{equation}
with
\begin{equation}
    r_\chi(t) \equiv \bigg(\frac{3\mathrm{k}T_\mathrm{c}(t)}{2\pi\mathrm{G}\rho_\mathrm{c}(t)m_\chi}\bigg)^{1/2}.
\end{equation}
In practice, there is no longer any real reason to prefer Eq.~\ref{approxrhoiso} to Eq.~\ref{rhoiso}.  The gravitational potential must be tabulated anyway at all heights in the star to obtain $v(r,t)$ for the capture calculation, and the integral in Eq.~\ref{rhoiso} can be evaluated quickly and easily using modern computers.  Most importantly, a reasonable estimate for $T_\mathrm{W}(t)$ can be directly calculated from the structure of the star (as we describe below).

The usefulness of the uniform sphere approximation lies in the length scale it defines, $r_\chi$.  This gives the approximate scale height of the WIMP distribution in the star, which can be compared with the WIMP mean free path
\begin{equation}
    l(r,t) \equiv \Big(\sum_i l_i(r,t)^{-1}\Big)^{-1}, \hspace{0.5cm} l_i(r,t)^{-1} \equiv \sigma_in_i(r,t)
\end{equation}
at the centre of the star to give the Knudsen number of the system,
\begin{equation}
    K(t) = \frac{l(0,t)}{r_\chi(t)}.
\end{equation}
The Knudsen number indicates whether the WIMPs travel a distance less than the scale size on average and transport energy locally ($K<1$), or typically travel out beyond $r_\chi$ before depositing energy non-locally ($K>1$).

In the extreme limit $K\rightarrow0$, the energy transport is completely local.  In this case WIMPs scatter about so often that they are in LTE with the nuclei, and the energy transport is exactly the case of LTE conductive transport by a gas of massive particles.  The exact form of the energy injection/removal rate at a given stellar radius (i.e. the contribution to the local luminosity), and the corresponding density structure for \emph{LTE}-distributed (rather than isothermally-distributed) WIMPs has been calculated by \citet{GouldRaffelt90a}.  These are
\begin{multline}
\label{LTEtransport}
    L_\mathrm{trans, LTE}(r,t) = 4\pi r^2 \kappa(r,t)n_{\chi,\mathrm{LTE}}(r,t)l(r,t) \\
    \quad\times\,\Big[\frac{\mathrm{k}T_\star(r,t)}{m_\chi}\Big]^{1/2}\mathrm{k}\frac{\mathrm{d}T_\star(r,t)}{\mathrm{d}r}
\end{multline}
and
\begin{multline}
\label{LTEdens}
    n_{\chi,\mathrm{LTE}}(r,t) = n_{\chi,\mathrm{LTE}}(0,t)\Big[\frac{T_\star(r,t)}{T_\mathrm{c}(t)}\Big]^{3/2} \\
    \times\exp\Big[-\int^r_0\frac{\mathrm{k}\alpha(r',t)\frac{\mathrm{d}T_\star(r',t)}{\mathrm{d}r'} + 
      m_\chi\frac{\mathrm{d}\phi(r',t)}{\mathrm{d}r'}}{\mathrm{k}T_\star(r',t)}\,\mathrm{d}r'\Big],
\end{multline}
where normalisation to $\int^{R_\star}_0 4\pi r^2 n_{\chi,\mathrm{LTE}}(r,t)\,\mathrm{d}r = N(t)$ dictates the value of $n_{\chi,\mathrm{LTE}}(0,t)$.  Noting that in general
\begin{equation}
\label{luminositygeneral}
    L(r,t) = 4\pi\int^r_0r'^2\rho(r',t)\epsilon(r',t)\,\mathrm{d}r',
\end{equation}
we see that
\begin{equation}
\label{epsLTE}
    \epsilon_\mathrm{trans, LTE}(r,t) = \frac{1}{4\pi r^2 \rho(r,t)}\frac{\mathrm{d}L_\mathrm{trans, LTE}(r,t)}{\mathrm{d}r}
\end{equation}
Note that our sign convention differs from that of \citet{GouldRaffelt90a}; Eqs.~\ref{LTEtransport} and \ref{epsLTE} both refer to the energy \emph{injection} rates, not the rate at which WIMPs remove energy (cf. the sign conventions in Eq.~\ref{epsWIMP}).

The factors $\alpha$ and $\kappa$ are the dimensionless thermal diffusivity and conductivity, respectively.  These vary throughout the star according to the relative abundances of the different atomic nuclei (and the distribution of WIMP-nucleus relative velocities, if the scattering cross-section is velocity-dependent due to a vector coupling to quarks; this is not the case for the neutralino).  They are obtained through numerical solution of the Boltzmann collision equation for any given gas mixture.  \citet{GouldRaffelt90a} found and tabulated the values of $\alpha$ and $\kappa$ for gases consisting of WIMPs and one other nucleus, varying the WIMP-to-nucleus mass ratio $\mu$ from 0 to 100.  Whilst the rigorous thing to do for each physical mixture would be to re-solve the Boltzmann equation with a composite collisional operator given as a linear combination of the operators applicable to the single-nucleus case, a good approximation is to simply take a weighted mean of the tabulated $\alpha$ and $\kappa$ values themselves.  That is,
\begin{equation}
    \alpha(r,t) = \sum_i \frac{\sigma_in_i(r,t)}{\sum_j\sigma_jn_j(r,t)}\alpha_i(\mu_i)
\end{equation}
and
\begin{equation}
    \kappa(r,t) = \Big\{l(r,t)\sum_i\big[\kappa_i(\mu)l_i(r,t)\big]^{-1}\Big\}^{-1}.
\end{equation}
For $\mu<100$, $\alpha$ and $\kappa$ can be found by interpolation in the tables of \citet{GouldRaffelt90a}.  For larger values of $\mu$, the authors found the limiting behaviour $\alpha\rightarrow2.5$ and $\kappa\rightarrow\frac{5}{32}\sqrt{2\pi\mu}$.  To get a smooth curve for both $\alpha$ and $\kappa$, we set the final point of the interpolation table to the limiting values at $\mu=120$, so that for $\mu>120$ no interpolation is required and the analytical limits are used.

\begin{figure}
\begin{center}
\includegraphics[width=\columnwidth, trim = 0 0 0 30, clip=true]{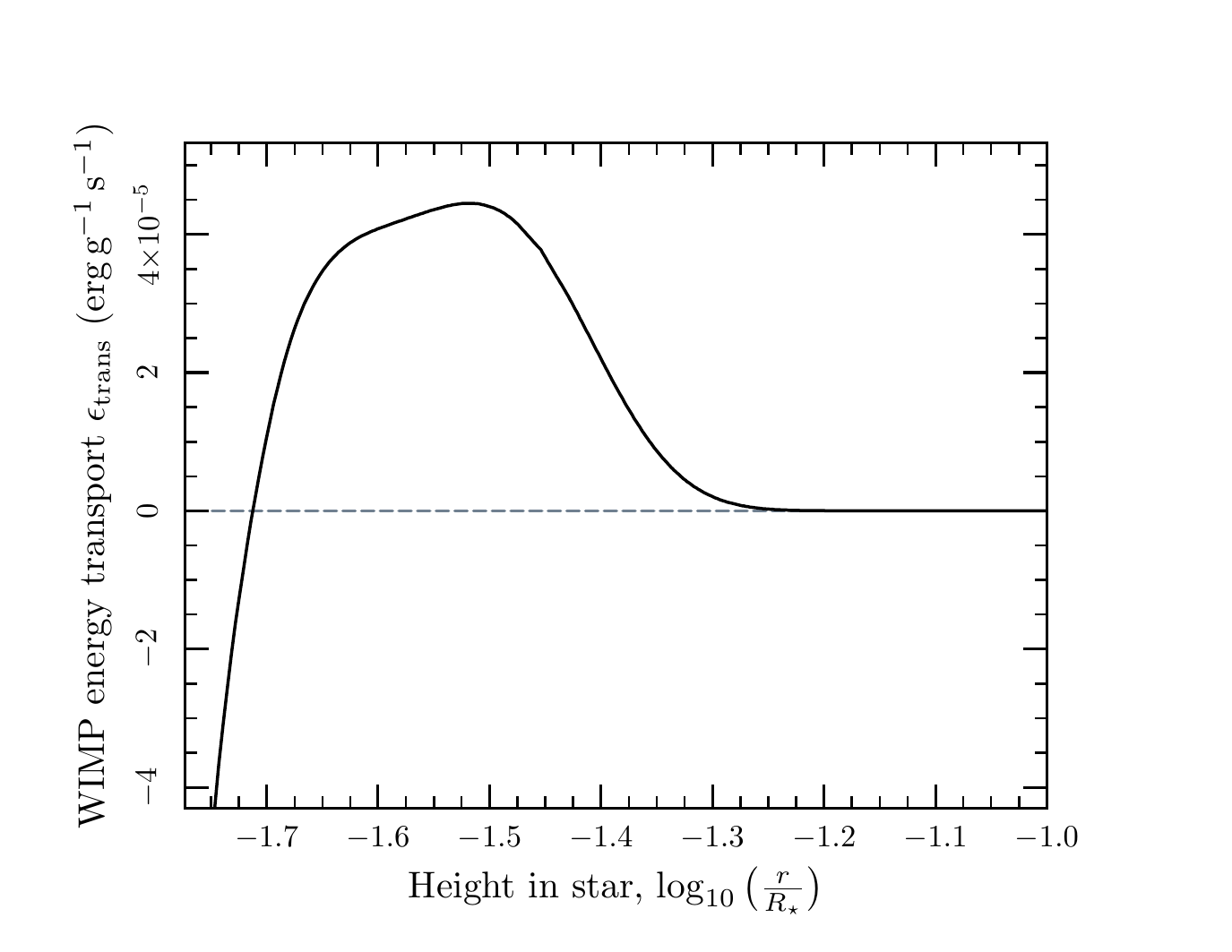}
\end{center}
\caption{A snapshot of the energy deposited by WIMP conductive energy transport with height in an example star (1\,M$_\odot$, $Z=0.01$, evolved in an isothermal halo with $\rho_\chi=10^{10}$\,GeV\,cm$^{-3}$, $v_\star=220$\,km\,s$^{-1}$ and $\bar{v}=270$\,km\,s$^{-1}$).  The snapshot was taken early in the star's evolution, before its structure had time to significantly adjust to the effects of WIMP annihilation.  Conduction by WIMP-nucleus scattering consumes energy from the very centre of the core and redeposits it further out.  The quantity shown here is the final form of $\epsilon_\mathrm{trans}$ (Eq.~\protect\ref{epstrans}), but it differs from $\epsilon_\mathrm{trans, LTE}$ only by some correction factors given in Eqs.~\protect\ref{knudsensupp} and \protect\ref{radialsupp}.}
\label{fig0}
\end{figure}

Notice that $L_\mathrm{trans, LTE}(R_\star,t)$ is always zero, as $\frac{\mathrm{d}T_\star(r,t)}{\mathrm{d}r}(R_\star)\approx0$; this reflects the fact that conduction by WIMPs never constitutes a net energy source nor sink in the star, and any energy the WIMPs remove from a hotter region is always returned in full to a cooler region.  As shown in Fig.~\ref{fig0}, $\epsilon_\mathrm{trans, LTE}$ is negative in the inner part of the star, increases with radius until it becomes positive, peaks, then drops again to asymptotically approach zero as $r\rightarrow R_\star$.  

The statement that there is no net energy outflow due to WIMP-nucleon scatterings is equivalent to there being no evaporation of WIMPs from the star.  That is, a captured WIMP never upscatters sufficiently energetically from a nucleus to become unbound and exit the system.  That this is the case in the LTE regime is immediately apparent: WIMPs do not travel far from the centre before re-scattering and sinking back to the core.  In the non-local regime, things are not so clear; when the WIMPs' mean free paths are much longer than the system scale height, evaporation could in principle be significant.  In this case, the evaporation rate must be equivalent to the net outward energy flux due to WIMP conductive energy transport \emph{in the isothermal (non-local) picture}.  That is,
\begin{align}
\label{evaporationcondition}
    L_\mathrm{trans, iso}(R_\star,t) & = m_\chi c^2E(t) \equiv L_\mathrm{evap}(R_\star,t) \\
    L_\mathrm{trans, iso}(r,t) & = 4\pi \int^r_0 r^2 \rho(r,t) \epsilon_\mathrm{trans, SP}(r,t,T_\mathrm{W})\,\mathrm{d}r \\
    L_\mathrm{evap}(r,t) & = 4\pi m_\chi\mathrm{c}^2 \int^r_0 r^2 R(r,t,T_\mathrm{W})\,\mathrm{d}r,
\end{align}
where $R(r,t,T_\mathrm{W})$ is the local WIMP evacuation rate per unit volume from a shell at height $r$, discussed in detail by \citet{Gould87a}.  The theoretical rate of energy conduction in the isothermal regime, 
\begin{multline}
\label{epsSP}
    \epsilon_\mathrm{trans, SP}(r,t,T_\mathrm{W}) = \frac{8\sqrt{\frac{2}{\pi}}\mathrm{k}^{3/2}}{\rho_\star(r,t)}n_{\chi,\mathrm{iso}}(r,t)
      [T_\star(r,t)-T_\mathrm{W}(t)] \\
    \times \sum_i\sigma_in_i(r,t)\frac{m_\chi m_{\mathrm{nuc},i}}{(m_\chi+m_{\mathrm{nuc},i})^2}\Big(\frac{T_\star(r,t)}{m_{\mathrm{nuc},i}}+\frac{T_\mathrm{W}(t)}{m_\chi}\Big)^{1/2}
\end{multline}
was developed by \citet{Spergel85}.  The equality in Eq.~\ref{evaporationcondition} uniquely determines $T_\mathrm{W}$, so given a a value of $E(t)$ or a particular choice of function $R(r,t,T_\mathrm{W})$, one simply implements an appropriate root-finding algorithm to get $T_\mathrm{W}$.  Knowing $T_\mathrm{W}$, one has all the information required to compute the isothermal density distribution (Eq.~\ref{rhoiso}).  Whilst the boundary condition itself (Eq.~\ref{evaporationcondition}) arises from the consideration of evaporation, and is important to include in order to obtain $T_\mathrm{W}$, in most cases of interest evaporation is negligible so $E(t)=0$.

Treating conductive energy transport by WIMPs when $K>1$, where the LTE conduction approximation breaks down entirely, is in general rather difficult.  In a second paper, \citet{GouldRaffelt90b} showed via explicit Monte Carlo solutions to the Boltzmann equation that there is no good way of analytically determining the WIMP energy transport for large $K$.  One would naively expect that as $K\rightarrow\infty$, the WIMPs would behave essentially isothermally, following the isothermal density structure (Eq.~\ref{rhoiso}) and transporting energy according to Eq.~\ref{epsSP}.  Whilst they do get rather close to the isothermal density structure, the energy transport via Eq.~\ref{epsSP} cannot be reconciled with the true Monte Carlo-derived $\epsilon_\mathrm{W}$ in any systematic way.  Eq.~\ref{epsSP} should therefore not be used as a description of WIMP conductive energy transport in practice, even when $K\rightarrow\infty$.  Its value is in providing a means of treating evaporation which is consistent with the isothermal density distribution, such that the characteristic temperature of the distribution naturally results in the correct evaporation rate (even if that happens to be zero).

In their earlier paper, \citet{GouldRaffelt90a} showed that as $K$ increases, the true conductive luminosity is suppressed relative to their analytical prediction (Eq.~\ref{LTEtransport}).  The breakdown  occurs in a clearly quantifiable way for increasing $K$, so one way to treat conductive energy transport by WIMPs in the non-local regime is to adopt the local expression, but with a `semi-empirical' luminosity suppression pre-factor in line with the suppression seen numerically.  The suppression shows a sigmoidal shape on a $\log K$ scale in \citeauthor{GouldRaffelt90a}'s results, 
\begin{equation}
\label{knudsensupp}
    \mathfrak{f}(K) \approx 1 - \frac{1}{1+\mathrm{e}^{-(\ln K - \ln K_0)/\tau}} = 1 - \frac{1}{1+\big(\frac{K_0}{K}\big)^{1/\tau}}.
\end{equation}
The relaxation scale must be about 0.4--0.5 to fit the numerical result well; with $\tau=0.5$ this function agrees exactly with the suppression function chosen by \citet{Bottino02}, so we use this value.  $K_0$ is the `crossing point' from the local to non-local regimes, where WIMP energy transport is most effective; \citeauthor{GouldRaffelt90a} found this to be $K_0\approx0.4$.

\citet{GouldRaffelt90a} also showed a similar suppression of the LTE WIMP conductive luminosity with radius.  This can be factored into the final expression for $L_\mathrm{T}(r,t)$ as a further multiplicative suppression factor $\mathfrak{h}(r,t)$, such that
\begin{equation}
    L_\mathrm{trans, final}(r,t) = \mathfrak{f}(K)\mathfrak{h}(r,t)L_\mathrm{trans, LTE}(r,t).
\end{equation}
The final form of the term appearing in Eq.~\ref{epsWIMP} and the stellar luminosity equation becomes
\begin{equation}
\label{epstrans}
    \epsilon_\mathrm{trans} = \frac{1}{4\pi r^2 \rho(r,t)}\frac{\mathrm{d}}{\mathrm{d}r}\big[\mathfrak{f}(K)\mathfrak{h}(r,t)L_\mathrm{trans, LTE}(r,t)\big],
\end{equation}
in analogy with Eq.~\ref{epsLTE}.  The radial suppression function appears as a simple cubic polynomial
\begin{equation}
\label{radialsupp}
    \mathfrak{h}(r,t) \approx \Big(\frac{r-r_\chi(t)}{r_\chi(t)}\Big)^3 + 1
\end{equation}
in \citeauthor{GouldRaffelt90a}'s numerical results.

For any given $K$, the degree to which $L_{\mathrm{trans}}(r)$ is suppressed relative to the LTE prediction can be thought of as a general indication of the goodness of the LTE approximation.  Since the density structure in the extremely non-local regime is at least moderately well-described by the isothermal approximation (Eq.~\ref{rhoiso}), the `goodness-of-LTE-assumption' function $\mathfrak{f}(K)$ can also be used as a way of interpolating between the isothermal and LTE (Eq.~\ref{LTEdens}) densities to give a semi-realistic density structure for all $K$:
\begin{equation}
\label{finaldens}
    n_{\chi,\mathrm{final}}(r,t) = \mathfrak{f}(K)n_{\chi,\mathrm{LTE}} + \big[1-\mathfrak{f}(K)\big]n_{\chi,\mathrm{iso}}
\end{equation}
Since $0\le\mathfrak{f}(K)\le1$ and both the LTE and isothermal densities are individually normalised, $n_{\chi,\mathrm{final}}(r,t)$ will stay correctly normalised for all $K$.  Since it is only $n_{\chi,\mathrm{LTE}}$ which enters into the expression for the final conductive energy transport, the density structure given by $n_{\chi,\mathrm{final}}$ is important only for determining the radial distribution of injected annihilation energy.

Because no simple measure of the overall efficiency of conductive energy transport by WIMPs presently exists in the literature, we define such a quantity as
\begin{equation}
\label{E}
    \mathfrak{E}(t) = \frac{\displaystyle\int^{R_\star}_0 r^2\dfrac{\rho_\star(r,t)}{\mu_\star(r,t)}\big|\dfrac{\epsilon_\mathrm{trans}}
      {\epsilon_\mathrm{other}}\big|\,\mathrm{d}r}{\displaystyle\int^{R_\star}_0 r^2\dfrac{\rho_\star(r,t)}{\mu_\star(r,t)}\,\mathrm{d}r},
\end{equation}
the dimensionless WIMP conductive effectiveness.  Here $\mu_\star(r,t)$ is the mean particle weight, $\epsilon_\mathrm{other}$ refers to all other energy terms (nuclear, gravitational and annihilation), and the denominator is simply a normalisation factor.  $\mathfrak{E}(t)$ is therefore the volume- and number-density-weighted, integrated ratio of WIMP-mediated energy transport to all other energy terms.  The weighting by number density is appropriate because WIMP energy transport is presumably more relevant in areas of higher nuclear density.  The absolute value arises because $\epsilon_\mathrm{trans}$ is a transport rather than a net source term (i.e.~takes on both positive and negative values).  Roughly speaking, in a star where WIMP conductive energy transport is the most important local source of luminosity, $\mathfrak{E}$ is greater than 1; where it is a sub-dominant contributor, $\mathfrak{E}$ is less than 1.  Whilst the weighting with local number density rightly biases $\mathfrak{E}(t)$ towards the stellar core, this means that it is possible for extremely effective, localised energy transport by WIMPs in the very centre of a star to dominate the overall stellar energetic effectiveness, without significantly altering the overall structure.  This is because in such a case, the enhanced transport only occurs in the most central parts of the core.

\subsection{The \textsf{DarkStars} code}

\textsf{DarkStars} includes WIMP capture based on Eq.~\ref{cap}, generalised from the solar capture routines of \textsf{DarkSUSY} \citep{darksusy}.  Exponential form-factor suppression (Eq.~\ref{expff}) is assumed for scattering off nuclei heavier than hydrogen.  In the basic version, the integral over incoming WIMP velocities can either be performed analytically assuming an isothermal velocity distribution (using Eqs.~\ref{heavystandardcap} and \ref{Hstandardcap}), or numerically over any arbitrary velocity distribution.  In Sect.~\ref{halo} we describe two additional velocity distributions which we have implemented in the code, one of which includes another analytical option for the velocity integral.

The capture routines are coupled to the EZ version \citep{Paxton04} of the \textsc{stars} stellar evolution code \citep{Eggleton71, Eggleton72, Pols95}, which uses relaxation to solve the hydrostatic equations of stellar structure over a 199-point adaptive radial mesh at each timestep.  We implement annihilation according to Eqs.~\ref{ann}--\ref{epsann}.  We determine the local WIMP density with Eqs.~\ref{rhoiso}, \ref{LTEdens} and \ref{finaldens}, obtaining $T_\mathrm{W}$ as the solution to Eq.~\ref{evaporationcondition}.  We include conductive energy transport via Eqs.~\ref{LTEtransport} and Eq.~\ref{epstrans}. 

The WIMP population is advanced at each timestep by solving Eq.~\ref{WIMPpop}.  We assume that since the stellar structure changes very slowly under the influence of the WIMPs in comparison to the evolution of $N(t)$, the evolution of the population between timesteps can be well described by the solution to Eq.~\ref{WIMPpop} in the special case where $C(t)$ and $A_\mathrm{c}(t) \equiv \frac{A(t)}{N(t)^2}$ are constant in time:  
\begin{gather}
    \label{popsolution}
    N(t+\Delta t) = \left\{
    \begin{array}{ll}
      \multicolumn{2}{l}{C(t)\tau_\mathrm{eq}(t)\bigg[\tanh\Big(\dfrac{\Delta t}{\tau_\mathrm{eq}(t)+t_\mathrm{equiv}}\Big)\bigg]
      ^\delta}\vspace{1mm}\\
      & \qquad\quad \text{when $C(t)\ne0$,} \vspace{2mm}\\
      \dfrac{N(t)}{1+N(t)A_\mathrm{c}(t)\Delta t} & \qquad\quad\text{when $C(t)=0$,}
    \end{array}\right.\\
    \label{taueq}
    \tau_\mathrm{eq}(t) \equiv \frac{1}{\sqrt{C(t)A_\mathrm{c}(t)}}, \\
    t_\mathrm{equiv} \equiv \tanh^{-1}\Bigg(\bigg[\frac{N(t)}{C(t)\tau_\mathrm{eq}(t)}\bigg]^\delta\Bigg), \\
    \delta \equiv \mathrm{sign}\bigg(\frac{\mathrm{d}N(t)}{\mathrm{d}t}\bigg).
\end{gather}
This is well-justified because both $C(t)$ and $A_\mathrm{c}(t)$ depend only upon the stellar structure, not directly upon the absolute WIMP population.  Here $\tau_\mathrm{eq}(t)$ is the emergent time-scale of equilibration between capture and annihilation, and $t_\mathrm{equiv}$ is an equivalent earlier time from which the approximate solution needs to be evolved for the current values of $C$ and $A_\mathrm{c}$.  Our approximation is an example of the general approach to solving stiff differential equations by separation into fast and slow subsystems known as coarse-graining, and allows a numerical solution to Eq.~\ref{WIMPpop} with timesteps of the order of those typically required for stellar evolution.

This scheme constitutes an explicit solution to Eq.~\ref{WIMPpop}, where each new stellar model is converged with the WIMP population at the previous timestep, which is calculated with capture and annihilation rates computed using the stellar structure of the previous model.  The models are therefore not completely self-consistent, as the WIMP population lags the stellar structure by one timestep.  Implementing Eq.~\ref{WIMPpop} in the internal implicit differencing scheme of the \textsc{stars} code would have required extensive revision of the internal solver.  As a consistency check, we have implemented a `reconvergence mode' similar to that described by \citet{Dearborn90}, where models are reconverged with the new WIMP population at every timestep, producing a fully self-consistent solution.  We have also experimented with rescaling the automatically-chosen timesteps to smaller values.  Except for some special cases which we describe in Sect.~\ref{ms}, the results do not change.

\begin{figure}
\begin{center}
\includegraphics[width=\columnwidth]{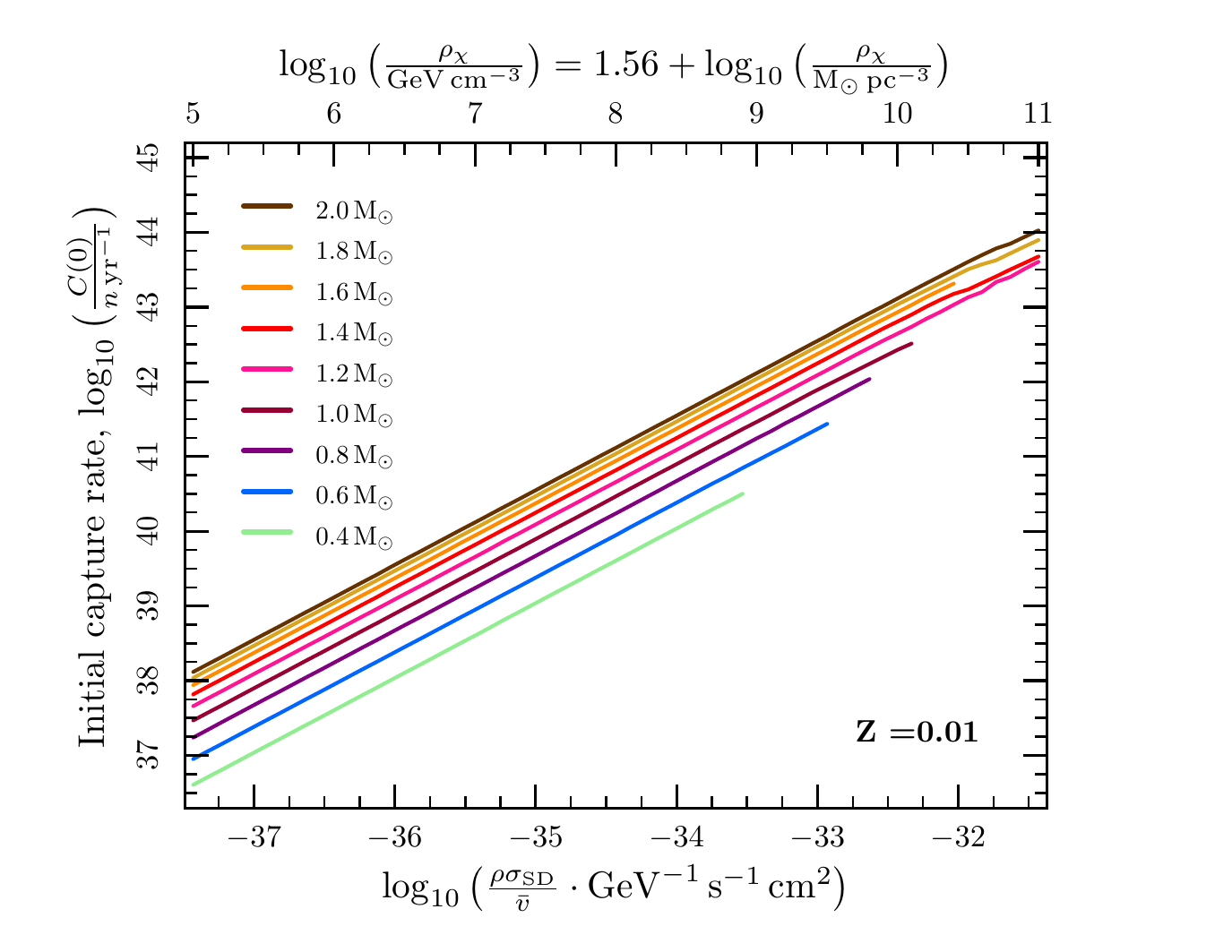}
\includegraphics[width=\columnwidth]{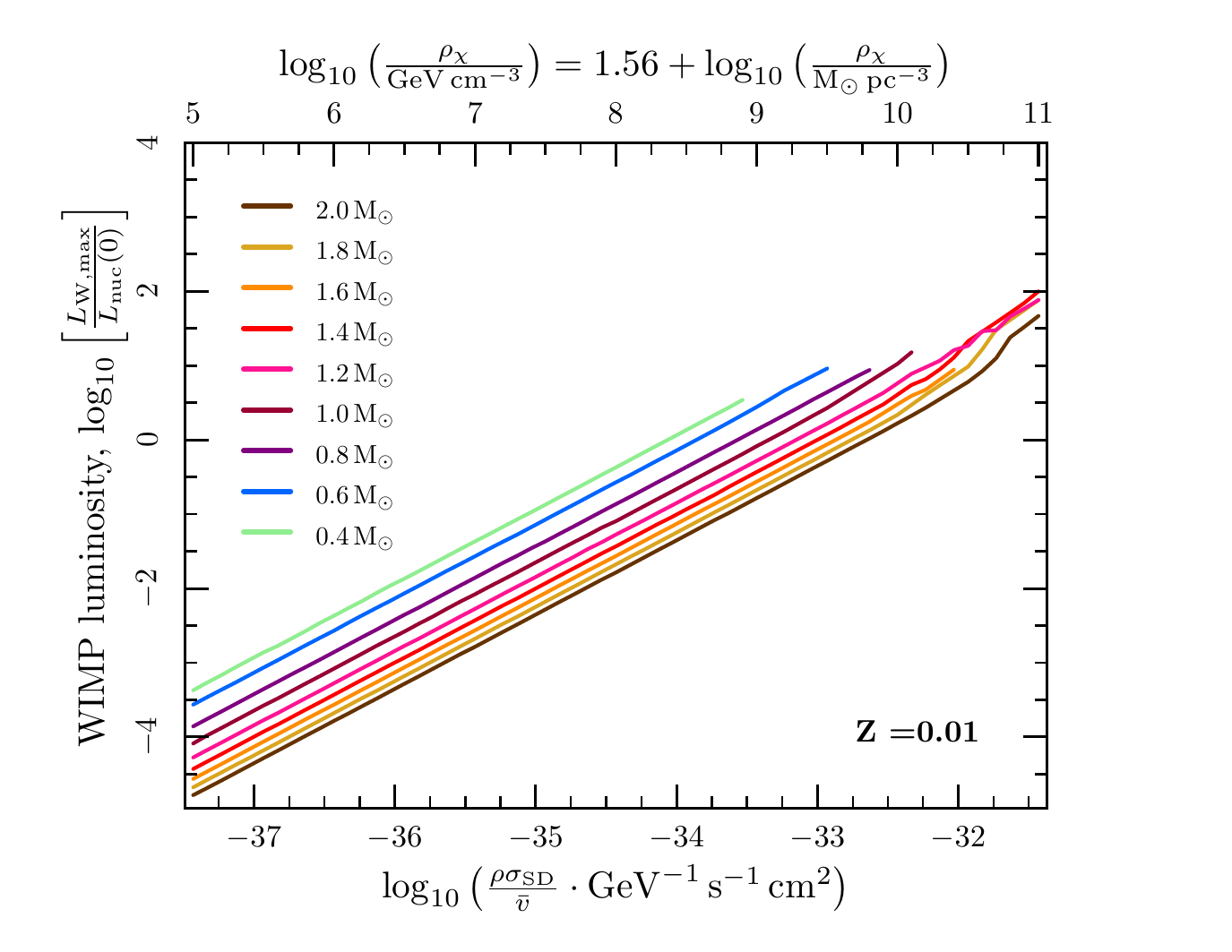}
\end{center}
\caption{Initial capture rates (top) and annihilation-to-nuclear luminosity ratios (bottom) achieved by stars evolved in differing dark matter densities.  Nuclear luminosities $L_\mathrm{nuc}(0)$ are initial values, whilst annihilation luminosities $L_\mathrm{W,max}$ are the maximum values achieved during a star's lifetime.  The dark matter velocity structure is that typically assumed by direct detection experiments, where the distribution is isothermal with solar values $v_\star=220$ and $\bar{v}=270$\,km\,s$^{-1}$. We refer to this canonical example as the `reference solar configuration' (RSC), where it should be understood that only the velocity structure, not the density, is reflective of the true solar situation.  These plots provide a simple conversion mechanism between capture rates, WIMP luminosities and equivalent RSC dark matter densities.}
\label{fig1}
\end{figure}

\begin{figure}
\begin{center}
\includegraphics[width=\columnwidth, trim = 0 0 0 0, clip=true]{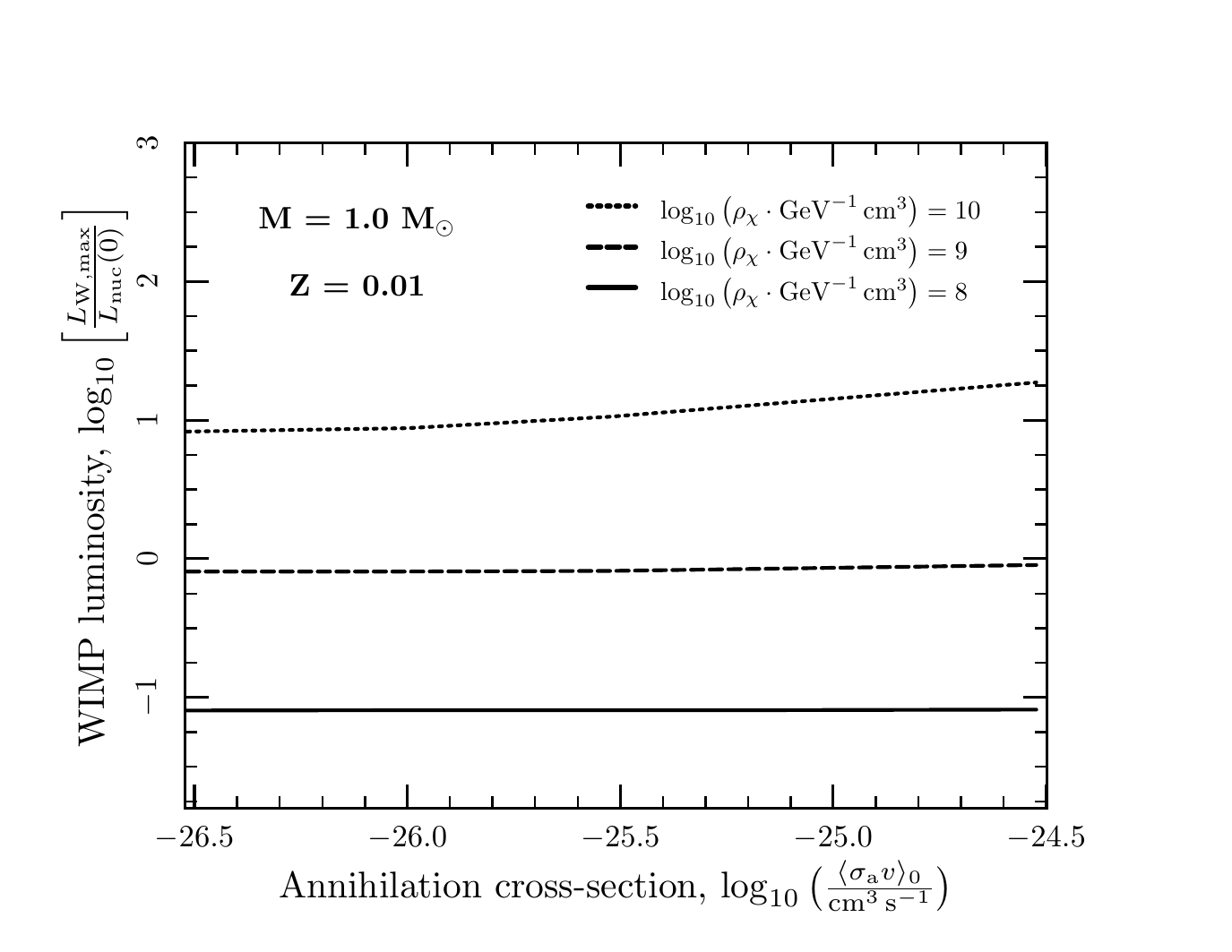}
\includegraphics[width=\columnwidth]{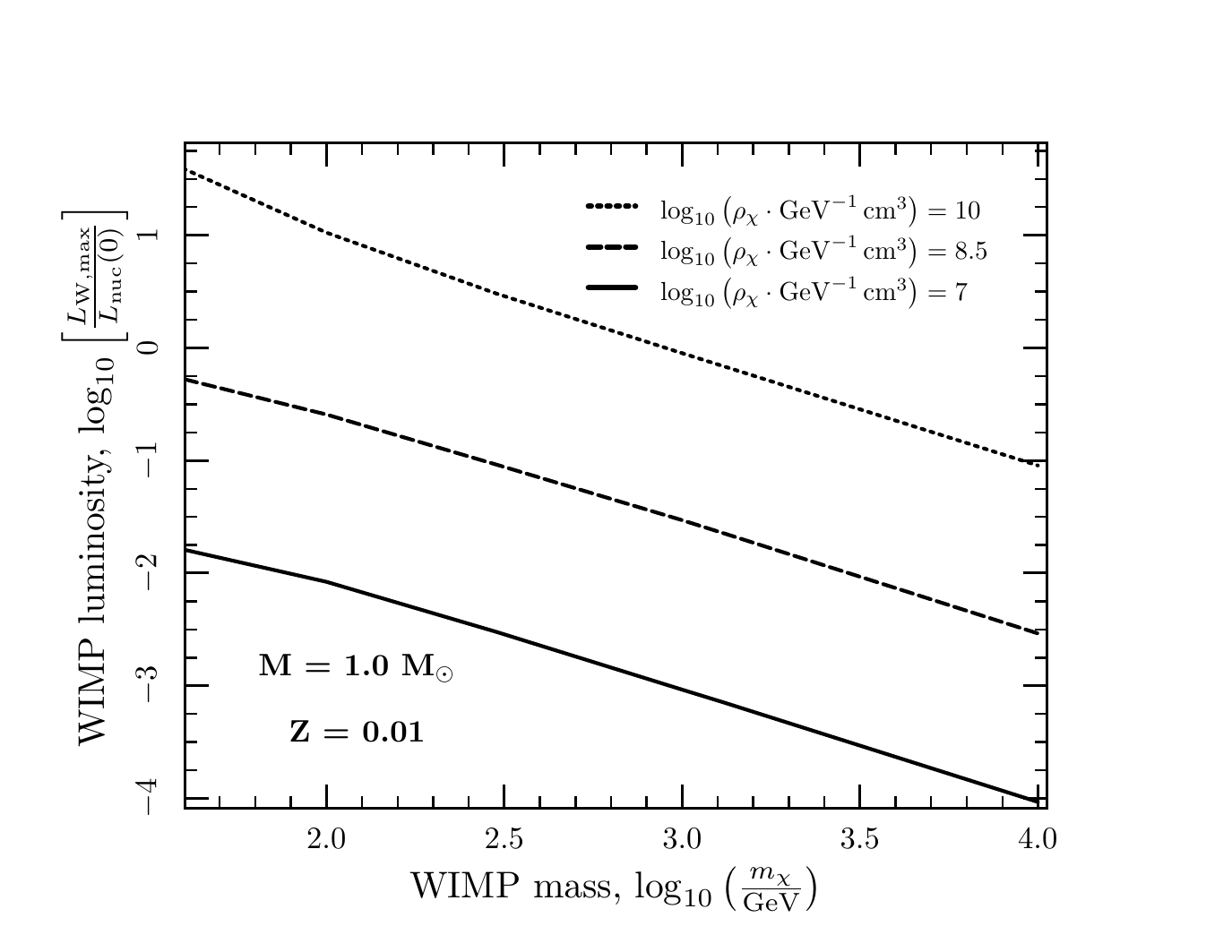}
\end{center}
\caption{WIMP-to-nuclear luminosity ratios achieved my a 1\,M$_\odot$ star with different annihilation cross-sections (top) and WIMP masses (bottom).  The dark matter halo configuration is the RSC.  Since capture rates do not depend upon the annihilation cross-section, the WIMP luminosity is independent of the annihilation cross-section in stars where capture and annihilation are in equilibrium at the time of maximum annihilation luminosity.  Because $v_\star$ and $\bar{v}$ are relatively large in the RSC, WIMP luminosities show a strong dependence upon the WIMP mass.}
\label{fig2}
\end{figure}

In general we limit timesteps to allow no more than a certain proportional change in the WIMP population per step. Typically we demand that the population does not change by more than the current value in one step, but for the more extreme situations in Sect.~\ref{gc} we reduce this by a factor of ten.  To aid initial convergence and prevent this prescription from demanding impossibly small timesteps early in the simulation, we begin simulations with populations of $10^{30-35}$ WIMPs.  This is many orders of magnitude less than the population required to have an effect upon the stellar structure.

We calculate capture by the 22 most relevant nuclei: $^{1}$H, $^{3}$He, $^{4}$He, $^{12}$C, $^{13}$C, $^{14}$N, $^{16}$O, $^{18}$O, $^{20}$Ne, $^{24}$Mg, $^{23}$Na, $^{27}$Al, $^{28}$Si, $^{32}$S, $^{40}$Ar, $^{40}$Ca, $^{56}$Fe, $^{58}$Ni, $^{60}$Ni, $^{206}$Pb, $^{207}$Pb and $^{207}$Pb.  The stellar code follows the abundances of $^{1}$H, $^{4}$He, $^{12}$C, $^{14}$N, $^{16}$O, $^{20}$Ne and $^{24}$Mg, and we assume that the remaining mass is distributed amongst the other 15 species according to their abundance ratios in the Sun.  The data on inter-elemental ratios comes from \citet*[but with the Ni abundance from \citealt{Scott09Ni}]{AGS05}, and on isotopic ratios from \citet[$^3$He/$^4$He]{Heber03}, \citet[$^{12}$C/$^{13}$C and $^{16}$O/$^{18}$O]{ScottVII}, preliminary results from work in progress ($^{58}$Ni/$^{60}$Ni) and \citet[via \protect\citealt{Anders89}, $^{208}$Pb/$^{207}$Pb/$^{206}$Pb]{Tatsumoto76}.  Atomic and nuclear masses are sourced from \citet{Audi03a} and \citet{Audi03b}.  The present code allows total heavy-element mass fractions $Z$ of 0.0001--0.03, which are paired with corresponding helium mass fractions of 0.24--0.30.

Since the annihilation rate goes as $n_\chi^2$ whilst the evaporation rate goes only as $n_\chi$, the evaporation rate is much smaller for the high ambient WIMP densities and capture rates we are typically interested in.  Even for the Sun, \citet{Gould87a} found that evaporation is insignificant unless the WIMP mass happens to be relatively closely matched with a nucleus found there in significant abundance.  The heaviest such element is iron, so evaporation can be considered negligible in the Sun for $m_\chi \gtrsim 60$\,GeV, which is the case for most WIMPs considered interesting today. (The limit given by \citeauthor{Gould87a} was $m_\chi \gtrsim 4$\,GeV, but this assumed that elements heavier than helium could be neglected because, at the time, a WIMP mass higher than $\sim$10\,GeV was considered unlikely).  We therefore simply obtain $T_\mathrm{W}$ with $E(t)=0$, neglecting evaporation.

To estimate the factor $\nu_\mathrm{loss}$ in Eq.~\ref{epsann}, we carried out explicit Monte Carlo simulations of WIMP annihilation in the Sun, along the lines of \citet{Blennow08}.  We considered a range of masses and annihilation channels, and included full three-flavour neutrino oscillations, neutrino interactions, stopping of muons and interactions of heavy mesons (e.g.\ $B$ mesons) in the Sun's core.  For essentially all annihilation channels except $\tau^+ \tau^-$, $\nu_\mathrm{loss}$ for a 100 GeV WIMP is 5--15\% of the rest-mass energy.  For heavier WIMPs, $\nu_\mathrm{loss}$ is reduced due to neutrino interactions in the star.  For annihilation to $\tau^+ \tau^-$, $\nu_\mathrm{loss} \simeq $ 35--40\% for a 100 GeV WIMP (and drops for heavier WIMPs).  Since the neutralino (our canonical WIMP) has very limited annihilation to $\tau^+ \tau^-$, we assume a flat neutrino energy loss of $\nu_\mathrm{loss}=10$\% for all annihilations.  We also neglect the slight dependence upon stellar structure and mass.

Capture integrals in \textsf{DarkStars} are performed with \textsc{quadpack} \citep{quadpack}, whilst simple integrals are done by the users' choice of Simpson's rule, Romberg integration or fifth-order Runge-Kutta with an adaptive step size.  Sufficiently well-behaved functions are interpolated using cubic splines.  For the others, we found the tensional spline routines of \citet{Renka93} and \citet{Testa99}, after a slight readjustment of the convergence parameters, invaluable.

Some provision has been made in the code for later allowing $R(r,t,T_\mathrm{W})\ne0$, alternative form factors and metal-free evolution if required.  \textsf{DarkStars} is available for public download from \url{http://www.fysik.su.se/~pat/darkstars/}.

Except where explicitly stated otherwise, we perform all simulations in this paper with a canonical WIMP mass of $m_\chi=100$\,GeV and an annihilation cross-section $\langle\sigma_\mathrm{a}v\rangle_0 = 3\times10^{-26}$\,cm$^{3}$\,s$^{-1}$, which arises from relic density considerations assuming WIMPs to be the dominant component of dark matter.  We use nuclear-scattering cross-sections corresponding to the maximally-allowed experimental values $\sigma_\mathrm{SI}=10^{-44}$\,cm$^{2}$ \citep{Angle08a,CDMS08} and $\sigma_\mathrm{SD}=10^{-38}$\,cm$^{2}$ \citep{Desai04,Behnke08}.

\begin{figure*}
\begin{center}
\begin{minipage}{\columnwidth}
\includegraphics[width=\columnwidth, trim = 0 0 0 30, clip=true]{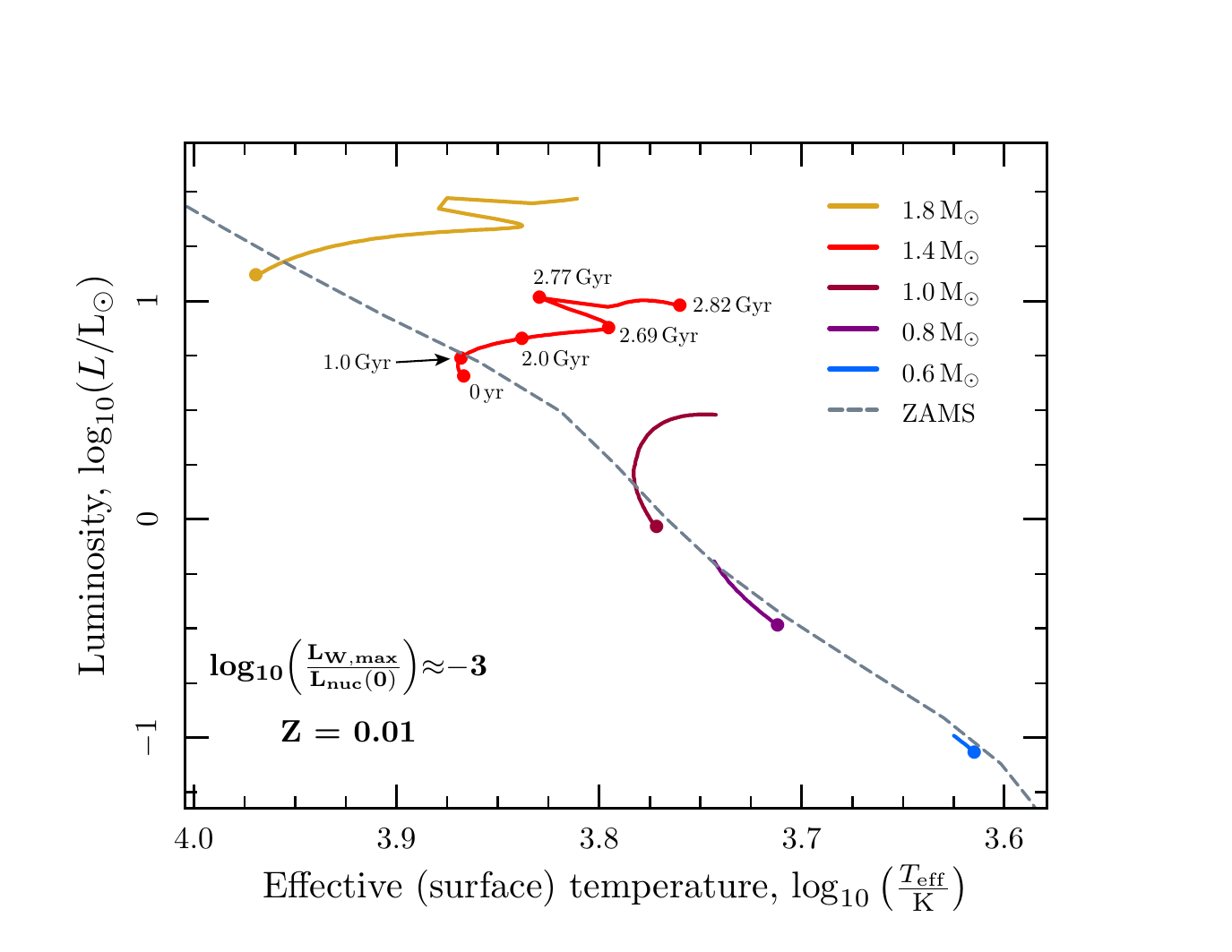}
\includegraphics[width=\columnwidth]{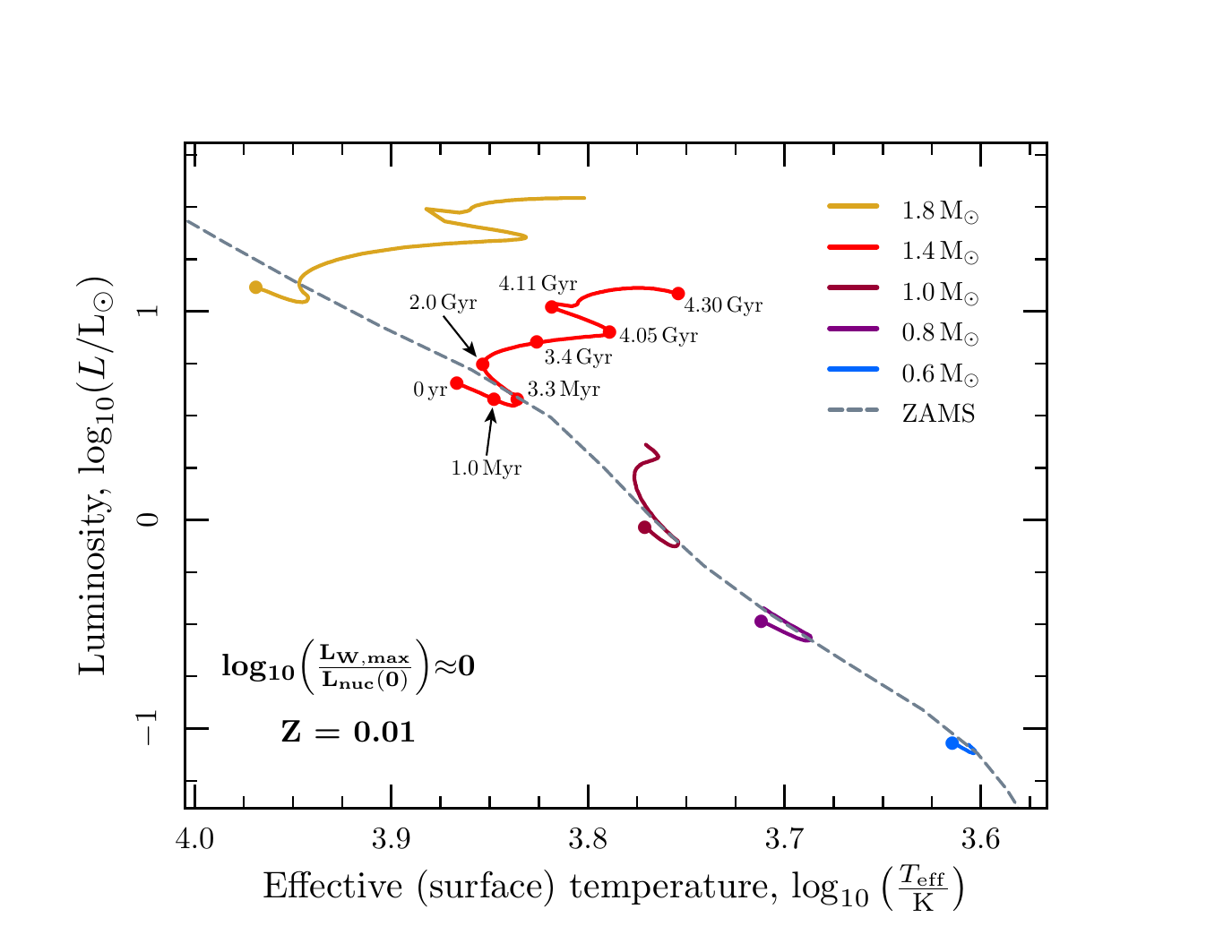}
\end{minipage}
\hspace{0.8\columnsep}
\begin{minipage}{\columnwidth}
\includegraphics[width=\columnwidth, trim = 0 0 0 30, clip=true]{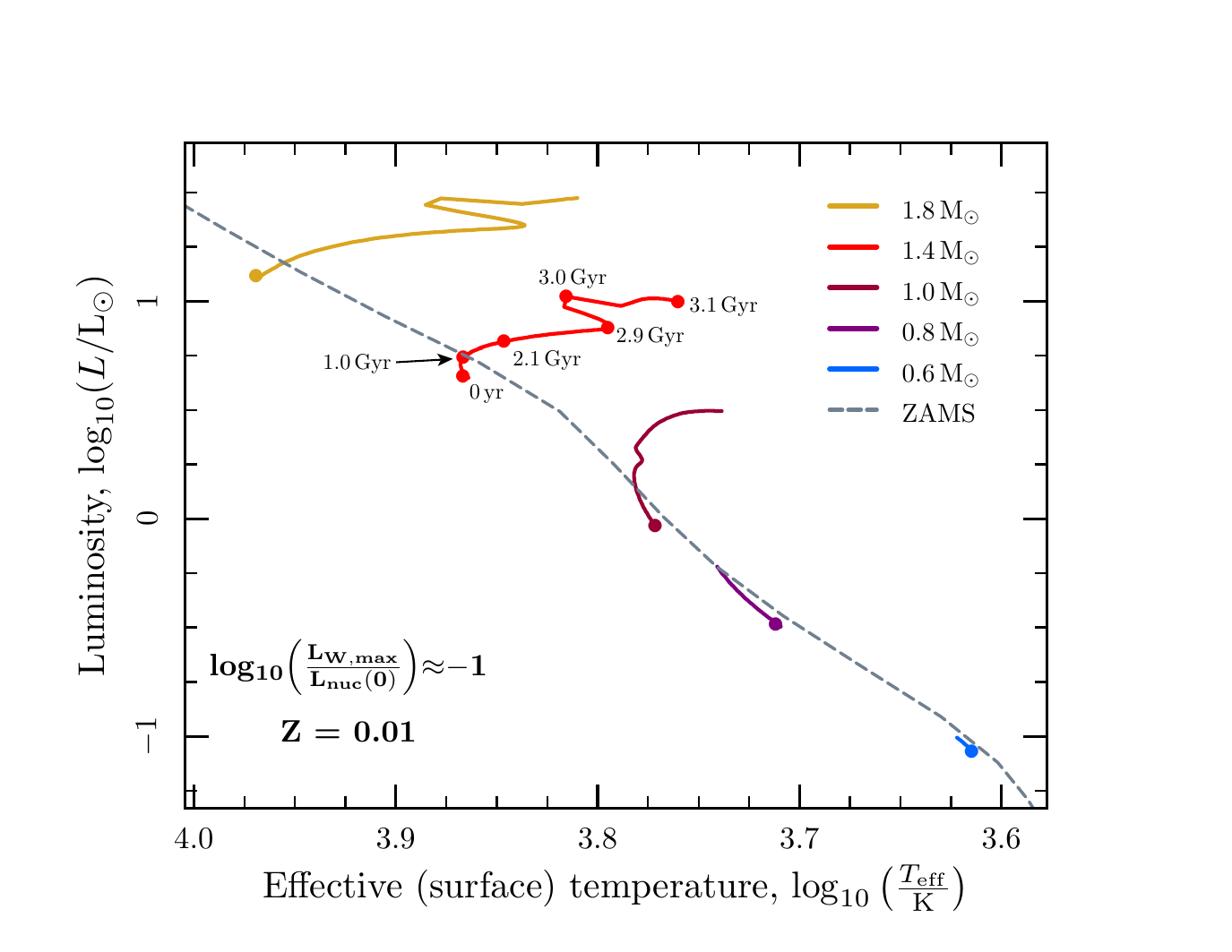}
\includegraphics[width=\columnwidth]{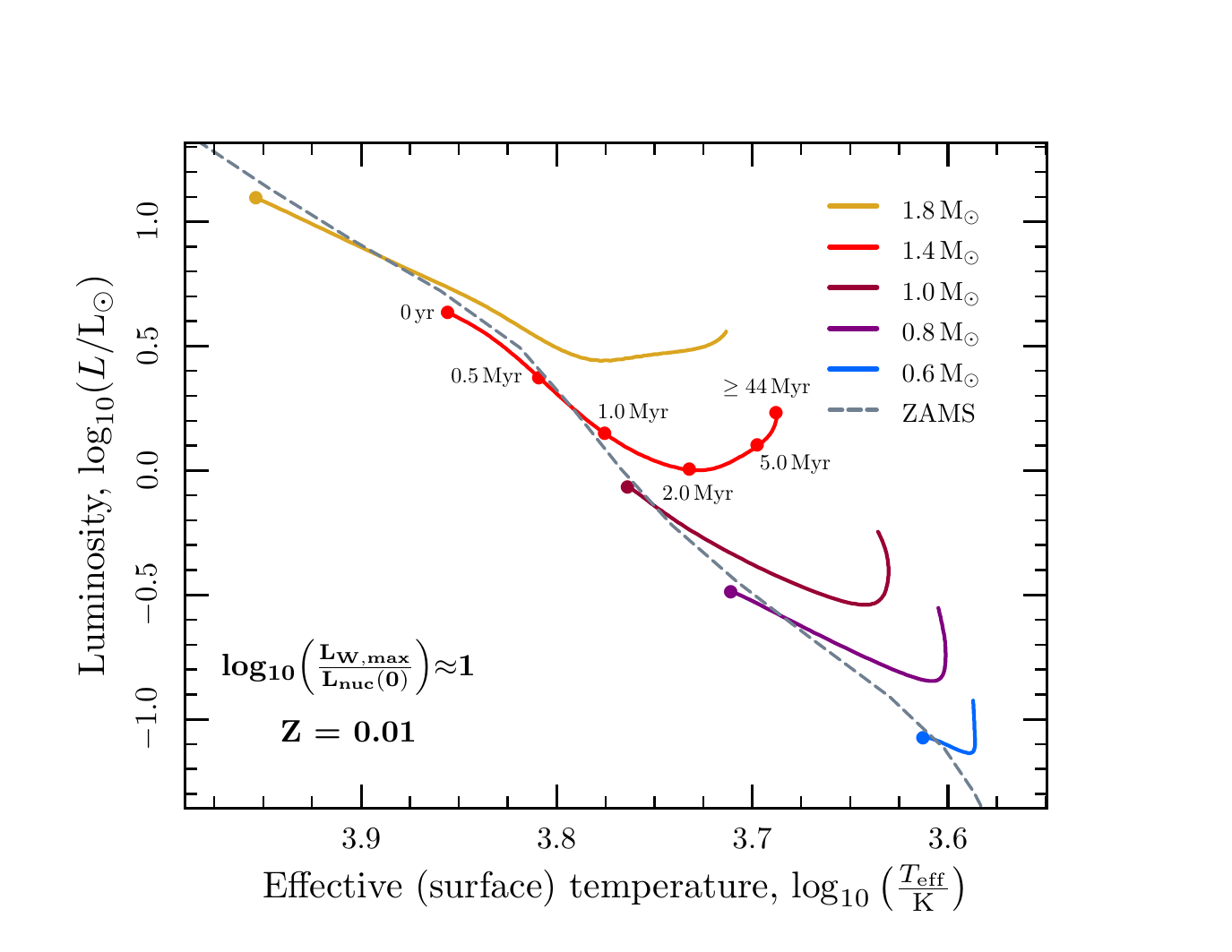}
\end{minipage}
\end{center}
\caption{Evolutionary tracks followed in the HR diagram by stars of various masses, when WIMPs provide different fractions of their total energy budgets.  Filled, unlabelled circles indicate the starting points of tracks, whilst labelled ones give indicative ages during the evolution of 1.4\,M$_\odot$ stars.  Tracks have been halted when the star exhausts the supply of hydrogen in its core or reaches the current age of the universe.  Stars with a greater luminosity contribution from WIMPs push further up the Hayashi track and spend longer there before returning to the main sequence.  Stars which come to be entirely dominated by WIMP annihilation (\emph{bottom right}) evolve quickly back up the Hayashi track and halt, holding their position in the HR diagram well beyond the age of the universe.}
\label{fig3}
\end{figure*}

\begin{figure*}
\begin{center}
\begin{minipage}{\columnwidth}
\includegraphics[width=\columnwidth, trim = 0 0 0 30, clip=true]{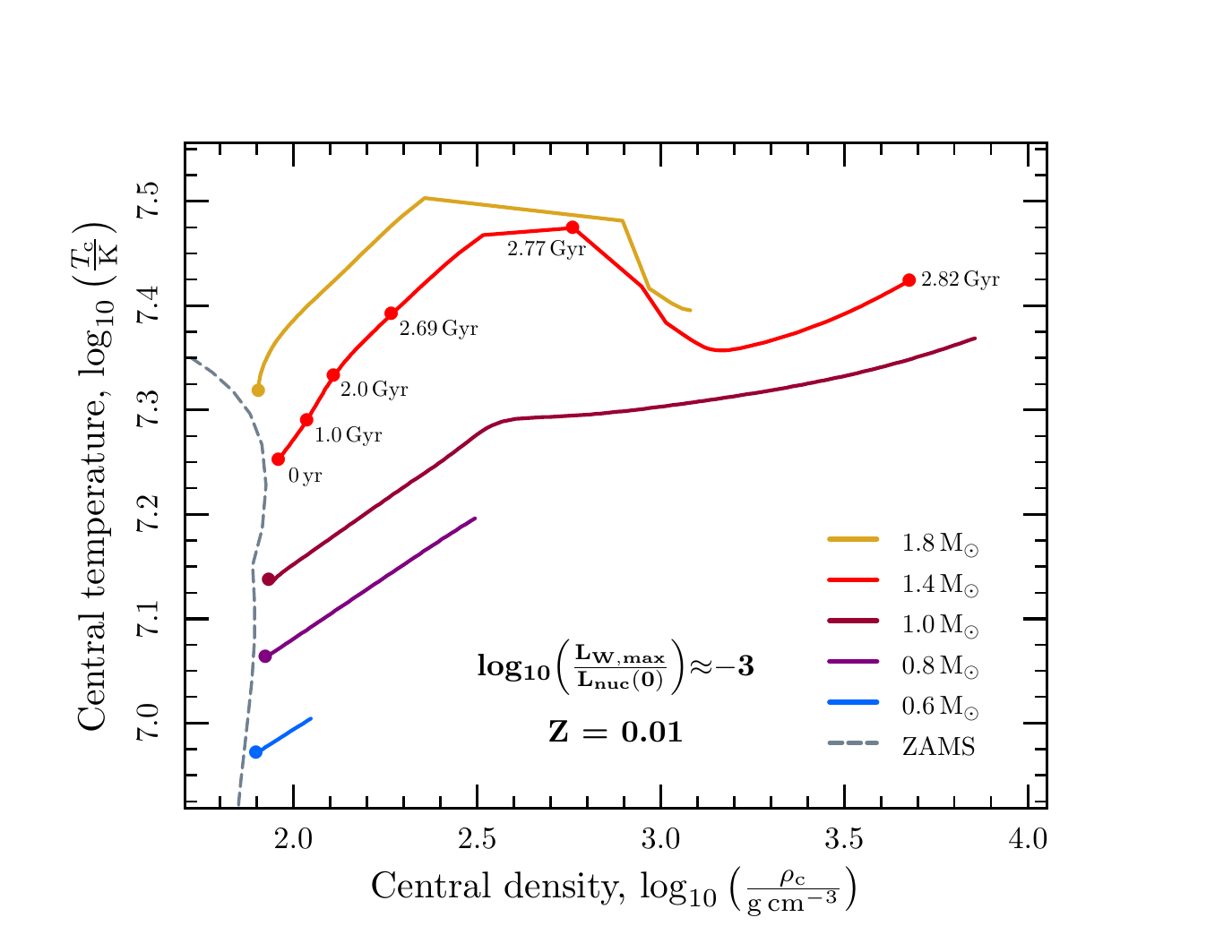}
\includegraphics[width=\columnwidth]{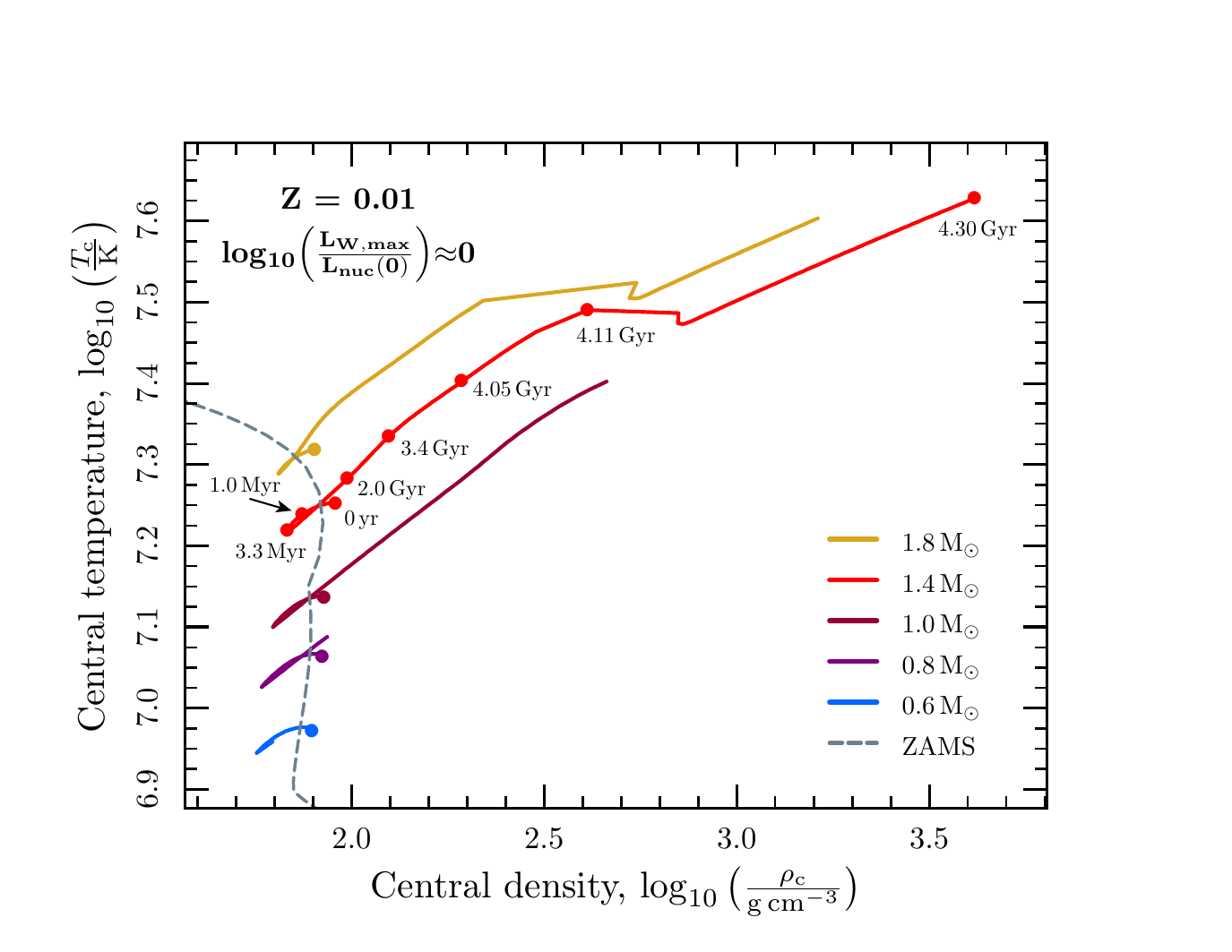}
\end{minipage}
\hspace{0.8\columnsep}
\begin{minipage}{\columnwidth}
\includegraphics[width=\columnwidth, trim = 0 0 0 30, clip=true]{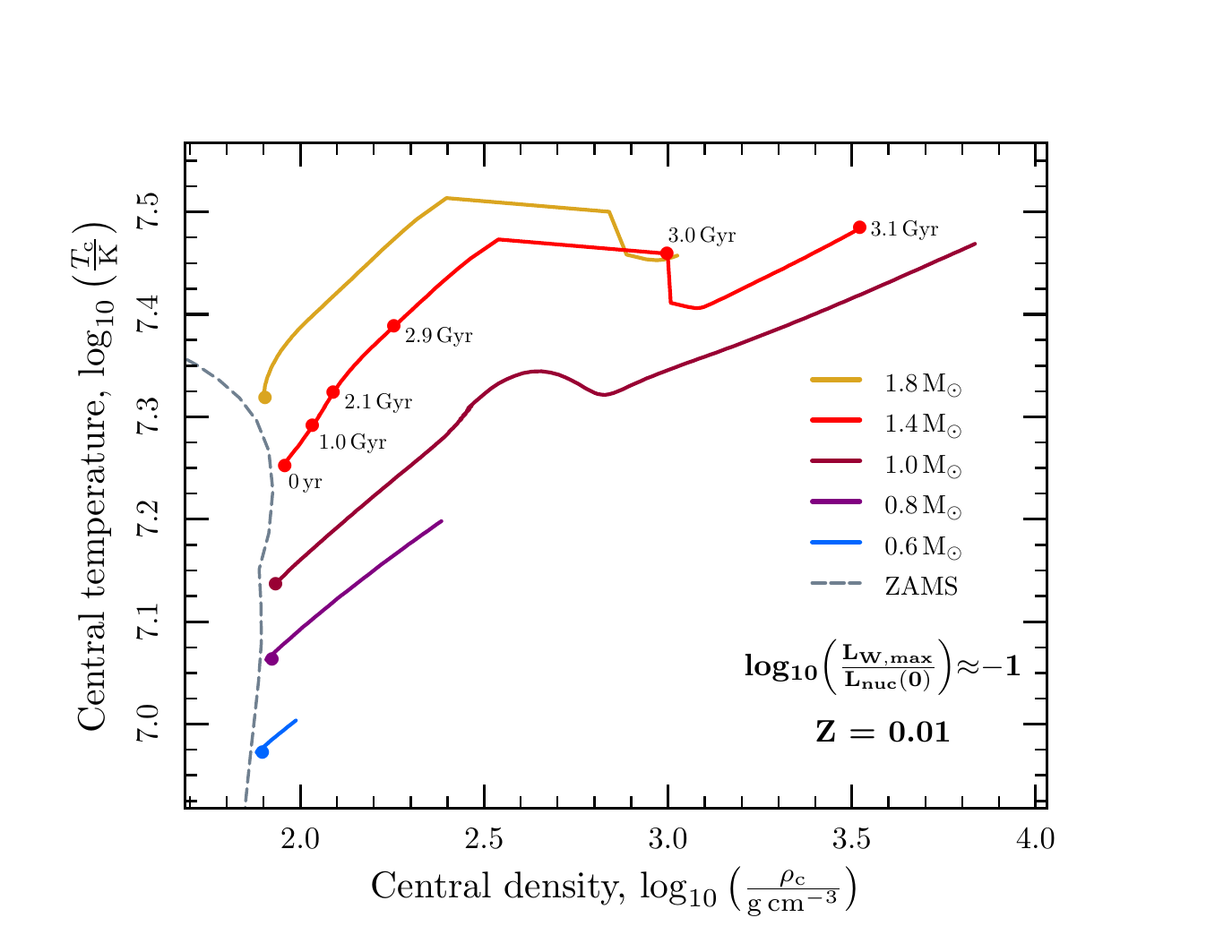}
\includegraphics[width=\columnwidth]{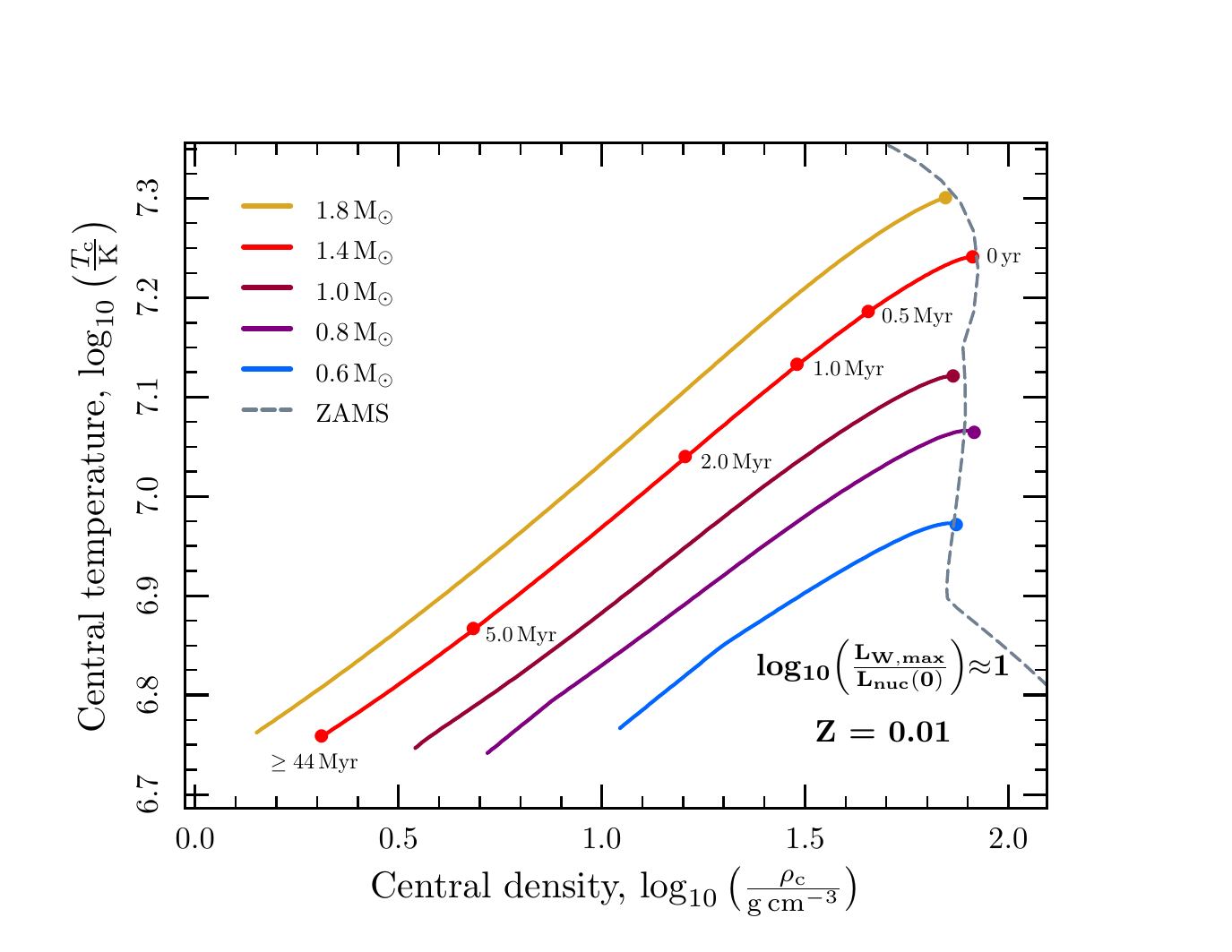}
\end{minipage}
\end{center}
\caption{Evolutionary tracks followed in the central equation-of-state diagram by the stars of Fig.~\protect\ref{fig3}.  Filled circles indicate the starting points of the tracks.  Dashed lines show the location at which hydrogen burning becomes the dominant energy source, which defines the zero-age main sequence (ZAMS).  To the bottom-left of this line, the core is too cool and diffuse to support the star by nuclear burning alone.  Stars with a greater luminosity contribution from WIMPs push further into this region as they ascend the Hayashi track and their cores cool and expand, and thus take longer to recontract and return to the main sequence.  The slight departure from smoothness apparent in some curves is simply due to finite temporal resolution of models.}
\label{fig4}
\end{figure*}

\section{Benchmark impacts on main sequence stars}
\label{ms}

To understand the general effects of dark matter accretion and annihilation upon main-sequence stars, we start by considering a set of benchmark stars in a reference halo of WIMPs. In later sections, we will expand on these results for more realistic scenarios. For the benchmark stars, we evolved a grid of models with $0.3\,\mathrm{M}_\odot\le M_\star \le 2\,\mathrm{M}_\odot$, $0.0003\le Z\le0.02$ and $5\le\log_{10}(\frac{\rho_\chi}{\mathrm{GeV\,cm}^{-3}})\le 11$.  The models were started from the zero-age main sequence (ZAMS), and evolved until one of the following stopping criteria was met:
\begin{enumerate}
  \item{The star left the main sequence (as indicated by the central hydrogen mass fraction $X_\mathrm{c}$ dropping below 10$^{-6}$).}
  \item{The star reached a stable equilibrium where all its energy was effectively provided by WIMP annihilation (as indicated by $X_\mathrm{c}$, $\log_{10}\rho_ \mathrm{c}$ and $\log_{10} T_\mathrm{c}$ changing by less than 10$^{-14}\frac{M_\star}{\mathrm{M}_\odot}$, 10$^{-10}$ and 10$^{-10}$ respectively over four consecutive timesteps).}
  \item{The age of the star exceeded the age of the Universe.}
\end{enumerate}
Except for main-sequence lifetimes, we present results at $Z=0.01$ and just give a brief discussion of the effects of metallicity in the text, since most of the properties we discuss did not show any major dependency upon metallicity.

For this grid we used the standard isothermal velocity distribution, with the default solar values of $v_\star=v_\odot=220$\,km\,s$^{-1}$ and $\bar{v}=\sqrt{3/2}v_\odot=270$\,km\,s$^{-1}$.   The resultant capture rates (at $t=0$) and ratios of annihilation-to-nuclear luminosity are presented in Fig.~\ref{fig1}.  For the sake of comparison, nuclear luminosities $L_\mathrm{nuc}$ are taken at zero age, whilst WIMP annihilation luminosities $L_\mathrm{W,max}$ are the maximum values the star achieves during its evolution.  Since stars start with almost no WIMPs, WIMP luminosity at zero age is essentially nil, and nuclear luminosity changes significantly after zero age as the WIMPs begin to influence the stellar structure.  As expected, capture rates and WIMP-to-nuclear luminosity ratios increase linearly with $\rho_\chi$, and lower-mass stars capture less but burn a greater ratio of WIMPs to nuclear fuel than their higher-mass counterparts.  Significantly, WIMP annihilation outstrips nuclear burning in a large area of the parameter space.  Capture rates increase slightly at lower metallicity because of the dominance of spin-dependent scattering and capture by hydrogen, but are outweighed by the increased nuclear luminosity, causing a small decrease in WIMP-to-nuclear burning ratios.

\begin{figure}
\begin{center}
\includegraphics[width=\columnwidth, trim = 0 0 0 30, clip=true]{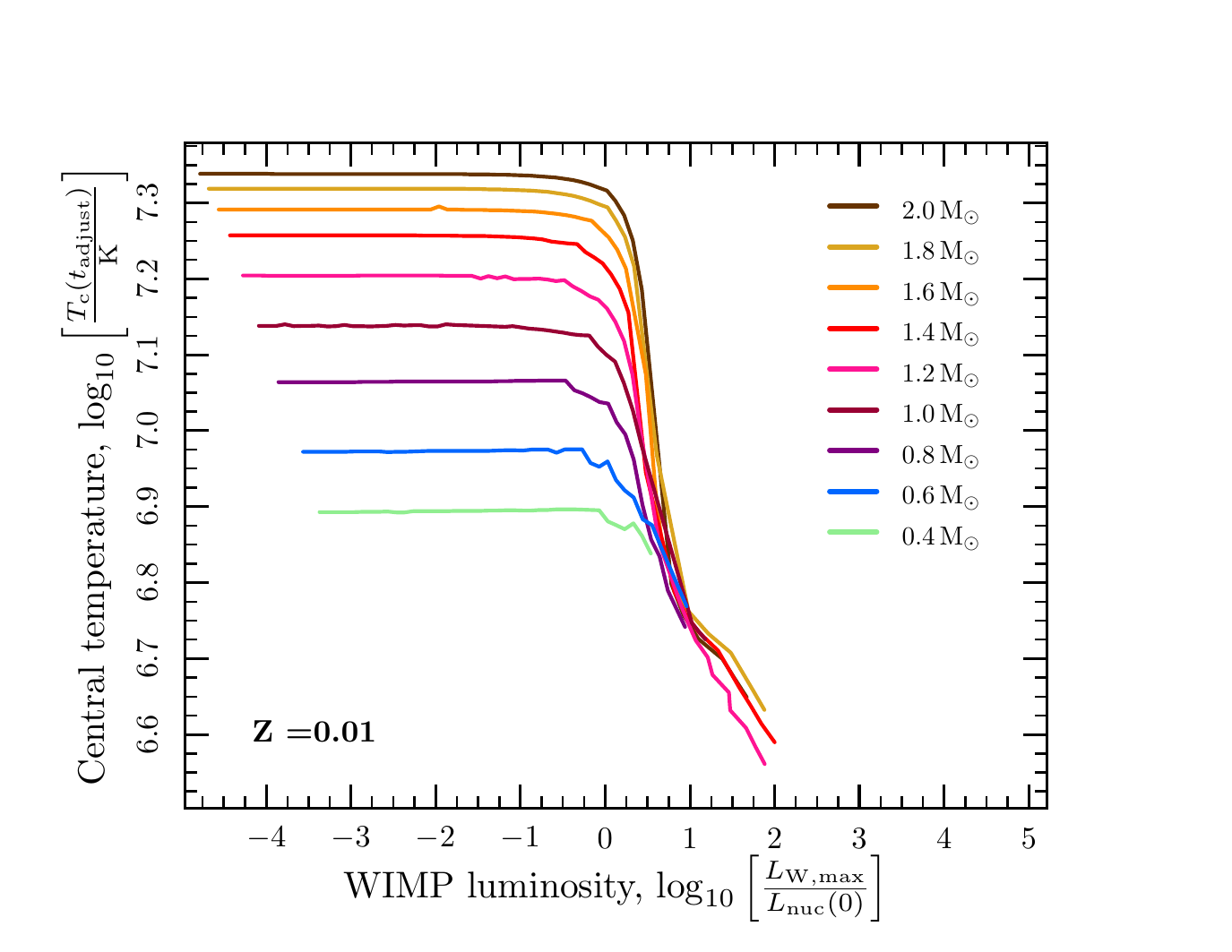}
\includegraphics[width=\columnwidth]{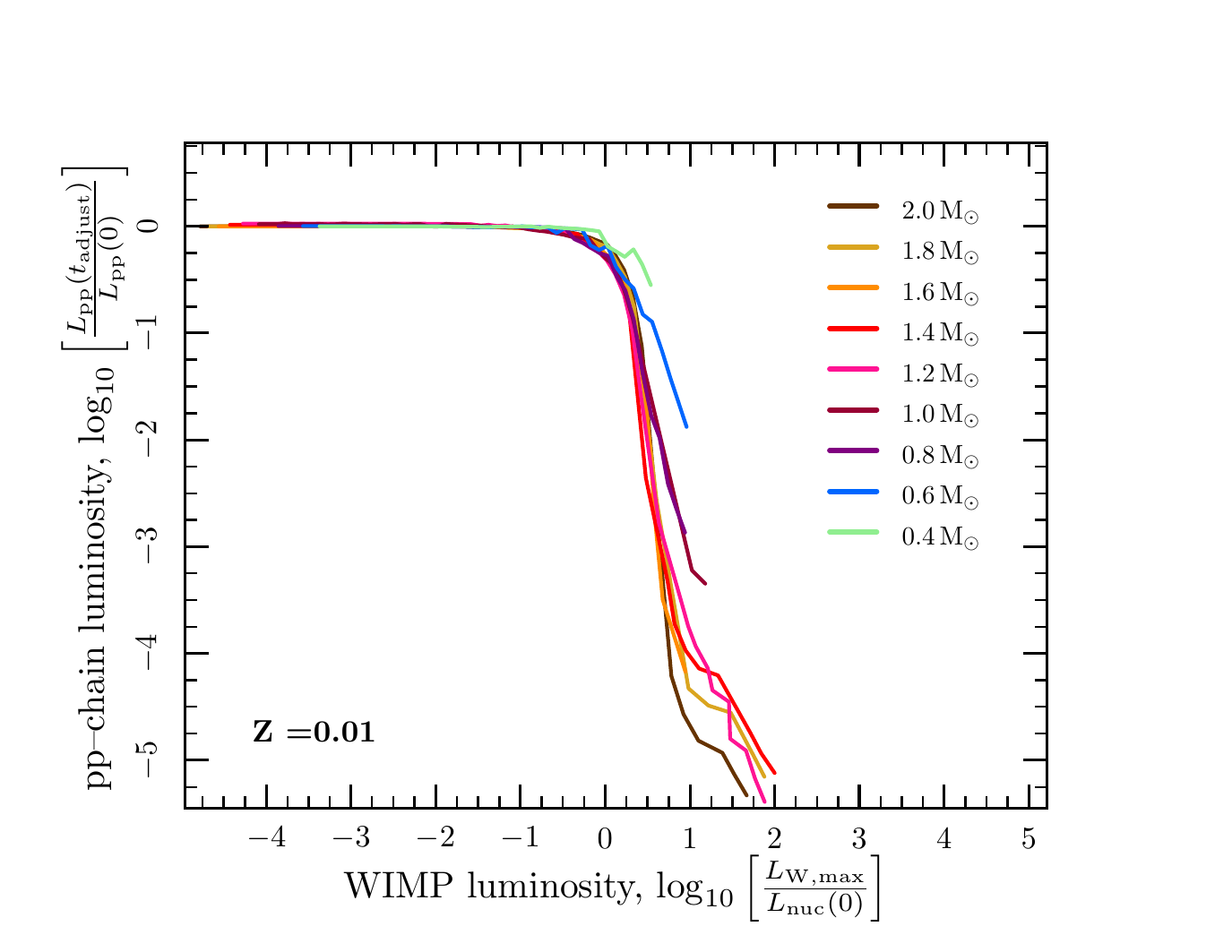}
\includegraphics[width=\columnwidth]{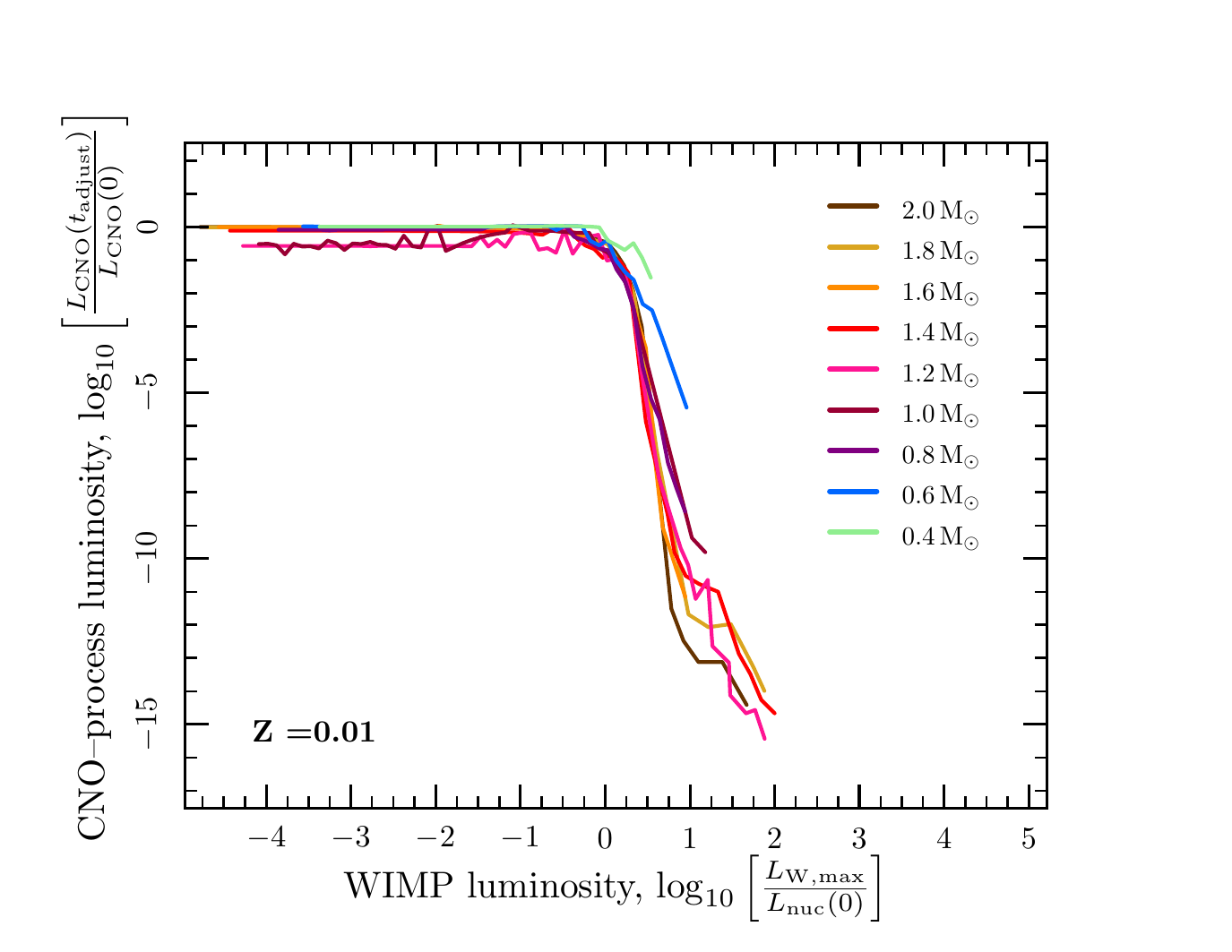}
\end{center}
\caption{Central temperatures (top) and total luminosities provided by hydrogen fusion via the pp-chain (middle) and CNO-process (bottom), as a function of the luminosity provided by WIMP burning.  Central temperatures and hydrogen-burning luminosities are as measured at $t_\mathrm{adjust}$, the point at which a star has just completed its initial adjustment to the presence of WIMPs in its core.  This corresponds to the time at which models' evolutionary paths have reached their bottom-leftmost points in Fig.~\ref{fig4}.  Rates of energy production from hydrogen burning are expressed as fractions of their initial values.}
\label{fig5}
\end{figure}

Because the primary factor governing the impact of WIMPs upon stellar evolution is simply $N(t)$, the benchmark results we present in this section will hold in general for other combinations of input particle and halo parameters, subject to an appropriate rescaling.  In particular, all scenarios which result in the same product of the capture rate and WIMP mass map to roughly the same ratio of WIMP-to-nuclear burning, which in turn maps in an essentially one-to-one manner to all physical changes in a star's structure and evolution.  (That is, ignoring `higher-moment' factors like WIMP thermalisation, distribution, conductive energy transport and capture-annihilation equilibration).  In this way, Fig.~\ref{fig1} acts as a conversion table between capture rates, WIMP luminosities and equivalent dark matter densities in the RSC.  (We use the term `WIMP luminosity' as shorthand for the ratio $\frac{L_\mathrm{W,max}}{L_\mathrm{nuc}(0)}$ when it is clear what we mean, or when the distinction is irrelevant).

Lines in Fig.~\ref{fig1} do not all extend to $\log_{10}\rho_\chi=11$.  This is because as WIMP luminosity becomes a more significant contributor to a star's energy generation, the stellar models become steadily more difficult to converge.  In many cases, we had to very carefully adjust the initial timesteps in order to converge models near the ends of tracks in Fig.~\ref{fig1}.  In some cases we either could not obtain initial convergence or could not properly maintain it until one of the criteria above was met.  Results from such models were discarded.

We performed control calculations on a single solar-mass star with different WIMP masses and annihilation cross-sections (Fig.~\ref{fig2}).  When capture and annihilation have equilibrated in a star, the WIMP luminosity effectively depends on the capture rate alone, which is independent of the annihilation cross-section (cf.~Eqs.~\ref{standardcap}, \ref{heavystandardcap} and \ref{Hstandardcap}), so we see that $\langle\sigma_\mathrm{a}v\rangle_0$ makes no difference to the amount of energy generated.  When equilibrium has not been achieved, this will not be the case.  Such an effect can be seen in the slight upturn of the WIMP luminosity in the uppermost curve of the upper panel in Fig.~\ref{fig2}.  In this case the very high $\rho_\chi$ and very low $\langle\sigma_\mathrm{a}v\rangle_0$ significantly change the stellar structure before equilibrium has been reached, causing $L_\mathrm{W}$ to peak prior to equilibrium.   The dependence of $L_\mathrm{W}$ on $m_\chi$ in Fig.~\ref{fig2} shows roughly an inverse square relationship, which is a result of using the full capture expressions in the RSC.  As can be seen from careful inspection of Eq.~\ref{Hstandardcap} for example, for small $\bar{v}$ and $v_\star$ such as those used in the context of the early universe by e.g. \citet{Iocco08a} and \citet{Freese08a}, the dependence disappears.

In Figs.~\ref{fig3} and \ref{fig4} we show evolutionary tracks in the HR and central equation-of-state diagrams of stars with different masses and WIMP luminosities.  At low WIMP luminosities, the evolution is essentially normal.  As WIMPs are allowed to provide more energy, the negative heat capacity of a star causes it to expand and cool.  The central temperature and density drop, nuclear burning reduces and the star moves some distance back up the Hayashi track.  The reduction in central temperatures and overall luminosities provided by pp-chain and CNO-process hydrogen burning are illustrated in Fig.~\ref{fig5}.  These values are taken at the time $t_\mathrm{adjust}$ when a star has completed its initial reaction to the presence of WIMPs, which corresponds to the central temperature and density reaching their minima and the star arriving at the bottom-leftmost point of its travels in Fig.~\ref{fig4}.  At very high WIMP luminosities, the stellar core expands and cools drastically, moving stars a long way back along the pre-main sequence and effectively shutting down nuclear burning all together.  Such an object becomes a fully-fledged dark star, powered entirely and perpetually by WIMP annihilation.

\begin{figure*}
\begin{center}
\begin{minipage}{\columnwidth}
\includegraphics[width=\columnwidth, trim = 0 0 0 30, clip=true]{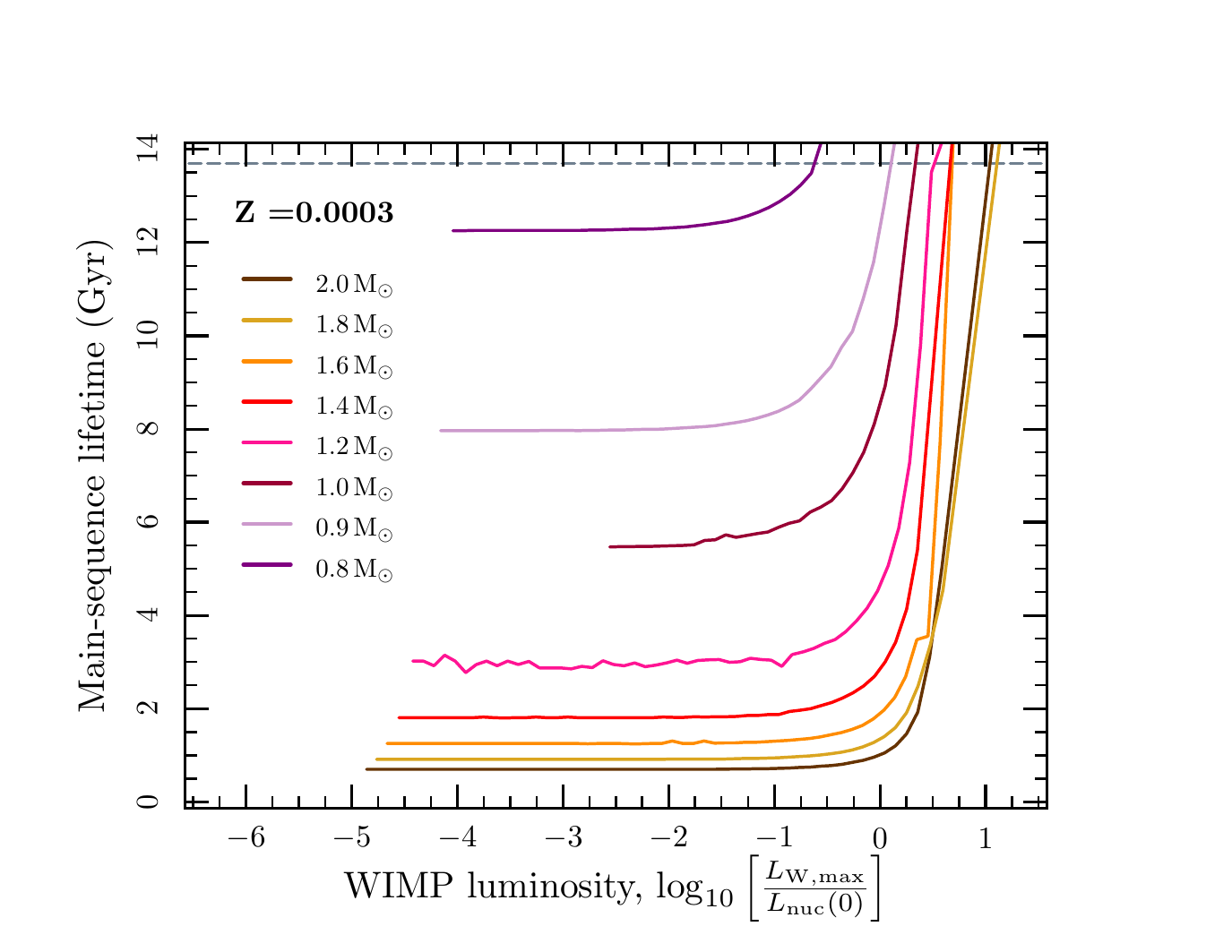}
\includegraphics[width=\columnwidth]{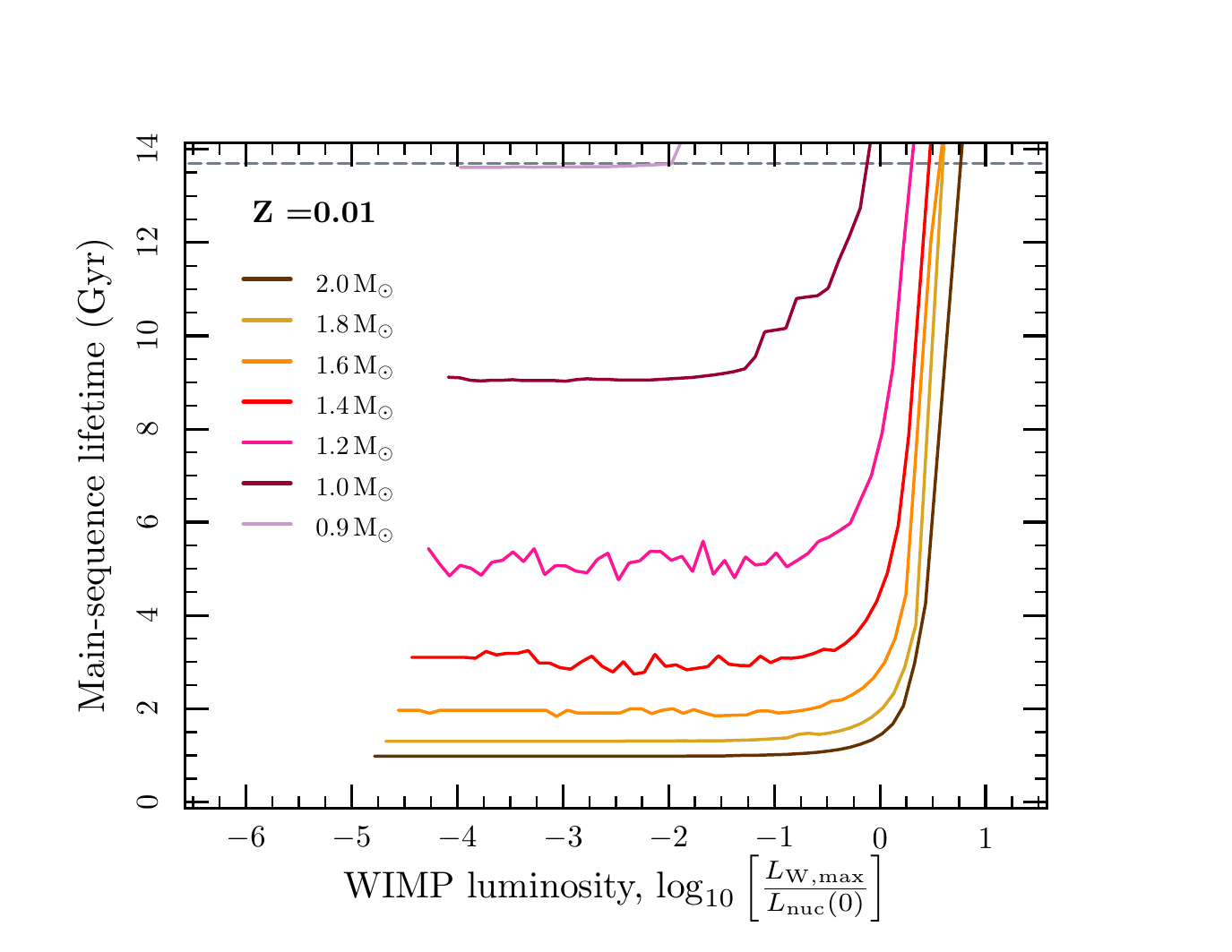}
\end{minipage}
\hspace{0.8\columnsep}
\begin{minipage}{\columnwidth}
\includegraphics[width=\columnwidth, trim = 0 0 0 30, clip=true]{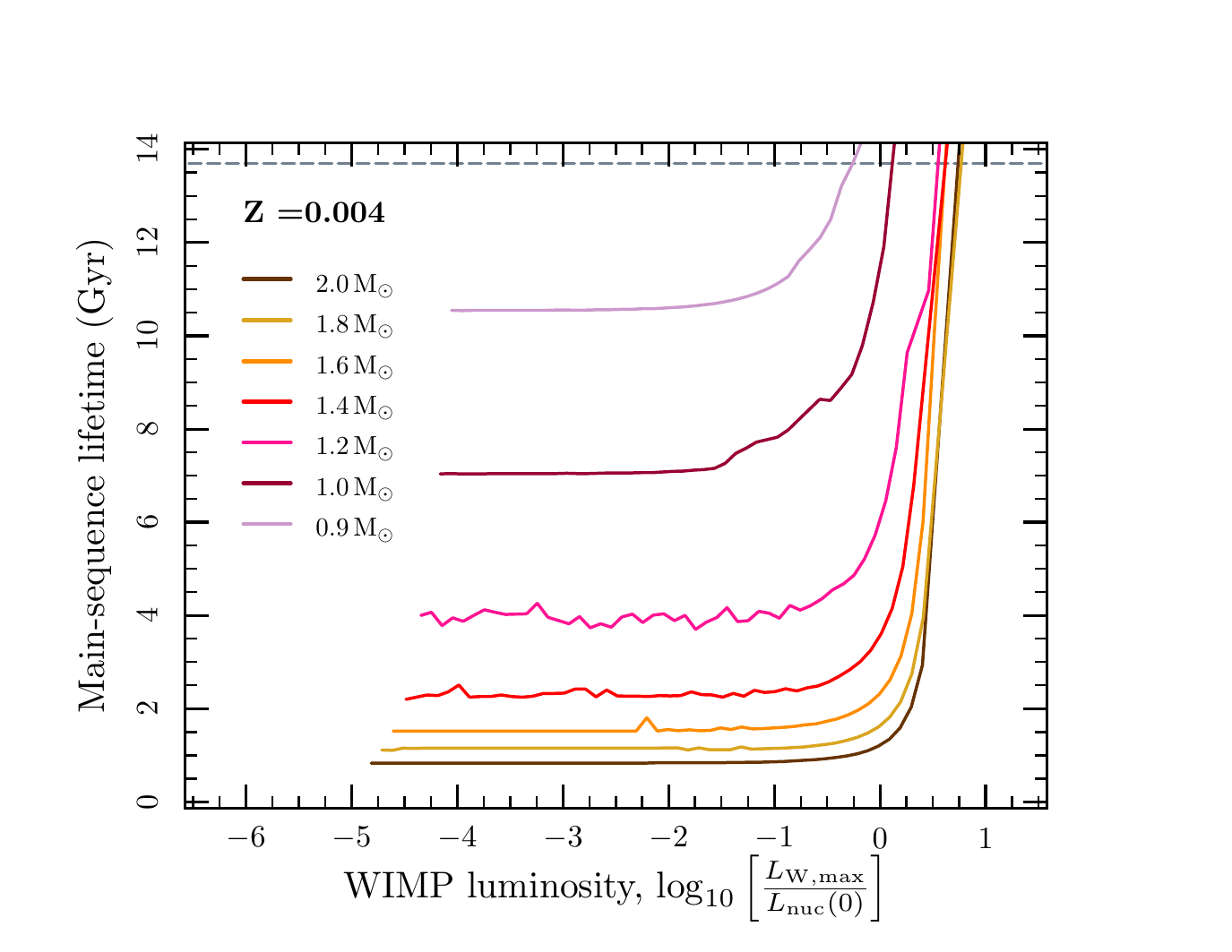}
\includegraphics[width=\columnwidth]{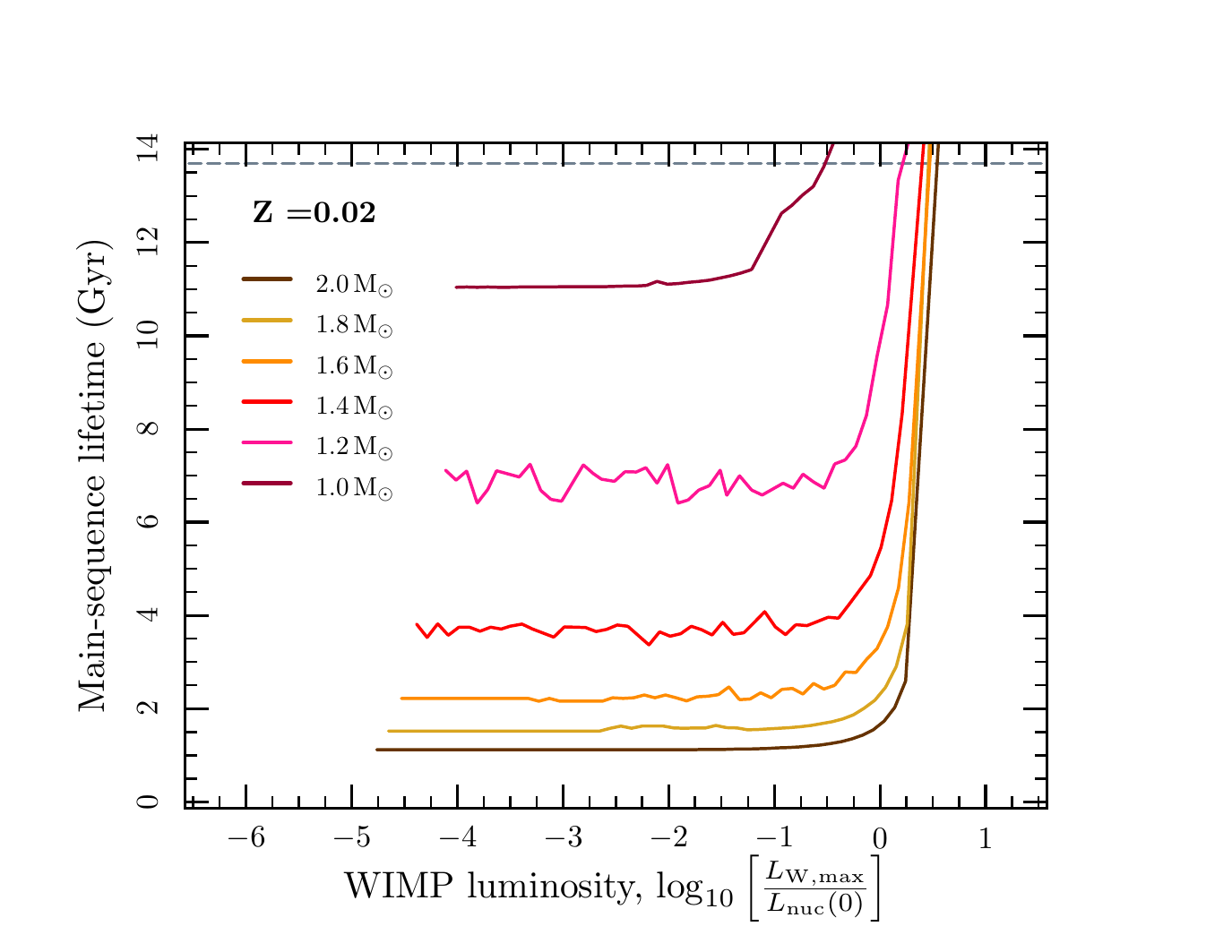}
\end{minipage}
\end{center}
\caption{Main-sequence lifetimes of stars with differing masses, metallicities and WIMP luminosities.  Dashed lines indicate the present age of the universe.  Stars with a greater fraction of their energy provided by WIMPs consume hydrogen more slowly, so spend more time on the main sequence.  For a given WIMP-to-nuclear burning ratio, the lifetimes of stars with lower masses and metallicities are more affected than their more massive, metal-rich counterparts.}
\label{fig6}
\end{figure*}

At intermediate WIMP luminosities, nuclear burning is suppressed rather than completely extinguished.  Its continued contribution to nuclear processing slowly raises the core temperature and density once more, in turn increasing the rate of nuclear reactions and accelerating the process.  The star burns hydrogen alongside WIMPs, and goes on to evolve through a hybrid WIMP-hydrogen main sequence.  Such evolution can be best seen in the bottom-left panel of Fig.~\ref{fig3}. Thanks to the energy input from WIMP annihilation, the time it takes such a star to consume its core hydrogen is lengthened, so its effective main-sequence lifetime is extended (Fig.~\ref{fig6}).  The increase in main-sequence lifetime is notable at all metallicities, but most prominent at low $Z$, essentially because normal main-sequence lifetimes are shorter at lower metallicity.  We did not see changes with metallicity in the central temperatures, pp-chain or CNO luminosities of the stars in our grid.

We should point out here that in the extreme case of a very large WIMP luminosity, it is highly questionable whether a star would have ever reached the main sequence at all, or if it might have simply halted during its initial descent of the Hayashi track.  The same might even be true of stars with intermediate WIMP luminosities, since it is not clear whether nuclear burning will win out over WIMP annihilation at exactly the same ages and capture rates when a star is evolved from the main sequence as when it is evolved from the pre-main sequence.  This behaviour has been seen explicitly by \citet{Iocco08b}.  We strongly suspect that the solutions we find for high WIMP luminosities are the same as those obtained with models begun from the pre-main sequence by \citet{Iocco08b}.  We are currently investigating this question in detail.

The greater influence of WIMP capture and annihilation upon low-mass stars is strongly apparent in Fig.~\ref{fig6}.  Even if WIMPs supply only a tenth the energy of nuclear burning, the lifetime of a 0.8\,M$_\odot$ star is increased by almost a billion years.  With the same ratio of WIMP-to-nuclear burning, the lifetime of a 2.0\,M$_\odot$ star is unchanged.  Considering that according to Fig.~\ref{fig1}, roughly an order of magnitude more dark matter is required for a 2.0\,M$_\odot$ star to even achieve the same WIMP-to-nuclear burning ratio as a 0.8\,M$_\odot$ star, lower stellar mass is clearly a highly favourable property in the observational search for dark stars.

\begin{figure*}
\begin{center}
\begin{minipage}{0.55\columnwidth}
\includegraphics[width=\columnwidth]{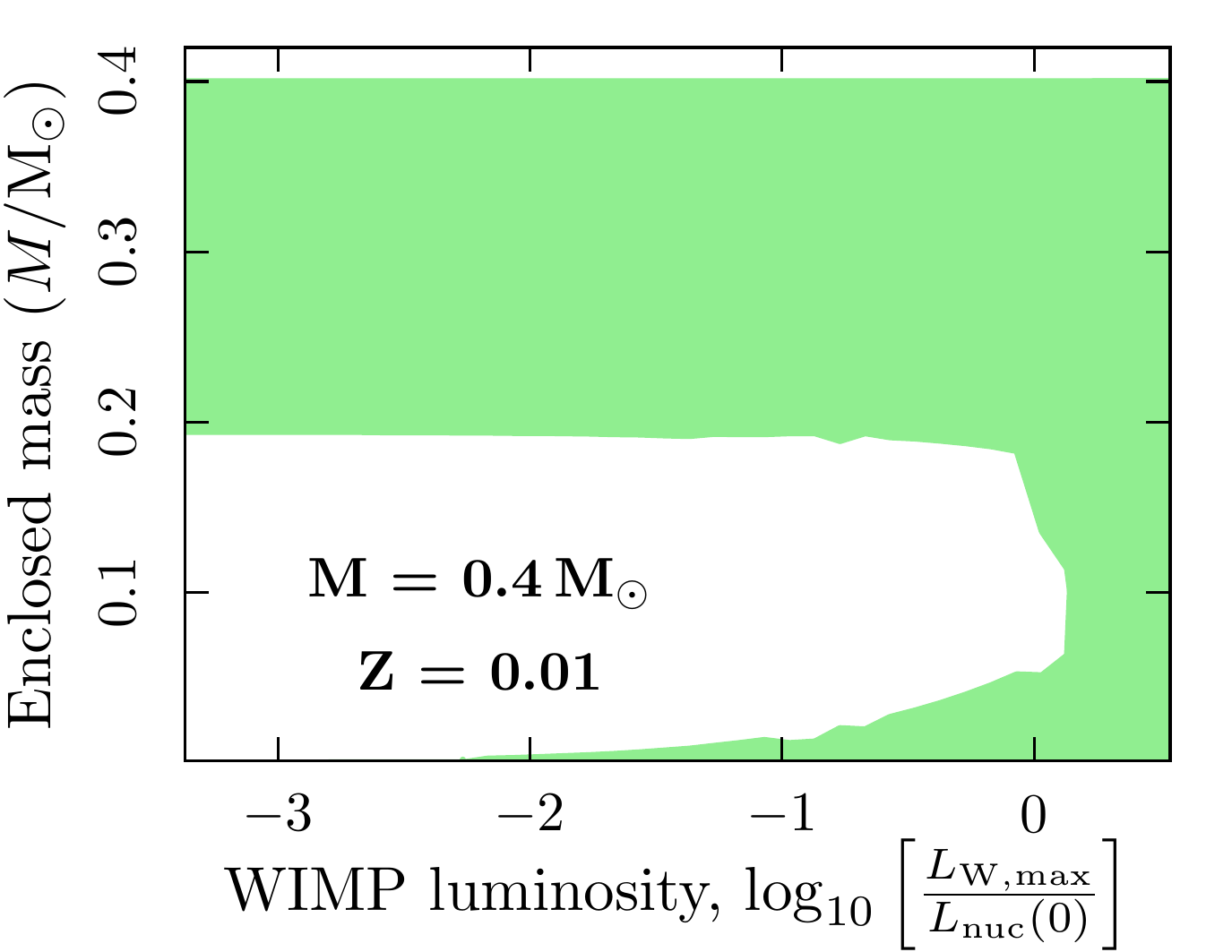}
\includegraphics[width=\columnwidth]{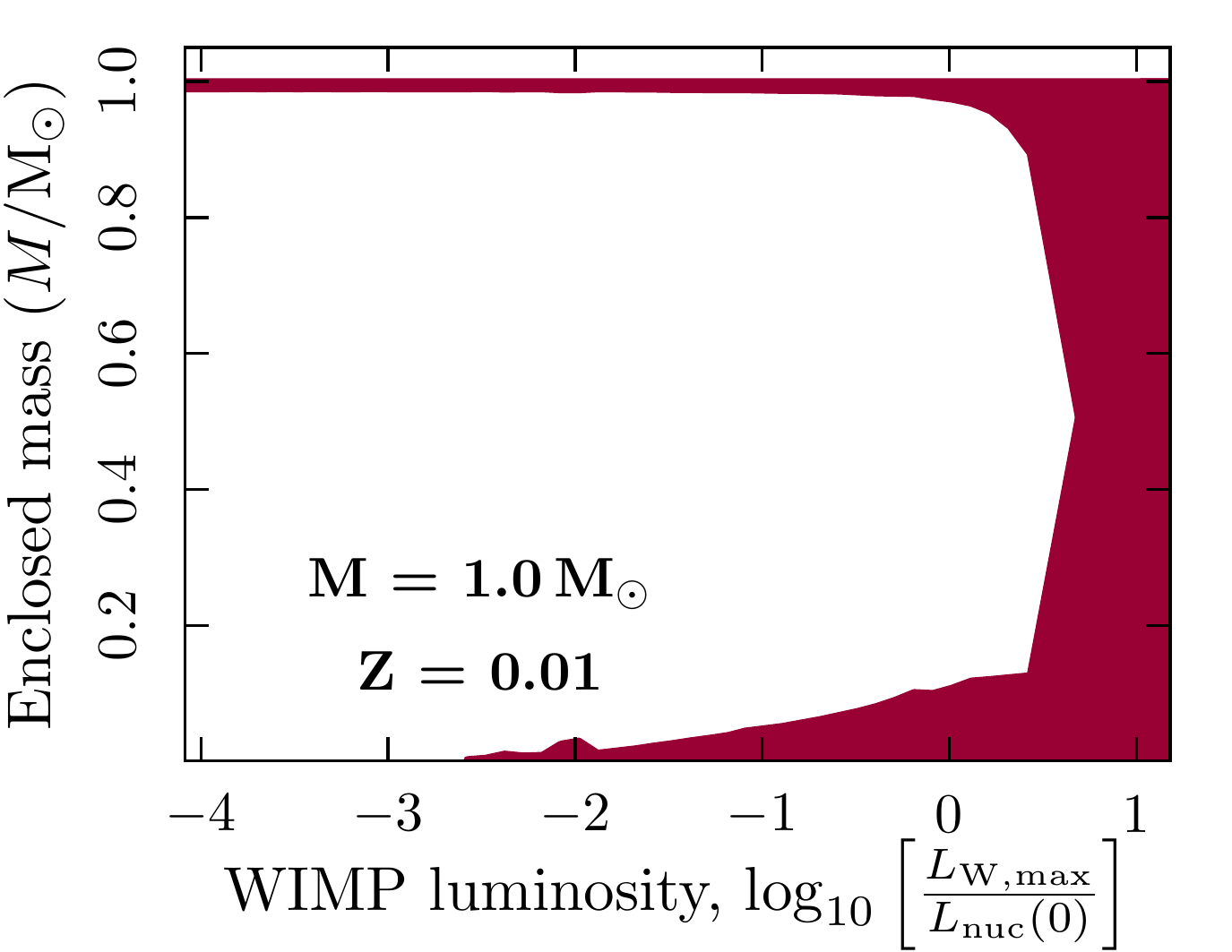}
\includegraphics[width=\columnwidth]{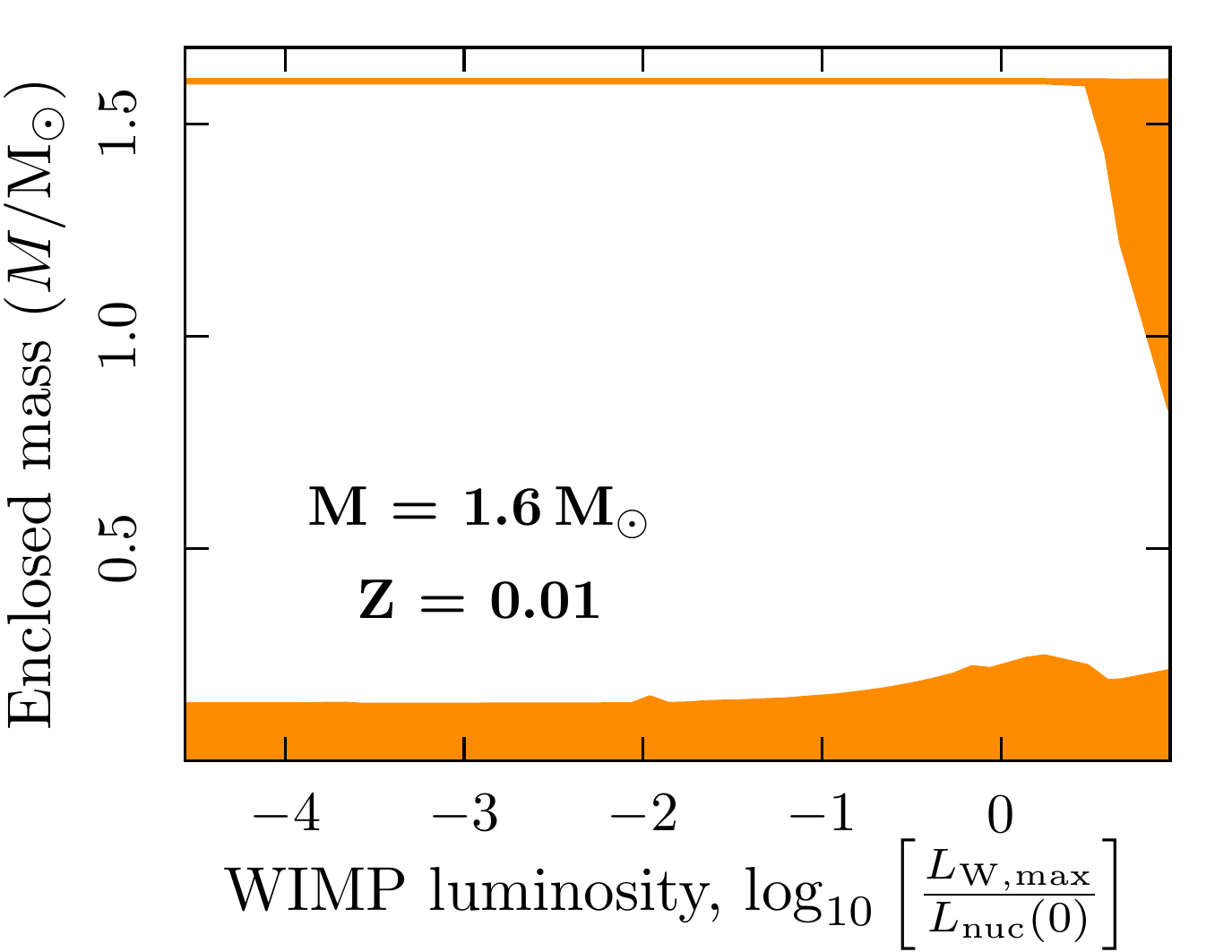}
\end{minipage}
\hspace{0.5\columnsep}
\begin{minipage}{0.55\columnwidth}
\includegraphics[width=\columnwidth]{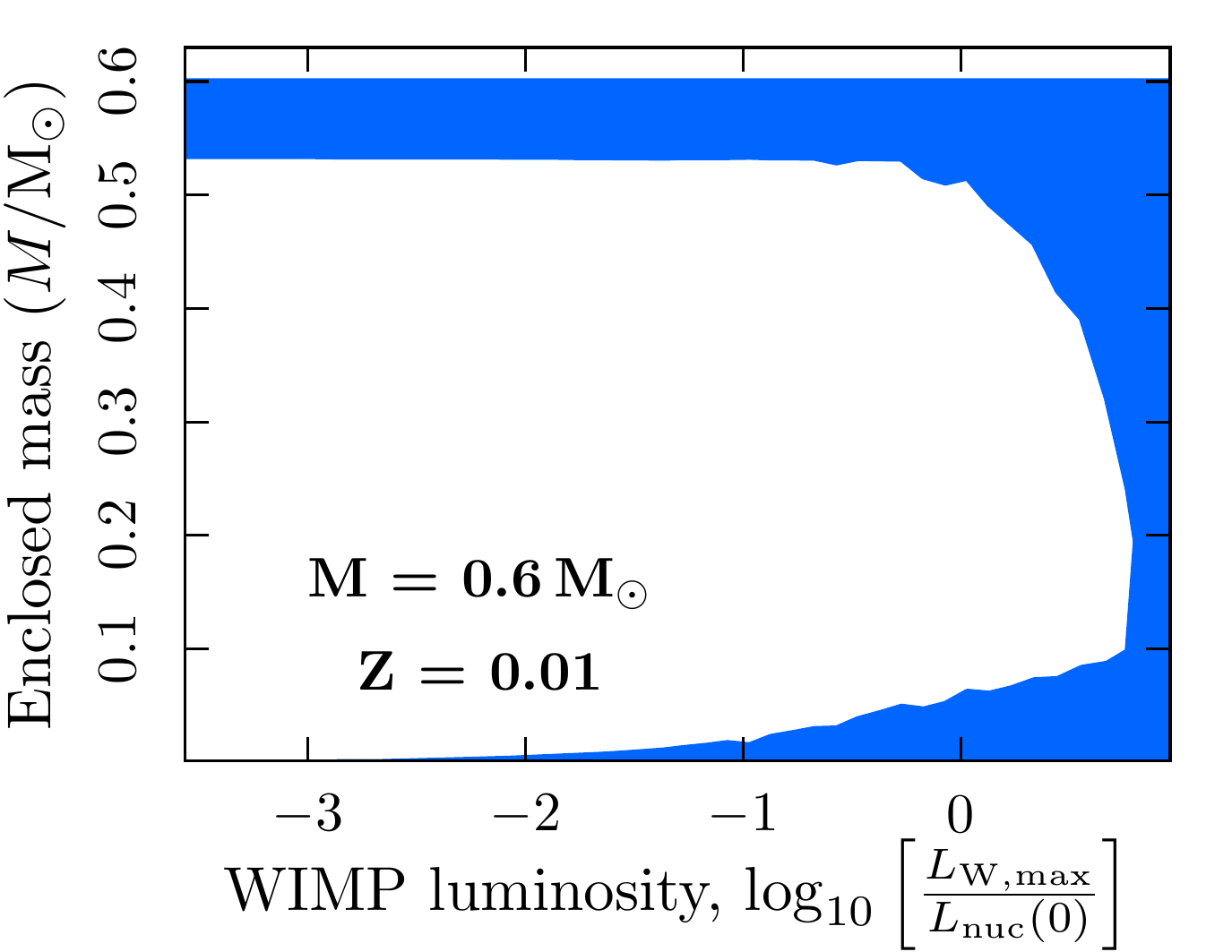}
\includegraphics[width=\columnwidth]{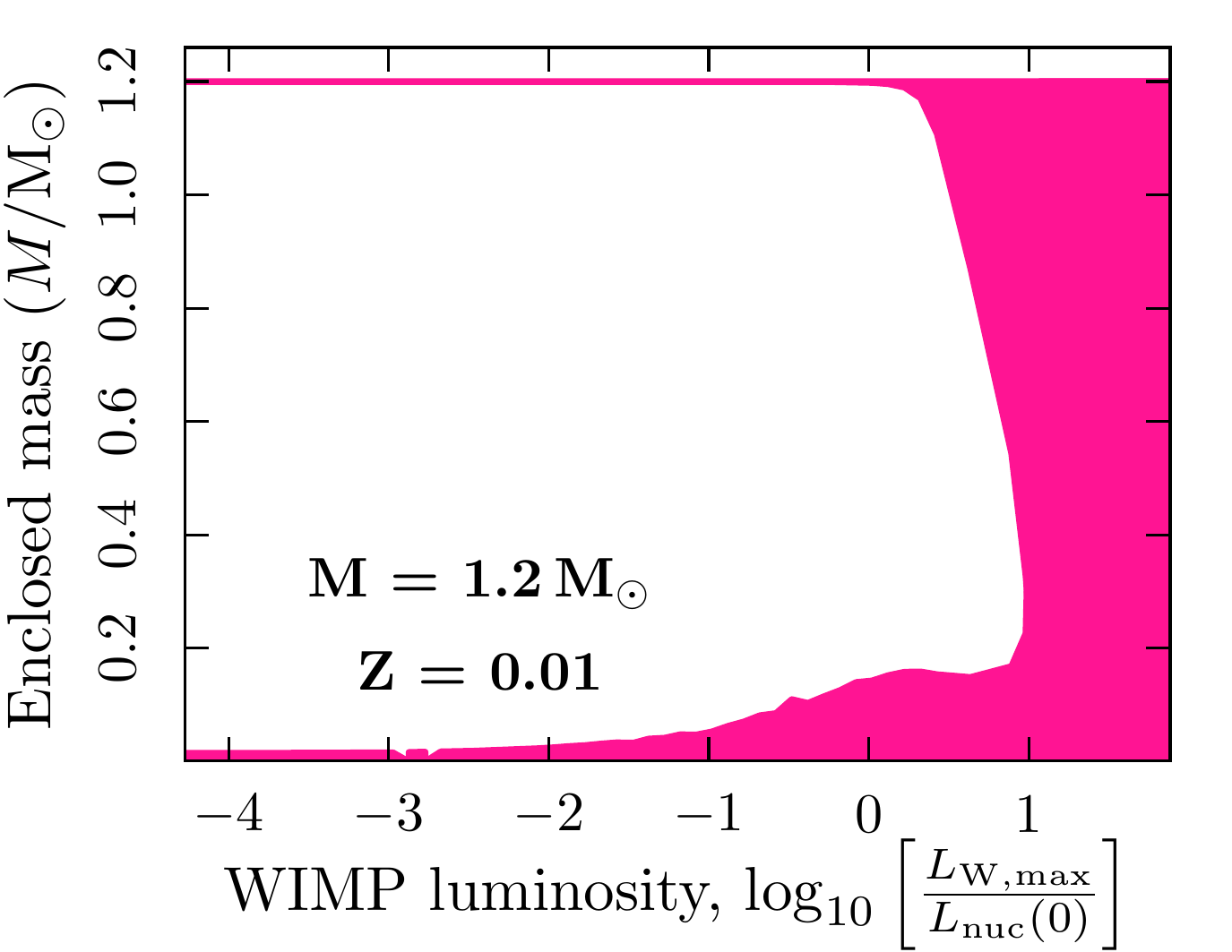}
\includegraphics[width=\columnwidth]{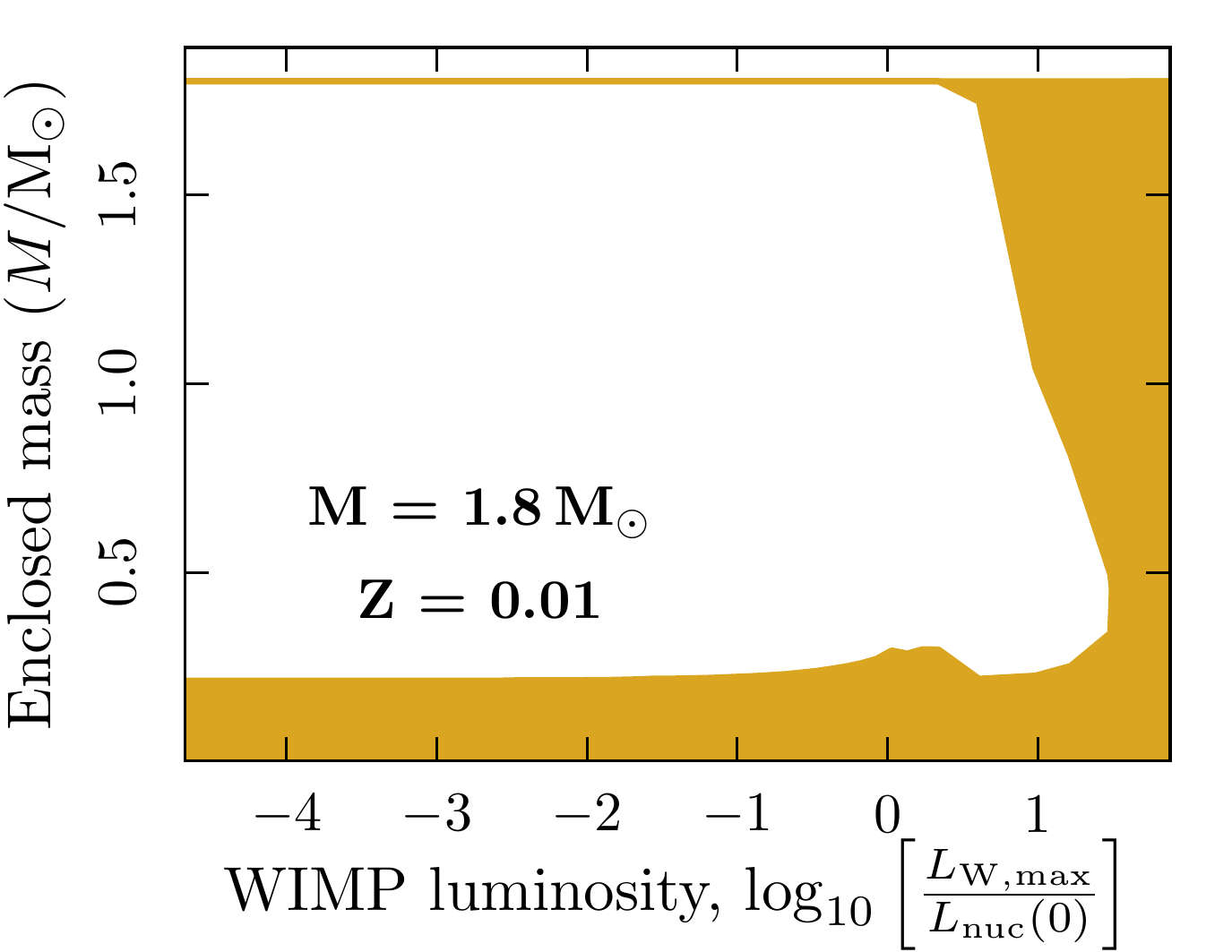}
\end{minipage}
\hspace{0.5\columnsep}
\begin{minipage}{0.55\columnwidth}
\includegraphics[width=\columnwidth]{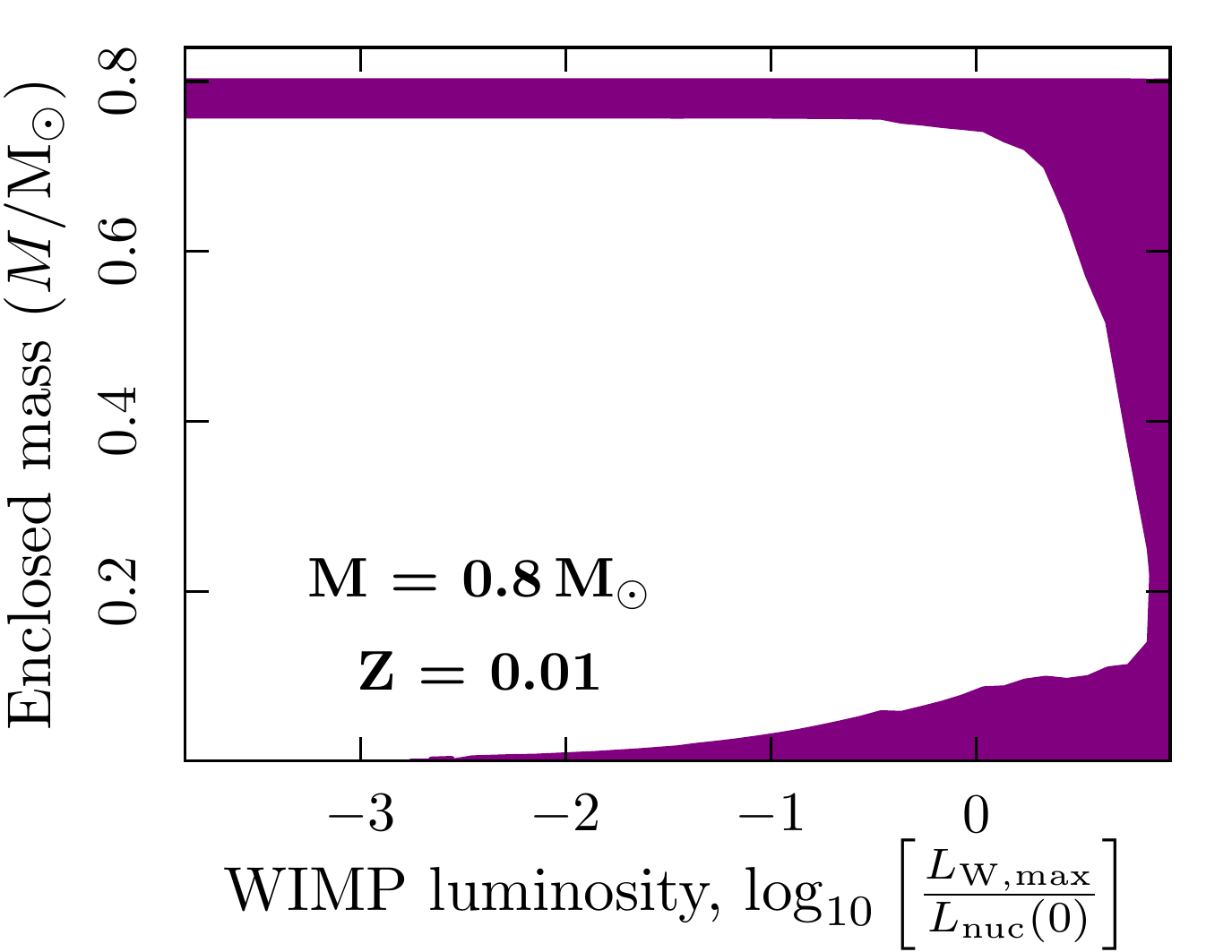}
\includegraphics[width=\columnwidth]{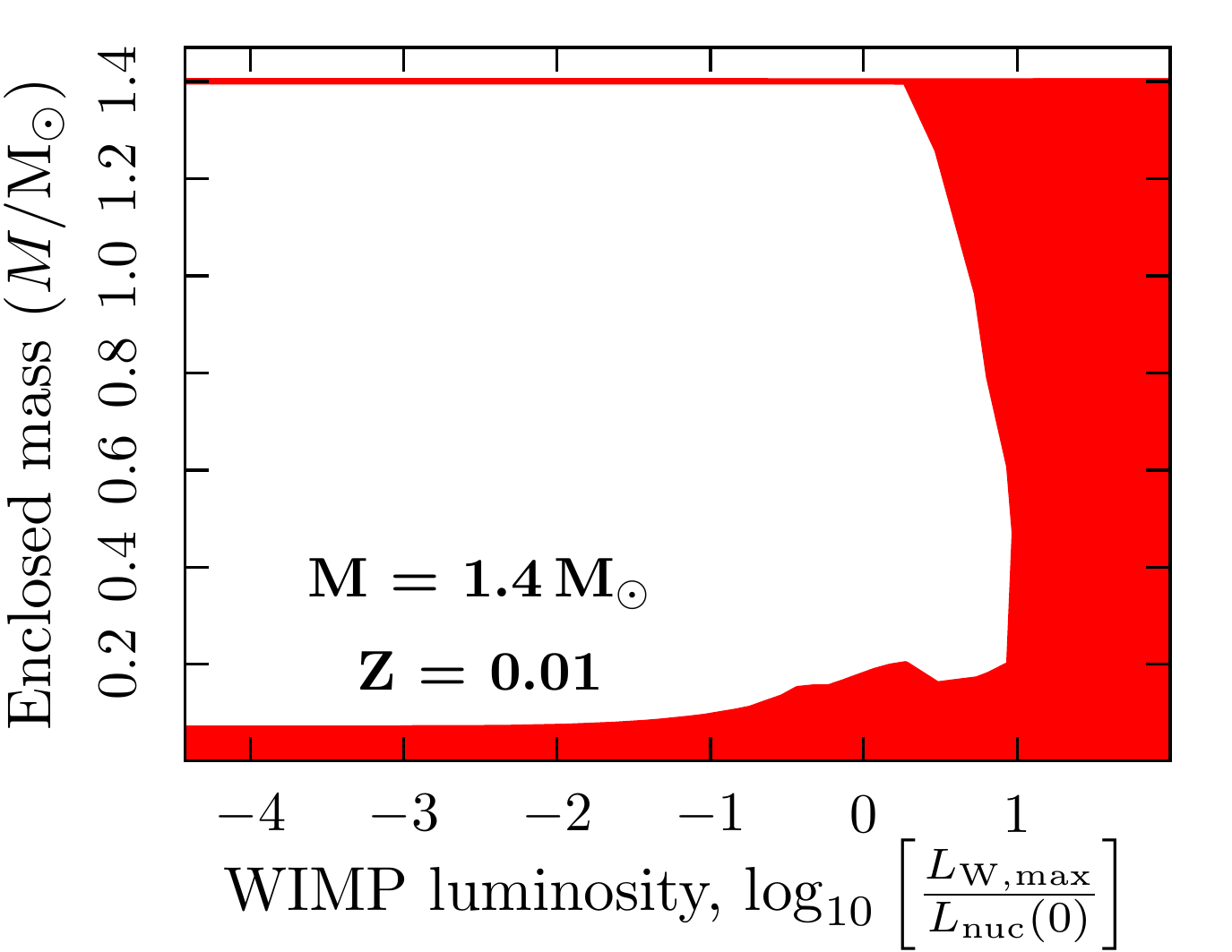}
\includegraphics[width=\columnwidth]{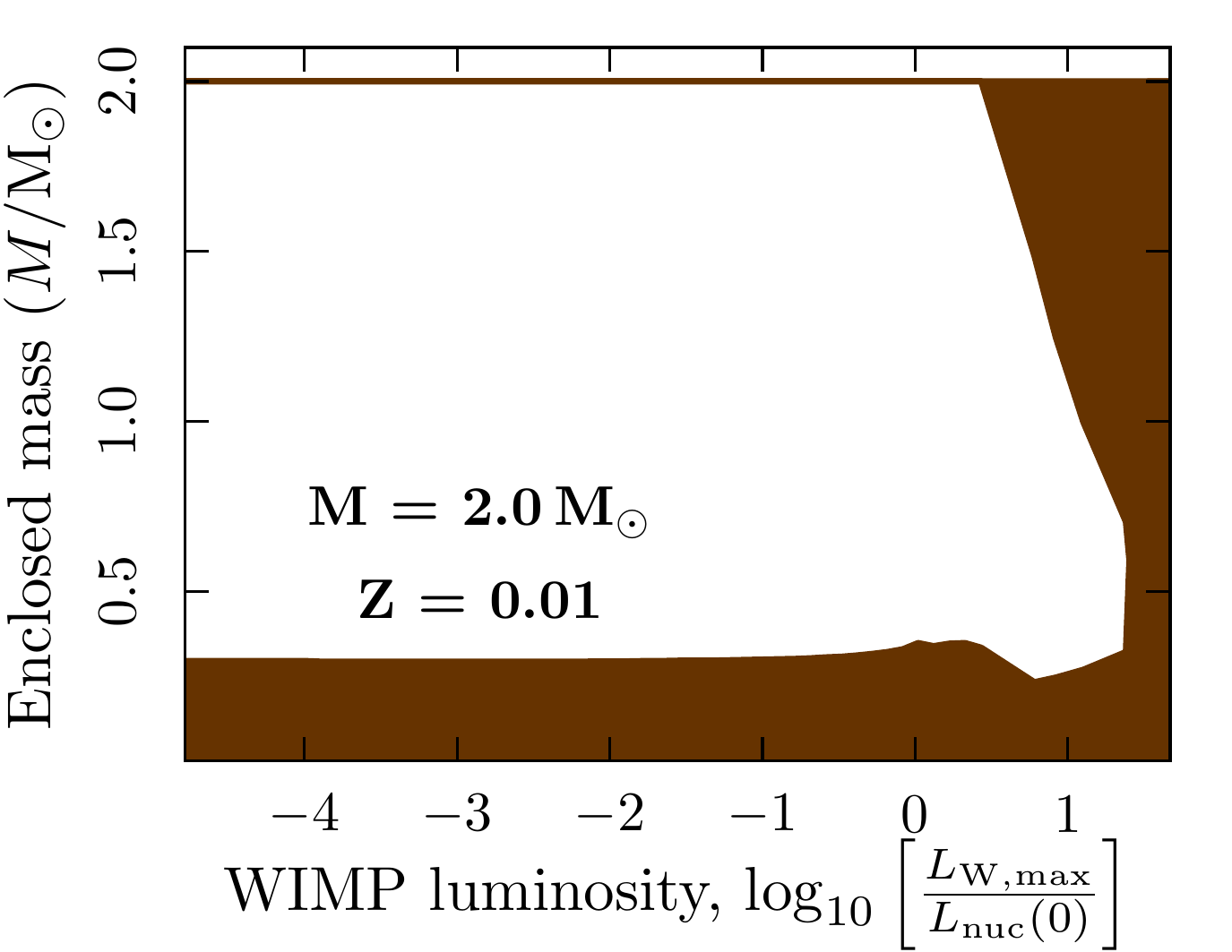}
\end{minipage}
\end{center}
\caption{The extent of convection at $t = t_\mathrm{adjust}$ in stars of different masses as the energy from WIMP annihilation in their cores is increased.  Shaded areas indicate regions in which the stellar energy transport is convective; elsewhere, transport is radiative.  Stars develop and extend their convective cores and envelopes as the WIMP luminosity is increased, eventually becoming fully convective at high values of $L_\mathrm{W,max}$.  Less massive stars require a smaller ratio of annihilation to fusion energy to alter their convective properties than heavier stars.  Plots in different panels extend over slightly different ranges of the WIMP-to-nuclear burning ratio, according to which models we were able to converge.  Not shown is the dependence upon metallicity: at lower metallicities, the onset of convection is deferred to higher WIMP luminosities, with effects greatest in higher-mass stars.}
\label{fig7}
\end{figure*}

In Fig.~\ref{fig7} we show the extent of convection at $t = t_\mathrm{adjust}$ in stars of various masses, as the WIMP luminosity is increased.  Because WIMP annihilation is far more concentrated at the centre of a star than nuclear burning, stars with higher WIMP-to-nuclear burning ratios exhibit steeper radiative temperature gradients in their cores.  This produces convective cores of increasing size, as the height over which the temperature gradient is superadiabatic increases.  In parallel, the overall cooling and expansion of the star results in cooler surface layers, increasing the H$^{-}$ concentration and opacity and resulting in progressively deeper surface convection zones.  At high enough WIMP luminosities the two zones meet and the star becomes fully convective.  At lower metallicities, the promotion of convection is deferred until significantly higher WIMP luminosities, with the effect strongest in higher-mass stars.  As an example, a 0.4\,M$_\odot$ star at $Z=0.02$ becomes fully convective at $\log_{10}[L_\mathrm{W,max}/L_\mathrm{nuc}(0)]\approx-0.1$, and requires $\log_{10}[L_\mathrm{W,max}/L_\mathrm{nuc}(0)]\approx0.3$ at $Z=0.0003$.  A 1.4\,M$_\odot$ star on the other hand requires $\log_{10}[L_\mathrm{W,max}/L_\mathrm{nuc}(0)]\approx0.8$ at $Z=0.02$, but $\log_{10}[L_\mathrm{W,max}/L_\mathrm{nuc}(0)]>2$ at $Z=0.0003$.

\begin{figure}
\begin{center}
\includegraphics[width=\columnwidth, trim = 0 0 0 20, clip=true]{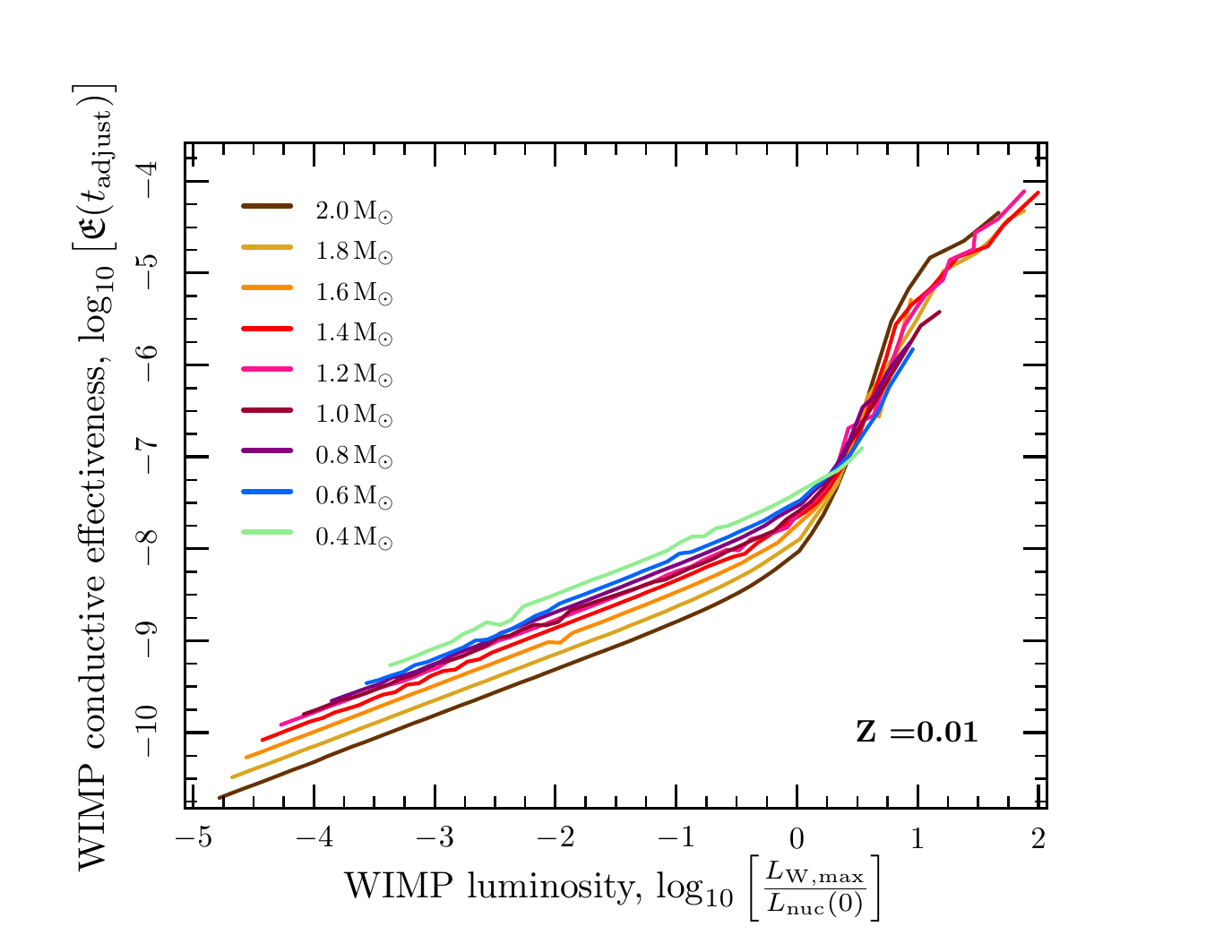}
\end{center}
\caption{The significance of conductive energy transport by WIMPs, as measured by the dimensionless WIMP conductive effectiveness (Eq.~\protect\ref{E}).  A value of 0 on the y-axis roughly indicates that conductive energy transport by WIMPs is as important as all other energy sources in the star combined.  The importance of conduction by WIMPs is significantly less than that of actual energy sources.  Not shown is the fact that $\mathfrak{E}$ becomes slightly larger at lower metallicity, with the effect increasing with stellar mass.  This suggests that conductive energy transport should probably be taken into account when simulating massive metal-free stars.}
\label{fig8}
\end{figure}

We plot the dimensionless WIMP conductive effectiveness $\mathfrak{E}$ (Eq.~\ref{E}) at $t = t_\mathrm{adjust}$ for the full range of WIMP luminosities and stellar masses in Fig.~\ref{fig8}.  Due to the Knudsen-dependent and radial suppression factors (Eqs.~\ref{knudsensupp} and \ref{radialsupp}), the contribution of WIMP conductive energy transport turns out to be small over most of the the parameter space.  Eq.~\ref{E} is, however, a rather coarse measure of the significance of conductive transport by WIMPs; a more detailed comparison would investigate its role in the effective opacity of nuclear matter in the star.  At lower metallicities, the values of $\mathfrak{E}$ become slightly larger, with the effect increasing with stellar mass.  WIMP energy transport might therefore be worth including in future studies of PopIII dark star evolution.

A number of numerical features complicate the interpretation of Figs.~\ref{fig5}--\ref{fig8}.  To be able to simulate a grid of $\sim$3500 stars and produce a manageable amount of output data, we chose only to save model data every tenth timestep.  This meant the resolution available for choosing the points at which to define $L_\mathrm{W,max}$ and $t_\mathrm{adjust}$ was not as high as it could have been.  The effect upon $t_\mathrm{adjust}$ was greater than on $L_\mathrm{W,max}$, as it was compounded by the fact that $t_\mathrm{adjust}$ is not as simply found as its definition would have one believe.  In cases where the equilibration time-scale is long, $t_\mathrm{adjust}$ can be comparable to or longer than $\tau_\mathrm{eq}$.  When this happens, $t_\mathrm{adjust}$ and $\tau_\mathrm{eq}$ begin to lose meaning, as the adjustment alters the capture rate, which feeds back on the adjustment.  This only occurs when $C(t)$ and $A(t)$ are of intermediate size, because even though $\tau_\mathrm{eq}$ is at its longest when $C(t)$ and $A(t)$ are very small, very little adjustment is necessary in this case so $t_\mathrm{adjust}$ is also very small.  The upshot of all this is that a small amount of noise appears in Figs.~\ref{fig5}, \ref{fig7} and \ref{fig8}.  For the sake of aesthetics, we recomputed individual evolutionary tracks in Figs.~\ref{fig3} and \ref{fig4} with data saved every timestep.

For 1.0 and 1.2\,M$_\odot$ stars, the reference CNO-process luminosity extracted at $t=0$ in Fig.~\ref{fig5} was somewhat overestimated, due to the initial relaxation of the stellar models.  The CNO process is only just present in 1.0-1.2\,M$_\odot$ stars, and extremely temperature-sensitive, so is very likely to be significantly altered during numerical relaxation.  The overestimation is the reason curves for 1.0 and 1.2\,M$_\odot$ do not tend properly to zero at low WIMP luminosities in the lower panel of Fig.~\ref{fig5}.

Despite extensive prior testing, we found that our stopping criterion ii) was sometimes not quite stringent enough.  Occasionally, models were halted which would have just managed to leave the main sequence in less than the age of the universe.  We removed a small number of stars we suspected this of having influenced from the grid.  As a result, the exact slopes of the steepest parts of the curves for higher masses in Fig.~\ref{fig6} are somewhat uncertain.

Some noise also exists in the plots of Fig.~\ref{fig6}, simply due to the finite temporal resolution of the models.  Timesteps typically become longer once a star nears the end of the main sequence, so some temporal `overshoot' can occur before criterion i) is triggered.  As always, our choice of the internal timestep scaling was a compromise between obtaining the smoothest results and being able to compute a reasonable number of models in a tractable timeframe (proper treatment of WIMP capture makes dark stellar evolution far more time-consuming than standard evolution).  

For a small window of WIMP densities, we also found that stars underwent seemingly random expansion and recontraction events during their evolution on the hybrid WIMP-hydrogen main sequence.  Suggestively, these windows correspond approximately to the areas where we were unable to find solutions using our static code \citep{Fairbairn08}.  The results of the static code suggest that this window can be thought of in the following way: for a particular ambient density of WIMPs, if one tries to obtain a static solution, one may find that two solutions exist for a WIMP-burning star.  The first solution corresponds to a star where the central temperature is rather close to that of a normal star of the same mass, such that the WIMPs are spread over a larger volume according to equation \ref{rhoiso}.  Because the WIMPs are spread over a larger volume, their annihilation is less centralised and the spatial distribution of their energy release into the star is closer to that of normal nuclear burning.  The core temperature of the star is therefore not reduced very much and the solution is self consistent.  The second solution occurs when the WIMPs are localised in a small region in the centre of the star because the central temperature is low.  The low central temperature is what would be expected if energy were input into a small region at the centre of the star, so this solution is also self-consistent.  This situation arises primarily for larger mass stars, partially due to the extreme temperature-dependence of the CNO process which dominates nuclear burning in such stars.

The existence of two solutions with the static code suggests that we might expect evolution in this region of parameter space to be unpredictable when looked at with a time-dependent code.  For example, stars might exhibit genuine periodic or chaotic variability, or could become numerically unstable.  By reducing the internal timestep scaling and employing the reconvergence mode, we established that the apparently random expansions and contractions are numerical artefacts caused by insufficient temporal resolution; given sufficiently small timesteps, the time-dependent code seems to follow a path intermediate to the two solutions appearing in the basic static code.  The borderline stability of the stellar structure in this region when treated with a code which assumes hydrostatic equilibrium (as \textsf{DarkStars} does) suggests that such stars might exhibit some true physical variability after all, but due to dynamical effects only captureable with a full hydrodynamic code.  If there is interesting variable behaviour in this region then, it probably occurs on a timescale smaller than can be resolved by \textsf{DarkStars}.  

The evolution of such stars over an entire lifetime seems largely unaffected by the excursions, so we are confident that the overall results of the grid still accurately reflect the general properties of main sequence dark stars.  However, the excursions do further complicate the task of automatically choosing $t_\mathrm{adjust}$, adding further noise to Figs.~\ref{fig5}, \ref{fig7} and \ref{fig8}.

\section{The Galactic dark matter halo}
\label{halo}

In order to simulate the accretion of dark matter by stars at the Galactic centre, one should understand its distribution in the Milky Way.  In particular, the density and velocity distribution of dark matter both play a role in determining the capture rate.

\subsection{Density}
Density profiles of dark matter halos have been a topic of computational study for over a decade \citep{nfw,moore,nfwsmooth,vialactea}.  As $N$-body simulations have been run on computers of ever-increasing speed, a standard lore for the expected distribution of dark matter in halos of all sizes has developed.  Two conclusions that seem to be universally accepted are that dark matter is denser in the centre of halo simulations, and that the logarithmic gradient of the density $\gamma=d\ln \rho / d \ln r$ is more negative in the outer parts of simulated halos than the inner parts.  A popular parametrisation is the `NFW profile,' which interpolates smoothly between asymptotic values of $\gamma=-1$ in the inner regions of the halo and $\gamma=-3$ in the outer regions \citep{nfw}.  \citet{nfwsmooth} have since suggested that a better profile would in fact be one where $\gamma$ varies smoothly with radius, and does not asymptote to any particular value at large nor small radii; such a profile has been advocated by various authors since the 1960s \citep[see][for an historical account]{Merritt06}.

Most simulations only include dark matter, since considering collisionless particles does not require the complicated hydrodynamics necessary to model baryons.  The presence of baryons is expected to change the distribution of dark matter: since the baryons are able to lose energy end sink into the middle of a galaxy, they create a potential well which subsequently pulls the dark matter into the central region.  This phenomenon of adiabatic contraction was first predicted by \citet{zel}, and formalised by \citet{blum}.  More recently, it has been realised that the non-circular nature of typical dark matter orbits reduces this effect, but does not remove it \citep{gnedin}.  In order to calculate the expected density profile of dark matter in a given galaxy, it is therefore necessary to take a dark matter halo from an $N$-body simulation of an appropriately-sized galaxy, then adiabatically contract it using the galaxy's observed baryonic profile.  This should be done taking into account the non-circular nature of the dark matter orbits \citep{gnedin,gustafsson}.

This prescription gives a more realistic estimate of the expected dark matter density in the central regions of a real galaxy than the results of a collisionless $N$-body simulation \citep[for a comparison of results obtained using this procedure with pure dark matter halos and those from simulations which also include baryons, see][]{gustafsson}.  The typical effect of such a contraction is to draw dark matter deeper into the central part of the galaxy, changing the inner slope of the density profile.

As one approaches the centre of a galaxy like the Milky Way from a large distance, the gravitational potential is first dominated by the diffuse dark matter halo.  Approaching the central bulge, the gravitational potential of the concentrated baryonic mass becomes more important.  If the current understanding of the dark matter distribution in the Milky Way is correct, the changeover occurs at a radius of the same order of magnitude as the solar position.  In the centremost regions, the supermassive black hole determines the gravitational dynamics.  The density of dark matter is thought to rise continuously towards the centre of the galaxy, but at radii much less than the solar position, its gravitational influence is always dwarfed by that of baryons or the central black hole.

In the central parsec, where the black hole starts to dominate the gravitational potential, the dark matter profile depends upon a number of factors.  One is simply the density of dark matter at larger radii, which forms an initial condition for the central density profile.  If the black hole forms in situ, it may create a miniature adiabatic contraction of the local dark matter profile, leading to a central spike where the density gradient is steeper than in the rest of the galaxy.  Immediately after its formation, the density of dark matter in such a spike can be extremely high \citep{gondolosilk}.  The spike is expected to diffuse away over time due to dark matter self-annihilation, loss of dark matter as it falls into the black hole and heating of the dark matter by gravitational interactions with stars.

These different effects can be incorporated into a diffusion equation that gives rise to a final prediction for the density of dark matter near the central black hole \citep{merritt05}.  In this work we will consider two dark matter density profiles, both approximations to the profiles presented by \citet{merritt05}.  These approximations correspond to profiles B and C used by \citet{lidia}.  The `NFW+spike' profile is a standard NFW $\gamma=1$ profile with a central spike which has diffused away over time, considerably reducing its density.  The `AC+spike' profile has undergone adiabatic contraction on galactic scales due to the presence of baryons, and also has a central spike which was allowed to diffuse away over time.  Both profiles can be parametrised by the expressions
\begin{align}
\label{profile}
\rho_\chi(r)&=\rho_\chi(100\rm pc)\left(\tfrac{100\rm pc}{r}\right)^{\gamma_1}\qquad && r > r_{out}\nonumber\\
\rho_\chi(r)&=\rho_\chi(r_{out})\left(\tfrac{r_{out}}{r}\right)^{\gamma_2}\qquad && r_{out}> r> r_{in}\\
\rho_\chi(r)&=\rho_\chi(r_{in})\qquad && r_{in} > r, \nonumber
\end{align}
where parameters are listed in Table~\ref{proftab}.  Note that following adiabatic contraction, the smoothly varying profile advocated by \citet{nfwsmooth} can become almost as steep in the central region as an equivalently-contracted NFW profile, depending upon the angular momentum of the dark matter particles.

\begin{table}
\caption{Parameters for the density profiles defined by Eqs.~\ref{profile}, which are approximations to the profiles presented by \protect\citet{merritt05}.
\label{proftab}}
\begin{tabular}{c@{\quad}c@{\quad}c@{\quad}c@{\quad}c@{\quad}c}
\hline\hline
Profile&$\rho_\chi(100\,\mathrm{pc})$&$\gamma_{1}$&$\gamma_2$&$r_\mathrm{out}$&$r_\mathrm{in}$\\
\hline
NFW+spike&$25\,\mathrm{GeV\,cm}^{-3}$& 1 & 1.85 &$7\cdot 10^4 \ R_\mathrm{BH}$&$10 \ R_\mathrm{BH}$\\
AC+spike&$360\,\mathrm{GeV\,cm}^{-3}$& 1.5 & 1.82 &$7\cdot 10^4 \ R_\mathrm{BH}$&$10 \ R_\mathrm{BH}$\\
\hline
\end{tabular}
\end{table}

\subsection{Velocities}
\label{halovel}
Having obtained some estimates of the possible densities of dark matter at the Galactic centre, we must also think about its velocity distribution, which has a strong bearing upon the number of particles captured by stars.

Various estimates of the velocity distribution exist in the literature.  For direct detection experiments such as CDMS, Xenon and COUPP to be able to easily compare results, the dark matter halo is typically assumed to be the isothermal RSC, with a radius-independent Keplerian velocity.  In this case, the velocity dispersion is set by the Keplerian velocity at the solar position ($\bar{v}=\sqrt{3/2}v_\odot=270$\,km\,s$^{-1}$), and acts as the width for a Gaussian distribution of velocities which is identical at every position.

As already mentioned, the highest resolution $N$-body simulations do not predict an isothermal halo, but rather one where the logarithmic density gradient close to the centre of the galaxy is much less than $-2$.  The Keplerian velocity in such a halo would therefore decrease to zero at the core, which would increase the rate at which dark matter would be accreted by stars.  In a real galaxy however, the dark matter is a subdominant component at these small Galactic radii, and the presence of stars and the central black hole increases the gravitational potential there.

Our default assumption is that the velocity distribution of dark matter is isotropic, spherically symmetric, Gaussian and has a dispersion set by the Keplerian velocity in the solar vicinity.  None of these assumptions are strictly correct, as we shall discuss shortly.  Our simplest attempt to improve the realism of the velocity distribution is to exclude velocities above the local Galactic escape velocity, as WIMPs with such velocities would presumably already have left the Galaxy some time earlier.  We therefore truncate the velocity distribution at the local escape velocity (in the rest frame of the Galaxy), which terminates the Maxwell tail of the distribution.  Given a generic WIMP velocity distribution $g_0(u)$, seen in the rest frame of the galaxy, the equivalent truncated distribution will be
\begin{equation}
\label{truncation}
    g_{\mathrm{gal},0}(u) = \frac{\rho_\chi}{m_\chi} \frac{g_0(u)\theta(u_\mathrm{gal}-u)}{\int^\infty_0 g_0(u')\theta(u_\mathrm{gal}-u')\,\mathrm{d}u'},
\end{equation}
where $u_\mathrm{gal}$ is the local escape velocity, and the integral ensures the new distribution remains correctly normalised.  For a star at rest with respect to the Galactic halo, this then brings the capture rate (Eq.~\ref{cap}) to the form
\begin{multline}
\label{trunccap2}
    C(t) = 4\pi D^{-1} \int^{R_\star}_0 r^2 \sum_i \int^{\mathrm{min}[u_\mathrm{gal},u_{\mathrm{max},i}(r,t)]}_0 \frac{g_0(u)}{u} \\
    \times w\Omega_{v,i}^-(w)\,\mathrm{d}u\,\mathrm{d}r.
\end{multline}
The normalisation factor $D$ is
\begin{equation}
    D = \frac{m_\chi}{\rho_\chi}\int^{u_\mathrm{gal}}_0 g_0(u)\,\mathrm{d}u.
\end{equation}
Working from Eq.~\ref{galacticframe}, in the case of an isothermal velocity distribution this becomes
\begin{equation}
\begin{split}
    D & = \frac{4}{\sqrt{\pi}} \Big(\frac{3}{2}\Big)^{3/2} \int^{u_\mathrm{gal}}_0 \frac{u^2}{\bar{v}^3} 
      \exp\Big(-\frac{3u^2}{2\bar{v}^2}\Big)\,\mathrm{d}u \\
    & = \mathrm{erf}\Big(\frac{u_\mathrm{gal}}{\bar{v}}\sqrt{\frac{3}{2}}\Big)-\sqrt{\frac{6}{\pi}}\frac{u_\mathrm{gal}}{\bar{v}}\exp\Big(-\frac{3u_\mathrm{gal}^2}{2\bar{v}^2}\Big).
\end{split}
\end{equation}

To find the capture rate in the frame of a star moving relative the the Galactic rest frame, we need to transform $g_{\mathrm{gal},0}(u)$ to some equivalent $g_{\mathrm{gal},\star}(u)$ via an appropriate Galilean transform, in analogy with the step from Eq.~\ref{galacticframe} to Eq.~\ref{stellarframe}, then consider what $u_\mathrm{gal}$ becomes in the frame of the star.  The maximum velocity any WIMP from a distribution cut off at $u_\mathrm{gal}$ can have in the galactic frame is obviously $u_\mathrm{gal}$.  In the frame of the star though, WIMPs coming from e.g.~the direction in which the star is moving through the halo can appear with much greater speed than those coming from `behind' the star.  If an incoming WIMP has speed $u_0$ and velocity in the direction of the unit vector $\pmb{e}_\mathrm{W}$ in the galactic frame, we see that its velocity $\pmb{u}_\star$ in the star's frame is 
\begin{equation*}
\begin{split}
    & & \pmb{u}_\star & = u_0\pmb{e}_\mathrm{W} - \pmb{v}_\star \\
    & \Rightarrow & u & \equiv |\pmb{u}_\star| = \sqrt{(u_0\pmb{e}_\mathrm{W} - \pmb{v_\star})^2} \\
    & \therefore & u & < \sqrt{(u_\mathrm{gal}\pmb{e}_\mathrm{W} - \pmb{v_\star})^2} 
      = (u_\mathrm{gal}^2 + v_\star^2 + 2u_\mathrm{gal}v_\star\cos\varphi)^\frac{1}{2}.
\end{split}
\end{equation*}
So in the star's frame, we have
\begin{equation}
    u_\mathrm{gal}\longrightarrow u_{\mathrm{gal},\star}(\varphi) = (u_\mathrm{gal}^2 + v_\star^2 + 2u_\mathrm{gal}v_\star\cos\varphi)^\frac{1}{2}, 
\end{equation}
where $\varphi$ is the angle in the galactic frame between the motions of the WIMP and star, such that $\varphi=0$ corresponds to a head-on collision.  This then poses something of a problem, as in the analogous case of the isothermal distribution, $\varphi$ had to be implicitly integrated over in the first place to obtain the integrand (Eq.~\ref{stellarframe}) for which this new cut-off velocity is the limit.  The solution to this is either to develop some sort of approximate averaging scheme, or to just do the integral over $\varphi$ explicitly, \emph{after} the integral over $u$.  In this case, the expression for capture of a truncated isothermal distribution of WIMPs by a moving star is
\begin{align}
    \label{pretrunccap}
    C(t) & = 8\sqrt\pi \Big(\frac{3}{2}\Big)^\frac{3}{2} \frac{\rho_\chi}{m_\chi} D^{-1} \int^{R_\star}_0 r^2 \sum_i \int_{-1}^1 
      \int^{u_{\mathrm{cut},i}}_0 \frac{u}{\bar{v}^3} \nonumber\\
    & \quad \times w\Omega_{v,i}^-(w) \exp\Big(-\frac{3}{2\bar{v}^2}(u^2 + v_\star^2 \\
    & \quad + 2uv_\star\cos\varphi) \Big)\,\mathrm{d}u\,\mathrm{d}(\cos\varphi)\,\mathrm{d}r, \nonumber
\end{align}
where
\begin{equation}
    u_{\mathrm{cut},i}(\varphi,r,t) \equiv \mathrm{min}[u_{\mathrm{gal},\star}(\varphi),u_{\mathrm{max},i}(r,t)].
\end{equation}

As in the non-truncated case, the velocity integral in Eq.~\ref{pretrunccap} can be performed analytically, yielding the truncated analogues of Eqs.~\ref{standardcap}--\ref{Hstandardcap}
\begin{multline}
\label{trunccap}
    C(t) = 4\pi D^{-1}\int^{R_\star}_0 r^2 \int_{-1}^1 \sum_i\,\big[M_i(u_{\mathrm{cut},i}(\varphi,r,t)) \\
    -M_i(0)\big]\,\mathrm{d}(\cos\varphi)\,\mathrm{d}r,
\end{multline}
where
\begin{gather}
    \begin{split}
    M_i(u) & = \frac{2\sigma_i n_i(r,t) \rho_\chi \mu_{+,i}^2E_{0,i}K^\frac{3}{2}}{m_\chi^2 \mu_i\sqrt\pi}\Bigg[\sqrt\pi Kv_\star\cos\varphi\Big(\Psi(H) \\
    & \quad \times \mathrm{e}^{K^2v_\star^2\cos^2\varphi/(K+H) - Kv_\star^2 - Hv(r,t)^2}(K+H)^{-3/2} \\
    & \quad - \Psi(G) \mathrm{e}^{K^2v_\star^2\cos^2\varphi/(K+G) - Kv_\star^2}(K+G)^{-3/2}\Big) \\
    & \quad + \mathrm{e}^{-2Kuv_\star\cos\varphi - K(u^2 + v_\star^2) - H(u^2 + v(r,t)^2)}(K+H)^{-1} \\
    & \quad - \mathrm{e}^{-2Kuv_\star\cos\varphi - K(u^2 + v_\star^2) - Gu^2}(K+G)^{-1}\Bigg],
    \end{split}\\
    \Psi(X) \equiv \mathrm{erf}\Big\{\frac{u(K+X) + Kv_\star\cos\varphi}{\sqrt{K+X}}\Big\}    
\end{gather}
for heavier elements, and
\begin{gather}
    \begin{split}
    M_\mathrm{H}(u) & = \frac{\sigma_\mathrm{H} n_\mathrm{H}(r,t) \rho_\chi K^{-\frac{1}{2}}}{2m_\chi\sqrt\pi}
      \Bigg[2\Big(\frac{\mu_{-,i}^2}{\mu_i} -Kv(r,t)^2 \\
    & \quad + K\frac{\mu_{-,i}^2}{\mu_i}\big[u^2 - uv_\star\cos\varphi + v_\star^2\cos^2\varphi\big]\Big) \\
    & \quad \times \mathrm{e}^{-2Kuv_\star\cos\varphi - K(u^2+v_\star^2)} + v_\star\cos\varphi\sqrt{\pi K} \\
    & \quad \times \Big(3\frac{\mu_{-,i}^2}{\mu_i} -2Kv(r,t)^2 + 2Kv_\star^2\cos^2\varphi\frac{\mu_{-,i}^2}{\mu_i}\Big) \\
    & \quad \times \mathrm{erf}\big[\sqrt{K}(u + v_\star\cos\varphi)\big]\mathrm{e}^{-v_\star^2K\sin^2\varphi}\Bigg]
    \end{split}
\end{gather}
for hydrogen.  This gives the machinery necessary to calculate the capture rate for a truncated isothermal (Gaussian) velocity distribution.

The fact that we \emph{do not} expect the velocity distribution to be Gaussian in reality is related to the fact that one only expects a Gaussian distribution in the limit of an extensive distribution of particles, such as an ideal gas.  For particles coupled to a long-range potential such as gravity, this is not the case.  It has been suggested in the literature that the Tsallis distribution, one designed specifically to model the departure from extensivity, is a better fit to the data than a Gaussian \citep{Hansen:2005yj}.

Furthermore, there are good reasons to believe that the radial distribution of dark matter will have a different width to the tangential distribution, since the orbits of dark matter particles are far from circular.  For a star on a non-circular orbit around the centre of the Galaxy, this could have important consequences.

Finally, it is interesting to test the truth of the assumption that the velocity dispersion is fully determined by the Keplerian velocity at the solar position.  The relationship between the Keplerian velocity, the velocity distribution and its anisotropy depends upon the shape of the potential well of the galaxy; the anisotropy in the velocity distribution can be obtained from the Jeans equations \citep{Fairbairn09}.

In order to quantify the departures from the isothermal halo model, we have examined data from the Via Lactea simulation \citep{vialactea}.  This $N$-body simulation contains more than $2\times10^{8}$ dark matter particles, and is one of the largest simulations of a Milky Way-size dark matter halo to date.  As suspected, the results do indeed show that all four of the simplifying assumptions involved in the isothermal halo model (isotropy, spherical symmetry, Gaussianity and a dispersion proportional to the Keplerian velocity at the solar position) are essentially incorrect.  To obtain a new velocity distribution from the data, we looked at the velocities of particles at different radii.  We attempted to fit the distributions with the Tsallis profile of \citet{Hansen:2005yj}, but found that although this does provide a better fit than a Gaussian, the following one-dimensional distribution is an even better fit:
\begin{equation}
\label{1dnbody}
   h_\mathrm{1D}(u_i) = \exp\Bigg\{-\bigg[\frac{1}{2}\Big(\frac{u_i}{\sigma_i}\Big)^2\bigg]^{\alpha_i}\Bigg\},
\end{equation}
where $i\in \{r,\theta,\phi\}$ and $u_i$ is the velocity in the $i$th direction in Galactic coordinates.  No normalisation prefactor has been included here.  The parameter $\alpha$ measures the departure from Gaussianity and is distinctly different for the one-dimensional velocity distributions in the radial and angular directions.  The values of $\alpha$ for the radial distribution ($\alpha_r$) and a composite tangential distribution ($\alpha_\mathrm{T}$, where $u_T^2\equiv u_\theta^2 + u_\phi^2$) can be found as a function of radius in \citet{Fairbairn09}.  The ratio of the velocity dispersion to the local Keplerian velocity ($v_\mathrm{Kep}$, or more precisely, the square root of the potential since the halo is not completely spherically symmetric) is also a function of radius and can be found in \citet{Fairbairn09}.

These distributions become less Gaussian as one approaches the Galactic centre, with $\alpha\ll 1$ in the region we are most interested in.  One should be careful not to take this result overly seriously though; not because the physics of the simulation is in doubt (above the $\sim$0.1\,kpc resolution scale), but rather because the simulation only includes dark matter.  The velocity dispersion at the centre of an NFW profile goes to zero, whereas in a real galaxy, the presence of baryons and the central black hole would be expected to change the velocity distribution quite significantly.  We do not make any strong claims as to the fitness of this non-Gaussian velocity distribution for modelling the very centre of the Galaxy; we employ it more with the goal of determining what degree of uncertainty exists in our capture results due to the velocity distribution.

In order to reasonably calculate the effect of the non-Gaussian distributions upon capture rates, we require a composite distribution of total velocity magnitudes in three dimensions.  Since the fitted values of $\alpha$ and $\sigma/v_\mathrm{Kep}$ become almost isotropic at low Galactic radii, for this purpose we can set them to the same values in the radial and angular directions.  We choose these values as the means of the fitted values in each spatial direction, for the smallest Galactic radius at which we fit the velocity distribution (1\,kpc).  This gives $\alpha=0.35$ and  $\sigma/v_\mathrm{Kep}=0.05$.  At the Galactic centre, $v_\mathrm{Kep}$ is dominated by the black hole, so
\begin{equation}
\label{nbodysigma}
  \sigma = 56.8\,\mathrm{km\,s}^{-1} \times \bigg(\frac{0.01\,\mathrm{pc}}{r}\bigg)^{1/2}.
\end{equation}
To obtain a three-dimensional distribution purely as a function of the velocity magnitude
\begin{equation}
   u\equiv\sqrt{u_r^2+u_\theta^2+u_\phi^2},
\end{equation}
we take the product of the 3 one-dimensional distributions, convert to local spherical polar coordinates and integrate over the angular directions.  This gives
\begin{equation}
   h_\mathrm{3D}(u_r,u_\theta,u_\phi) = 4\pi u^2\exp\bigg[-\frac{1}{(2\sigma^2)^\alpha}(u_r^{2\alpha}+u_\theta^{2\alpha}+u_\phi^{2\alpha})\bigg].
\end{equation}
When $\alpha=1$, this clearly produces a Gaussian distribution of velocity magnitudes.  For $\alpha\ne1$, we can express $u_r^2$, $u_\theta^2$ and $u_\phi^2$ in terms of one another and $u^2$, then expand to second-order in a binomial series to produce
\begin{equation}
    (u^2 - u_i^2 - u_j^2)^\alpha = u^{2\alpha} - \alpha u^{2(\alpha-1)}(u_i^2 + u_j^2),
\end{equation}
where $i$ and $j$ are different members of the set $\{r,\theta,\phi\}$.  This gives
\begin{equation}
    u_r^{2\alpha}+u_\theta^{2\alpha}+u_\phi^{2\alpha} = (3-2\alpha)u^{2\alpha}.
\end{equation}
Normalising over $u\in[0,\infty)$ gives the final three-dimensional `$N$-body' velocity distribution
\begin{equation}
\label{3dnbody}
    h_\mathrm{3D}(u) = \frac{3(3-2\alpha)^\frac{3}{2\alpha}}{2^\frac{3}{2}\Gamma(1+\frac{3}{2\alpha})}
    \frac{\rho_\chi}{m_\chi}\frac{u^2}{\sigma^3} \exp\Bigg\{-(3-2\alpha)\bigg[\frac{1}{2}\Big(\frac{u}{\sigma}\Big)^2\bigg]^\alpha\Bigg\}.
\end{equation}
Whilst this expression breaks down at third order in the binomial expansion, we expect it to be a reasonable approximation given the level of uncertainty in choosing realistic values of $\alpha$ and $\sigma/v_\mathrm{Kep}$.  We now have a three-dimensional velocity distribution as a function of the local Keplerian velocity, which can be inserted into Eq.~\ref{cap} to obtain the capture rate.  Putting Eq.~\ref{3dnbody} into Eq.~\ref{trunccap2} instead, one obtains the capture rate from an equivalent truncated distribution, which is also a function of the local escape velocity.

\section{The Galactic potential}
\label{potential}

To calculate dark matter capture rates, it is extremely important to know the stellar velocity $v_\star$ through the dark matter halo.  In order to correctly truncate the velocity distribution of the dark matter, one also needs to know the local Galactic escape velocity $u_\mathrm{max}$.  

The orbital velocities are rather simply obtained.  Within a few tenths of a parsec, the Galactic potential is dominated by the central black hole.  All the elliptical orbits we will consider lie within one fiftieth of a parsec, so they can be treated as exactly Keplerian about a point-mass black hole.

Calculating the escape velocities is more arduous, as we need to integrate the potential experienced by a test particle exiting the Galaxy.  It is important to consider not only dark matter but also the presence of baryons, which dominate the potential from around 0.5\,pc to several kpc.  To model the baryon density of the Milky Way we use the same prescription as \citet{gustafsson}, assuming a central bulge of stars with density $\rho\propto r^{-\gamma}e^{-r/\lambda}$.  We assume a thin disc of matter with surface density 
\begin{equation}
\sigma_{\mathrm{disc}}(r) =\frac{cM_{\mathrm{disc}\infty}}{2\pi\left(r^2+c^2\right)^{\frac{3}{2}}}.
\end{equation}
We choose the free parameters to match observations of the Milky Way: $\gamma=1.85$, $\lambda=1 \,\mathrm{kpc}$, $c=5 \,\mathrm{kpc}$ and the total disc and bulge masses $M_{\mathrm{disc}\infty} = 5M_{\mathrm{bulge}}=6.5\times10^{10}M_{\odot}$ \citep{Kent:1991me, Zhao:1995qh, Dehnen:1996fa, Klypin:2001xu}.  We assume that the extent of the disc is 15\,kpc.  We use an NFW profile with a scale radius of 20\,kpc for the dark matter, normalised to 0.3\,GeV\,cm$^{-3}$ at the location of the solar system, and find the local escape velocity by integrating the energy loss along a radial path exiting the Galaxy.

\section{Impacts on stars at the Galactic centre}
\label{gc}

Armed with detailed estimates of the stellar orbits, local escape velocity, density of dark matter and its velocity distribution at the Galactic centre, we can now realistically evaluate the potential impact of WIMPs upon stellar evolution there.  We ran two further grids of evolutionary models, both at $Z=0.02$ and over $0.3\,\mathrm{M}_\odot\le M_\star \le 1.5\,\mathrm{M}_\odot$.  We computed models in both grids using the NFW+spike and AC+spike profiles from Sect.~\ref{halo}.

\subsection{Circular orbits}

\begin{figure}
\begin{center}
\includegraphics[width=\columnwidth, trim = 0 0 0 30, clip=true]{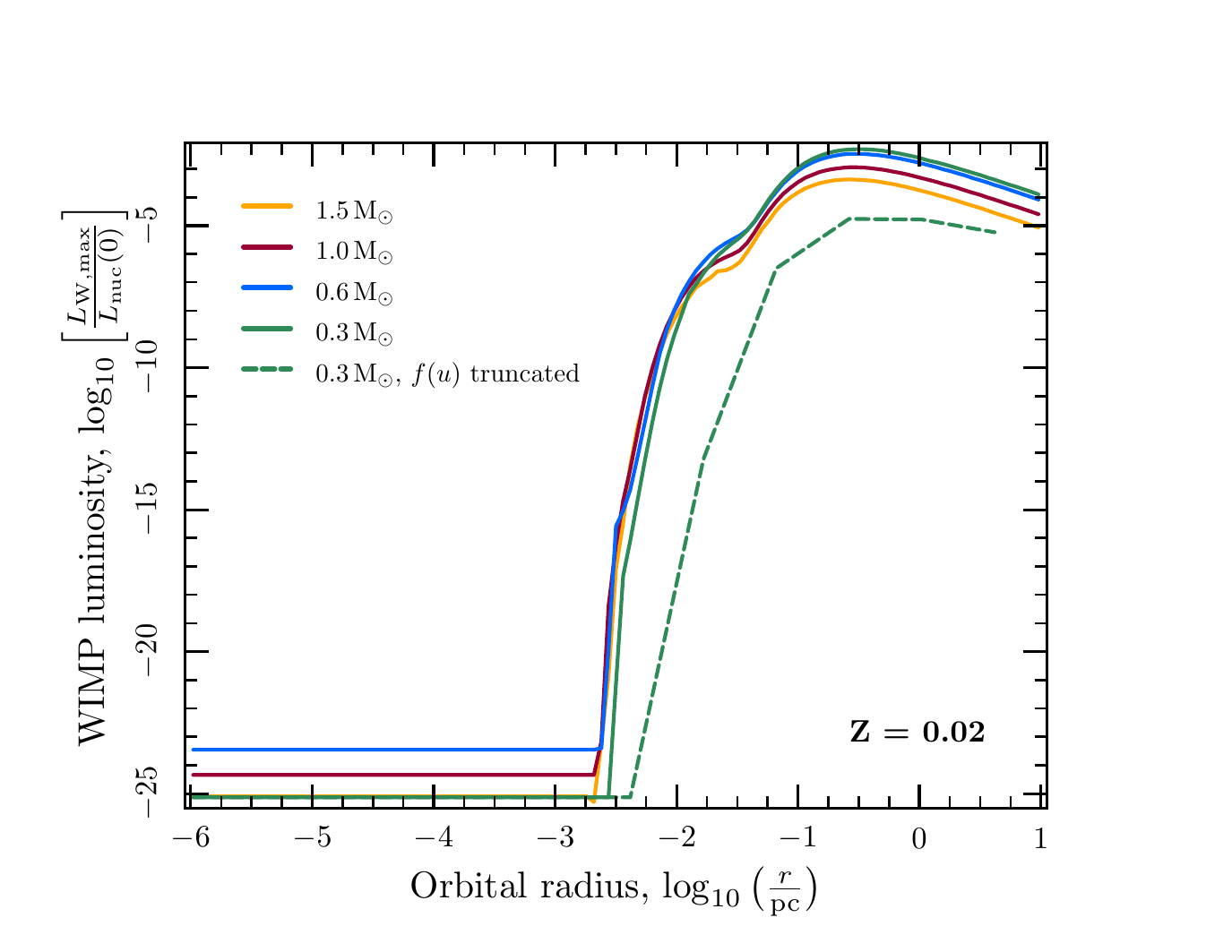}
\end{center}
\caption{WIMP luminosities achieved by stars orbiting circularly about the central black hole.  The dark matter velocity distribution is isothermal with dispersion $\bar{v}=270$\,km\,s$^{-1}$, and the density profile follows the adiabatically contracted profile (AC+spike).  The impact of instead using a velocity distribution truncated at the local Galactic escape velocity is shown with a dashed curve.  Capture is maximised at a radius of approximately 0.3\,pc, but no circular orbit produces capture rates high enough to translate into a WIMP luminosity which can produce any significant changes in stellar evolution.  WIMP luminosities produced with the alternative density profile (NFW+spike) are even lower, so are not shown.}
\label{fig9}
\end{figure}

The first grid covered single stars on circular orbits, with orbital radii extending from 10\,pc to 10$^{-6}$\,pc.  For this grid we used the standard, non-truncated version of the isothermal velocity distribution (Eq.~\ref{stellarframe}).  Results for models computed with the AC+spike density profile are shown in Fig.~\ref{fig9}.  As orbits are made smaller, capture rates rise because stars encounter higher densities of dark matter.  This effect is balanced by a reduction in capture caused by stars' increasing circular velocities as they orbit closer to the black hole.  The capture rate and resultant WIMP luminosity peaks at an orbital radius of $\sim$0.3\,pc.  Inwards of this the velocity effect dominates and capture is highly suppressed.  The constant WIMP luminosities at very small radii in Fig.~\ref{fig9} are entirely due to the initial populations of WIMPs that the models were started with.  

We evolved some supplementary models (dashed curve in Fig.~\ref{fig9}) with the truncated isothermal distribution (Eq.~\ref{truncation} applied to Eq.~\ref{galacticframe}), to see if capture might be boosted to interesting levels by removing unphysical WIMP velocities.  On the contrary, the truncation of the isothermal WIMP velocity distribution caused a strong reduction in capture rates.  Stars moving as quickly as those on circular orbits near a black hole must capture predominantly from the Maxwell tail of the isothermal distribution, so truncation denies them many of their best capture candidates.  The opposite is true of a star in the RSC, which captures from the centre of the distribution and benefits (slightly) in capture rate if the distribution is truncated and renormalised.

If it follows a circular orbit in an isothermal WIMP halo, even the lowest-mass single star, placed at the optimal radius in the most optimistic density profile, cannot accrete enough WIMPs to bring annihilation luminosity to within two orders of magnitude of its nuclear luminosity.  We do not show WIMP luminosities resulting from the NFW+spike density profile, as they are even less interesting due to the profile's lower central density.

\subsection{Elliptical orbits}
\label{elliptics}

The primary assumptions of the previous grid were that stars always follow circular orbits, and that the halo is isothermal.  We know these to be untrue in reality, so in the second grid we considered the effects of elliptical orbits and a non-Gaussian velocity distribution.

\begin{table}
\label{orbittable}
\begin{center}
\caption{Orbits considered in Sect.~\ref{elliptics}, along which we evolved stars in Figs.~\ref{fig10}--\ref{fig12}.}
    \begin{tabular}{llcc}
    \hline\hline
    Orbit class & $e$ & $r_\mathrm{min}$ (pc) & $r_\mathrm{max}$ (pc) \\
    \hline
    $P=50$\,yr & 0 & $9.49\times10^{-3}$ & $9.49\times10^{-3}$\\
    & 0.5 & $4.74\times10^{-3}$ & $1.42\times10^{-2}$\\
    & 0.9 & $9.49\times10^{-4}$ & $1.80\times10^{-2}$\\
    & 0.99 & $9.49\times10^{-5}$ & $1.89\times10^{-2}$\\
    & 0.999 & $9.49\times10^{-6}$ & $1.90\times10^{-2}$\\
    & 0.9998 & $1.90\times10^{-6}$ & $1.90\times10^{-2}$\\
    $P=10$\,yr & 0 & $3.24\times10^{-3}$ & $3.24\times10^{-3}$ \\
    & 0.5 & $1.62\times10^{-3}$ & $4.87\times10^{-3}$\\
    & 0.9 & $3.25\times10^{-4}$ & $6.17\times10^{-3}$\\
    & 0.99 & $3.25\times10^{-5}$ & $6.46\times10^{-3}$\\
    & 0.999 & $3.25\times10^{-6}$ & $6.49\times10^{-3}$\\
    & 0.9995 & $1.62\times10^{-6}$ & $6.49\times10^{-3}$\\
    $r_\mathrm{max}=0.01$\,pc & 0 & $1.00\times10^{-2}$ & $1.00\times10^{-2}$\\
    & 0.5 & $3.33\times10^{-3}$ & $1.00\times10^{-2}$\\
    & 0.9 & $5.26\times10^{-4}$ & $1.00\times10^{-2}$\\
    & 0.99 & $5.03\times10^{-5}$ & $1.00\times10^{-2}$\\
    & 0.99967 & $1.65\times10^{-6}$ & $1.00\times10^{-2}$\\
    \hline
    \end{tabular}
\end{center}
\label{tab2}
\end{table}

\begin{figure*}
\begin{center}
\begin{minipage}{1.1\columnwidth}
\includegraphics[height = 310pt, trim = 95 30 80 45, clip=true]{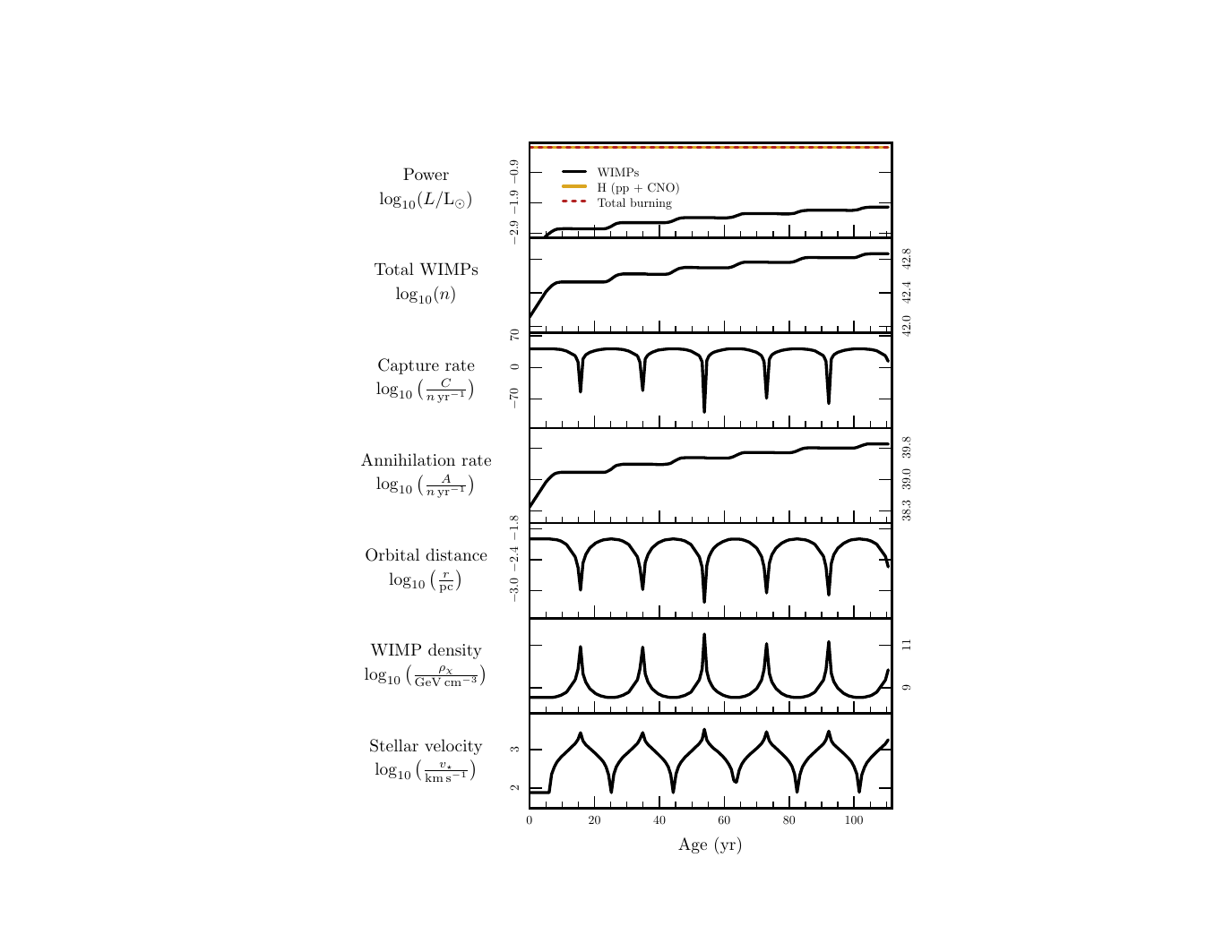}
\end{minipage}
\hspace{0.8\columnsep}
\begin{minipage}{0.9\columnwidth}
\includegraphics[height = 310pt, trim = 160 30 80 45, clip=true]{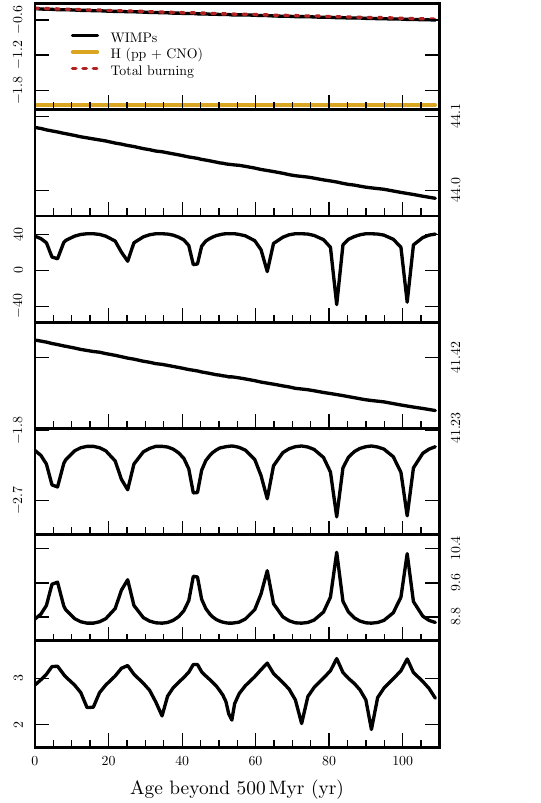}
\end{minipage}
\end{center}
\caption{Evolution of a 1\,M$_\odot$, $Z=0.02$ star on a highly elliptical orbit close to the central back hole, followed for 5 orbits in the beginning of its lifetime (left) and at an age of half a billion years (right).  The orbit is that listed in the final line of Table~\protect\ref{tab2}.  Dark matter velocities follow an isothermal distribution with $\bar{v}=270$\,km\,s$^{-1}$, truncated at the local escape velocity, and densities follow the AC+spike profile.  WIMP capture occurs exclusively around apoapsis despite this being the point at which the ambient WIMP density is lowest, thanks to the stars very low orbital velocity.  Early in the evolution, before capture and annihilation have equilibrated and the WIMP population has stabilised, the total population, annihilation rate and resultant WIMP luminosity undergo punctuated increases each time the star goes through a period of capture.  In the evolved star on the right, the equilibrium population of WIMPs provides a buffer against the transient nature of capture, and the evolution is essentially smooth.  Because explicitly following the evolution on an elliptical orbit is highly time-consuming, between these two plots the star was allowed to evolve on an artificial circular orbit with the same initial mean capture rate as exhibited in the left panel.  Because of the finite temporal resolution, some of the peaks at periapsis are not properly resolved; because capture here is effectively zero, this has no effect upon the evolution.}
\label{fig10}
\end{figure*}

This grid consisted of single stars on orbits with various ellipticities, within three classes: orbits with periods $P=10$\,yr, orbits with $P=50$\,yr and orbits where the maximum star--black hole separation was 0.01\,pc.  The galactocentric distances at periapsis and apoapsis ($r_\mathrm{min}$ and $r_\mathrm{max}$ respectively) of each of the orbits we considered for this grid are given in Table~\ref{orbittable}.  The maximum eccentricity in each class was chosen so as to ensure that stars did not come within five Scwharzschild radii of the centre of the black hole, ensuring that relativistic corrections to the orbit were not critical.

The early evolution of one of the stars from the grid is shown in the left panel of Fig.~\ref{fig10}.  The initial flat sections of curves are where the star was held at a constant galactocentric radius for the first 3 timesteps to allow the model to properly relax.  Capture and annihilation occur in punctuated stages, clearly correlated with the orbital period.  Strikingly, the times of greatest capture are in fact when the star is \emph{farthest} from the centre of the Galaxy, at apoapsis.  This is because it has had a chance to slow down relative to the DM halo, and achieve a significant capture rate for a time before plummeting back down towards the black hole.  By the time it reaches periapsis, the star is moving so quickly that capture is essentially zero, regardless of how high the dark matter density is.

Because following the dark evolution of a star on such an orbit is extremely time-consuming, we evolved the models in this grid for just 5 full orbits each, then calculated the average capture rates achieved over this time.  We explicitly assume that given identical initial mean capture rates, the long-term evolution of a star on a short-period elliptical orbit would be the same as the evolution of one which evolves on a circular orbit.  That is, we assert that because of the one-to-one mapping between capture rate, WIMP luminosity and evolutionary effects discussed in Sect.~\ref{ms}, the evolution of stars on elliptical orbits in the Galactic centre can be predicted by assigning them an equivalent RSC star from the grid of models we presented earlier.  Recall now our assumption in Sect.~\ref{condanddistro} that WIMPs instantaneously thermalise with the stellar material.  This is almost certainly not a good approximation on the timescale of just 5 orbits.  We therefore probably overestimate annihilation rates and how closely they track capture during such stars' early years, since annihilation takes longer to catch-up with capture when instantaneous thermalisation is not assumed.  As our primary goal is to predict the long-term evolution on the basis of the initial capture rate, we don't expect this approximation to have a large impact here.  We plan to directly test all these assumptions in later papers, using single simulations with very long runtimes and explicitly simulating the thermalisation process.

Under these assumptions, we evolved the star of Fig.~\ref{fig10} for a further half a billion years with a constant equivalent RSC density of $3\times10^9$\,GeV\,cm$^{-3}$, to let it fully adjust to the effects of its captured WIMPs.  We then put it back on the same elliptical orbit, where it was allowed to evolve for a further 5 orbits (right panel of Fig.~\ref{fig10}).  As an evolved dark star, it continues to capture in the same punctuated manner as during its early years.  Now that the star has built up an equilibrium population of WIMPs though, its total population and annihilation rate are essentially immune to the transient nature of capture.  The small decrease in the total population and annihilation rate over the 100\,yr shown here is just due to a slight mismatch between the chosen equivalent RSC density and the actual mean capture rate on the elliptical orbit.

\begin{figure}
\begin{center}
\includegraphics[width=\columnwidth, trim = 0 0 0 30, clip=true]{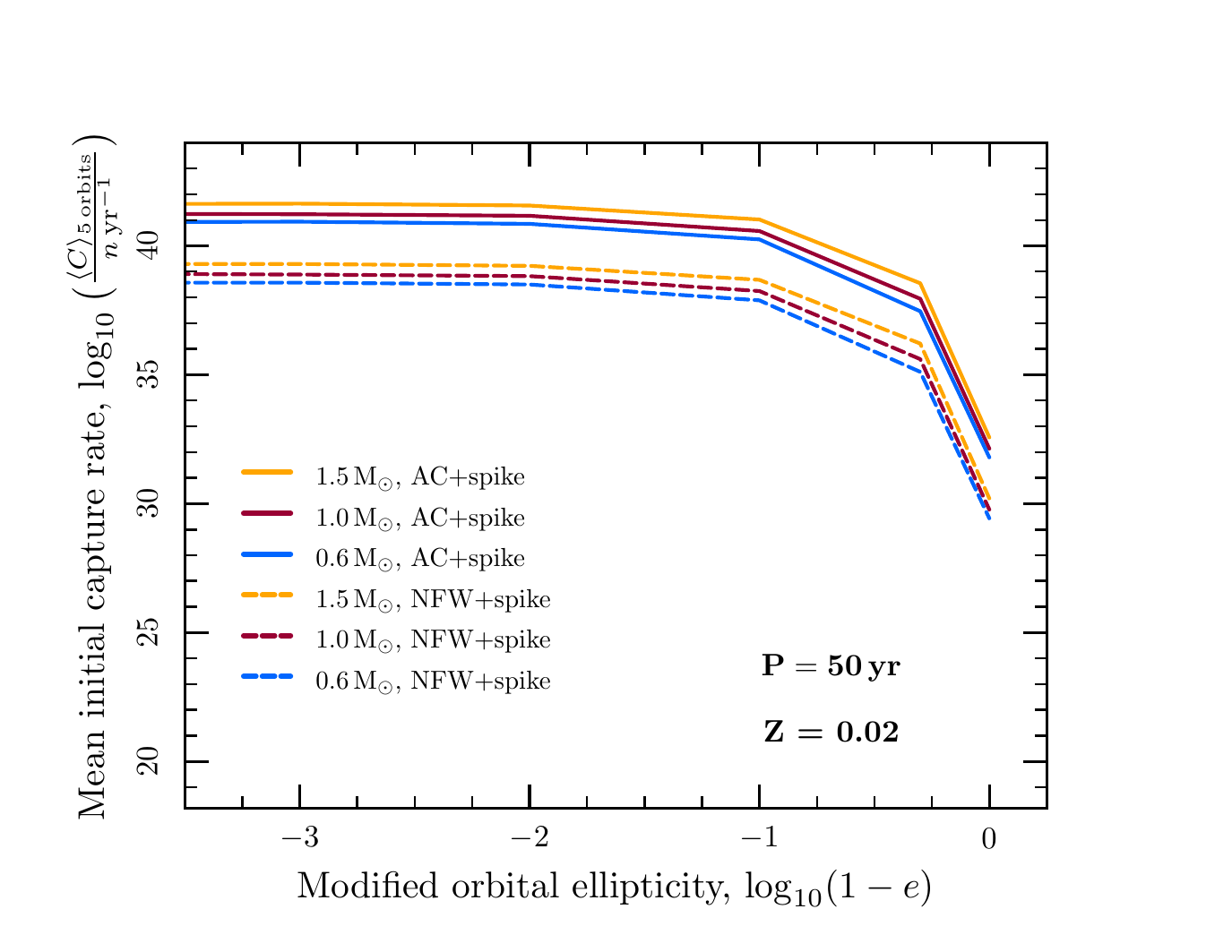}
\includegraphics[width=\columnwidth]{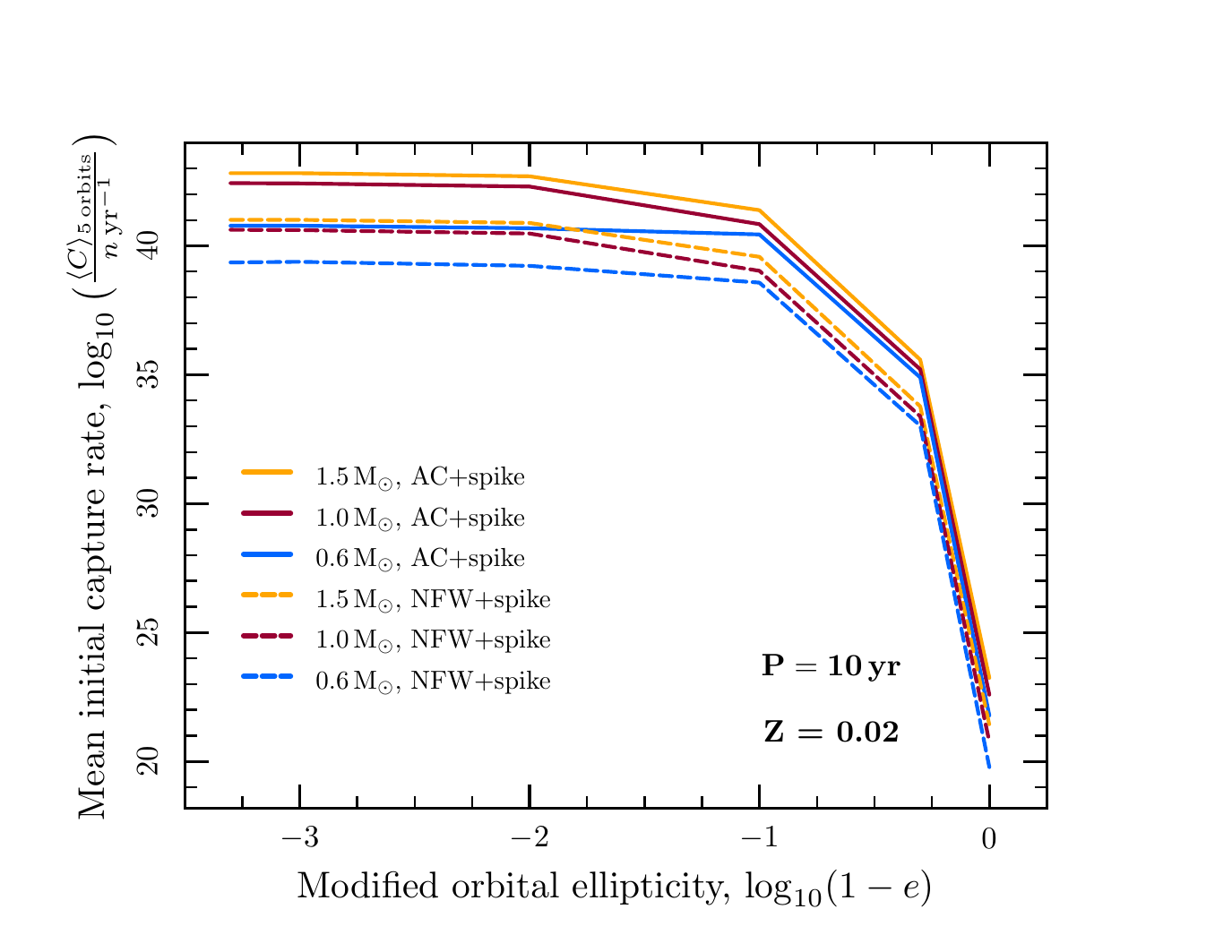}
\includegraphics[width=\columnwidth]{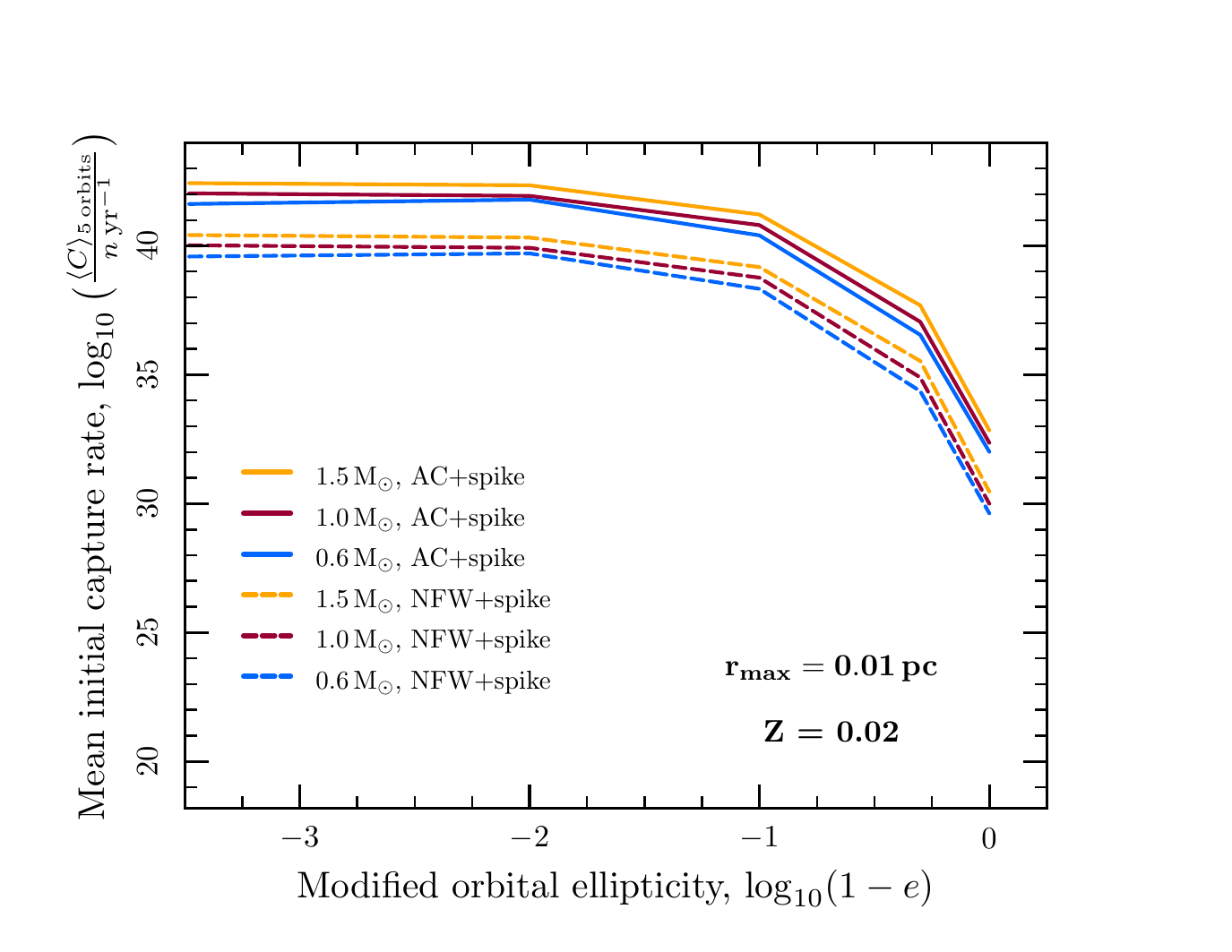}
\end{center}
\caption{Mean capture rates achieved by stars on elliptical orbits with periods of 50 years (top) and 10 years (middle), as well as orbits where the star-black hole separation is 0.01\,pc at apoapsis (bottom).  Dark matter velocities follow an isothermal distribution with dispersion $\bar{v}=270$\,km\,s$^{-1}$, truncated at the local Galactic escape velocity.  Capture rates are boosted by up to 20 orders of magnitude when stars follow elliptical rather than circular orbits.}
\label{fig11}
\end{figure}

\begin{figure}
\begin{center}
\includegraphics[width=\columnwidth, trim = 0 0 0 30, clip=true]{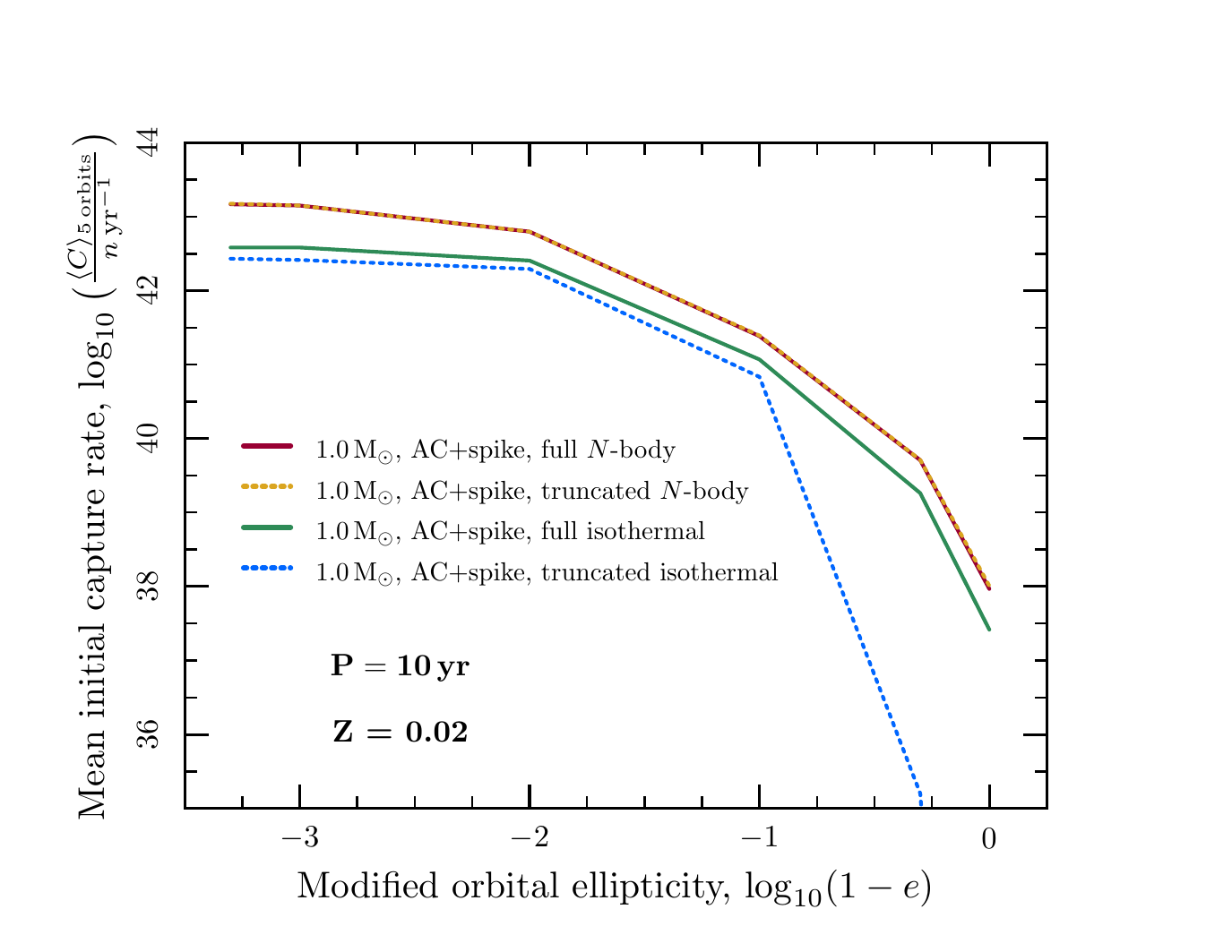}
\end{center}
\caption{Mean capture rates achieved by a 1\,M$_\odot$ star on elliptical orbits through dark matter halos with different velocity distributions.  All halos follow the AC+spike density profile.  The $N$-body velocity distribution results in globally higher capture rates than the standard isothermal distribution.  Truncating the velocity distribution at the local Galactic escape velocity decreases capture rates from the isothermal distribution, but has no impact upon capture from the $N$-body distribution.}
\label{fig12}
\end{figure}

\begin{figure}
\begin{center}
\includegraphics[width=\columnwidth, trim = 0 0 0 30, clip=true]{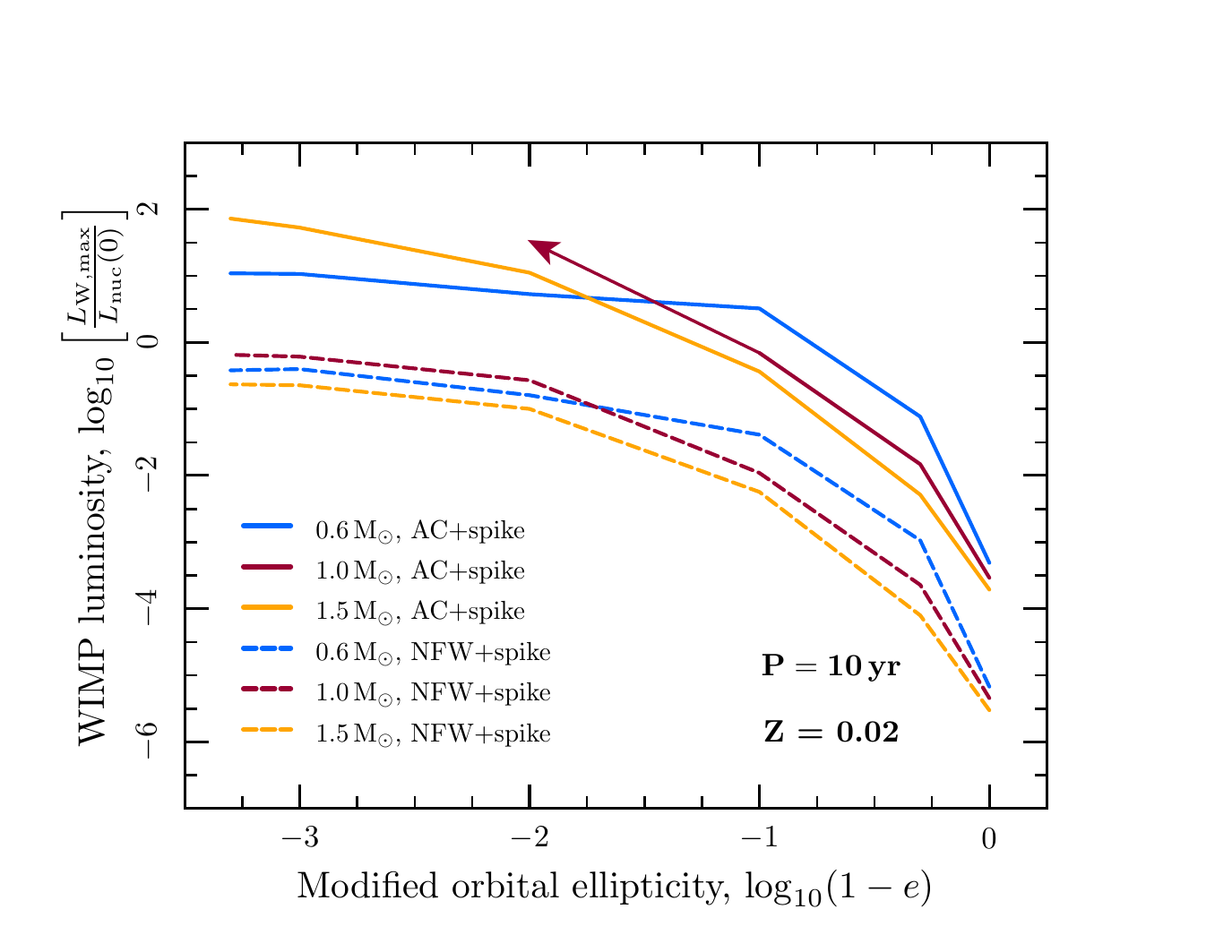}
\end{center}
\caption{WIMP luminosities achieved by stars on orbits with 10-year periods around the Galactic centre.  Annihilation can provide up to 100 times the power of nuclear fusion in stars on realistic orbits.  If the Galactic halo has undergone adiabatic contraction (AC+spike), annihilation rivals nuclear fusion in stars on any orbit with an eccentricity greater than about $e=0.9$, for all masses less than or equal to 1.5\,M$_\odot$.  If not (NFW+spike), stars of a solar mass or less approach break-even between fusion and annihilation energy on orbits with $e\gtrsim0.99$.  These curves have been obtained by applying the boosts seen in Fig.~\protect\ref{fig12} to the capture rates of Fig.~\protect\ref{fig11}, and then interpolating within the results shown in Fig.~\protect\ref{fig1} to obtain the resulting WIMP-to-nuclear burning ratios.  The arrow indicates that the 1\,M$_\odot$, AC+spike curve is expected to continue in this direction, but there is no reliable way to convert capture rates to WIMP luminosities in this region because the capture rates and implied WIMP luminosities are beyond the range of convergence of the benchmark models in Sect.~\protect\ref{ms}.  In the case of the least interesting orbits (circular orbits in an NFW+spike density profile), capture rates and WIMP luminosities are in fact below the limits of the grid in Sect.~\protect\ref{ms}, so WIMP luminosities are extrapolations based on the assumption that the scaling with ambient WIMP density is the same at $e=0$ as at $e=0.5$.}
\label{fig13}
\end{figure}

In an attempt to prevent very rapid changes in the capture rate, we limited timesteps to those which would prevent the ratio $\frac{\rho_\chi}{v_\star}$ shifting by more than 30 per cent in a single step.  Because the evolution code has trouble with convergence when timesteps are reduced too far, we also had to impose a \emph{lower} limit to timesteps of 0.5--1.7\,yr in order to prevent the above criterion breaking convergence as stars passed through periapsis.  Luckily, periapsis turns out to be the least important part of the orbit for the calculation of the mean capture rate and the actual evolution, so the only consequence of this is aesthetic: some of the peaks and troughs in Fig.~\ref{fig10} are not fully resolved.

Fig.~\ref{fig11} shows the mean capture rates achieved by all stars in this grid of models, assuming a truncated isothermal velocity distribution.  Thanks to the additional capture window opened at apoapsis, stars following elliptical orbits have their capture rates boosted by up to 20 orders of magnitude beyond what they would have achieved on the equivalent circular orbit.  The more elliptical the orbit, the slower the star is moving at apoapsis, so the more dark matter it is able to capture.  Whilst stars in the two orbital classes with constant periods reach apoapsis further from the black hole with increasing orbital ellipticity, the reduction in capture rate caused by the resulting decrease in dark matter density at apoapsis is completely outweighed by the increase in capture brought on by the reduced star-WIMP relative velocities.  The ellipticity boost is most marked for shorter orbital periods, as stars on shorter-period orbits have the most to lose by following circular orbits (due to their very high circular velocities), yet the most to gain by following elliptical orbits (because they sample regions of higher dark matter density).

Referring to Fig.~\ref{fig1}, the stars of Fig.~\ref{fig11} evolved in the AC+spike density profile can all achieve break-even between WIMP and nuclear luminosity if their orbits have a period of 10\,yr and $e\gtrsim0.99$.  A 0.6\,M$_\odot$ star can achieve this goal even on a 50-year orbit, with ellipticity as low as $e=0.9$, whilst a 1\,M$_\odot$ star can do the same for $e\gtrsim0.99$.  With the NFW+spike density profile and a truncated isothermal halo, stars will generally never achieve the same level of energy output from WIMP annihilation as nuclear burning (at least not if $m_\chi\ge100$\,GeV; cf.\ Fig.~\ref{fig2}).

In Fig.~\ref{fig12}, we show the impacts upon capture of going beyond the isothermal halo approximation.  Here we illustrate the capture rates achieved by a 1\,M$_\odot$ star on 10-year orbits through both truncated and non-truncated versions of the $N$-body velocity distribution, and compare with the corresponding capture rates from isothermal distributions.  Regardless of the orbital ellipticity, capture is at least a factor of 3 (i.e.~half an order of magnitude) higher from the standard version of the $N$-body distribution than from the standard isothermal distribution.  This difference blows significantly when the distributions are truncated at the local escape velocities and renormalised; whereas the truncation reduces capture from the isothermal distribution (particularly at low ellipticities), it leaves capture from the $N$-body halo entirely unaffected.

The truncated $N$-body distribution boosts capture rates at high ellipticities by a factor of 3--5 over the truncated isothermal distribution, significantly increasing the range of orbits over which we might expect to see dark stars.  In Fig.~\ref{fig13} we estimate the long-term behaviour of stars on the 10-year orbits by combining the capture rates seen in Fig.~\ref{fig11}, the boosts seen in Fig.~\ref{fig12} and the results from our benchmark simulations in Sec.~\ref{ms}.  Here we have made the approximation that boosts do not depend upon the stellar mass or halo density profile; that is, we adjusted the capture rates of all stars by the difference between the truncated $N$-body and truncated isothermal curves in Fig.~\ref{fig12}, depending only upon the orbits which stars followed.  We then interpolated within the data of Fig.~\ref{fig1} to obtain WIMP luminosities from the adjusted capture rates, which included interpolating further amongst the curves of Fig.~\ref{fig1} to obtain data for 1.5\,M$_\odot$ stars.

Fig.~\ref{fig13} shows that the effects of WIMPs can be drastic, with the energy produced by WIMP annihilation outstripping that of nuclear fusion by up to a factor of 100.  Indeed, some of the adjusted capture rates imply a WIMP luminosity even greater than that of any star we were able to reliably evolve in the grid of benchmark models (indicated by an arrow pointing in the direction one would expect the curve to continue in).  We now see from Fig.~\ref{fig13} that even the NFW+spike density profile can actually produce stars where annihilation comes close to breaking even with nuclear burning ($P\approx10$\,yr, $e\gtrsim0.99$ and $M_\star\lesssim 1$\,M$_\odot$).  With the AC+spike profile, the same is true of stars with masses of up to 1.5\,M$_\odot$, following orbits with eccentricities $e\gtrsim0.9$.  If one were to perform the same conversion on capture rates achieved by stars on other orbits, postulating similar boosts as seen on 10-year orbits and assuming an AC+spike density profile, stars of $M_\star\lesssim 1$\,M$_\odot$ would also achieve break-even for $e\gtrsim0.9$ with orbital periods of up to 50 years. 

Finally, we point out that the magnitude of these results depends upon the chosen WIMP mass (Fig.~\ref{fig2}).  As such, one expects the final WIMP luminosities shown in Fig.~\ref{fig13} to be even further boosted for WIMPs lighter than 100\,GeV, but suppressed for higher masses.  The dependence of the capture rate and final WIMP-to-nuclear burning ratio upon the WIMP mass becomes weaker for smaller velocity dispersions, at least in an isothermal halo.  Whilst we have not explicitly investigated how this picture changes in a halo with a non-Gaussian velocity distribution, we would expect similar behaviour.  With velocity dispersions given by Eq.~\ref{nbodysigma}, we therefore expect WIMP luminosities in the $N$-body distribution to have some dependence upon the WIMP mass, though not so pronounced as seen in Fig.~\ref{fig2}.  This dependence should theoretically allow one to place limits upon the WIMP mass in the event of either a positive or null detection of dark stars at the Galactic centre.

\subsection{Binaries and higher-multiplicity stellar systems}

Although we have not calculated the WIMP luminosities achievable by systems consisting of more than one star, our results with single stars make it easy to comment on this scenario.  If a binary system's internal orbital plane was partially aligned with its gross orbit about the Galactic centre, the motions of its component stars would at various stages counteract the motion of the system about the central black hole.  At certain times the component stars would have far smaller velocities relative to the WIMP halo than if they were orbiting as single stars on a similar orbit, resulting in increased capture rates.  The systems most effective at achieving this boost would be those with the highest orbital speeds, as these would produce the greatest reduction in the relative velocity between stars and WIMPs.  This effect would therefore be strongest in short-period, low-separation, higher-mass binary systems.  Since the effects of WIMP annihilation are most marked in low-mass stars though, the optimal configuration would be a close binary with a maximal mass difference.  A system consisting of $\sim$1M$_\odot$ and $\sim$4M$_\odot$ partners orbiting with a period of 5\,hr would have an orbital velocity of $\sim$700\,km\,s$^{-1}$, enough to have a profound impact upon capture rates shown in Fig.~\ref{fig10}, for example.  In favourable cases, binary systems could mimic the effects of highly eccentric orbits upon capture rates, allowing stars on almost circular orbits to achieve significant capture rates, and further boosting capture rates from elliptical orbits. Similar effects could be expected in systems consisting of three or more stars, though their stability on orbits close to the Galactic Centre might be doubtful; indeed, even binaries might not survive for long near the central black hole \citep{Perets09}.

\subsection{Observational constraints and prospects}

Single stars on circular orbits capturing dark matter from an isothermal halo cannot achieve WIMP luminosities any greater than 1 per cent of their nuclear luminosities.  Even if the WIMP velocities were not isothermal, but instead followed the $N$-body distribution, Fig.~\ref{fig12} indicates that capture would not be boosted by more than an additional factor of 5.  From the results of Sect.~\ref{ms}, we know that $\log_{10}[L_\mathrm{W,max}/L_\mathrm{nuc}(0)] \lesssim -1$ would not result in any significant change to a star's structure or evolution.  Whilst this level of WIMP luminosity would create small convective cores in some stars, potentially interesting for asteroseismology of Galactic centre populations some time in the distant future, we find it very unlikely that main-sequence dark stars would exist outside binaries on any circular orbits in the Milky Way.  Stars following elliptical orbits are not only far more likely to be dark stars, but are also considerably more common at the Galactic centre.

The central 30\,pc of the Milky Way not only contains a $3-4\times 10^6 M_\odot$ black hole, but also two of the densest star clusters in the Galaxy, including the Arches cluster.  In the late 1980s, an unusual star with broad H\,\textsc{i} and He\,\textsc{i} emission lines was detected less than 0.5\,pc from the central compact radio source, SgrA*, which is thought to be associated with the central black hole \citep{forrest86,allen90}.  Because the centre of the Galaxy is shrouded with dust, observations can only take place in the infrared, so the normal spectral information used to identify stars is not available.  Over the next few years, an increasing number of such stars were discovered, appearing to be helium-rich blue supergiants and Wolf-Rayet stars, with masses of up to 100 M$_\odot$ \citep{krabbe91,krabbe95}.  The presence of such young stars close to the Galactic centre was difficult to understand, as it was unclear how a gas cloud could condense to form a star in a region so close to the central black hole due to the extreme tidal forces there.  This problem came to be known as the `Paradox of Youth' \citep{sanders92,morris93}.  Recent work has increased the number of known OB stars in the central parsec of the Galaxy to close to 100, excluding the central square arcsecond \citep{paumard06}.  These young stars appear to form two counter rotating discs, suggesting that they are possibly associated with different star formation events in dense accreting matter \citep{levin03, Genzel03, paumard06}.  

The very fact that we know the mass of the central black hole itself is due to the discovery and subsequent tracking of stars even closer than the young He\,\textsc{i} stars: a cluster of stars was discovered in the 1990s within the central square arcsecond, and dubbed the `SgrA* stellar cluster' \citep{genzel97}.  The stars in this cluster (referred to as S stars, E stars or SO stars apparently depending upon the native language of the lead researcher) move extremely close to the central black hole.  The star which has been observed moving closest to the central black hole is called S14, E2 or SO-16, and was seen within 45 AU, or only 600 Schwarszchild radii, of the black hole \citep{Ghez05}.  The kinematics of the SgrA* cluster is such that stars are on randomly oriented, highly elliptical orbits, rather than circular ones confined to a single disc.

Near-infrared observations of the S stars made with the Keck telescope and VLT revealed that their atmospheres did not contain CO \citep{genzel97}, setting a lower bound on their surface temperatures.  Further work led to the conclusion that the S stars are in fact simply 10--15\,M$_\odot$ main-sequence stars \citep{Ghez05, Martins08}.  The short main-sequence lifetimes of such high-mass stars ($\sim$10$^7$\,yr) presents a further Paradox of Youth at the Galactic centre, not explainable by star formation in an accretion disc due to the random orientation of the S-star orbits.

Many have tried to explain the presence of the S stars.  One idea is that they formed far from the Galactic centre (perhaps in the Arches cluster) and subsequently migrated inwards \citep{gerhard01}.  Another is that they formed in situ during an earlier era when the density was much larger than it is today \citep{levin03}.  They could also be old stars which look young because they have collided with other stars \citep{Genzel03}, a scenario reminiscent of the blue straggler phenomenon in globular clusters.  A rather convincing explanation is that they were originally members of binaries belonging to one of the outer discs, which were perturbed either by interactions between the discs \citep{Lockmann08} or by interactions with other massive objects \citep{Perets07}.  The three-body interaction then caused one star from each binary to become tightly bound to the black hole, and the other to be ejected as a hypervelocity star.

What implications might dark stars have for this picture?  We have already seen that stars burning dark matter have significantly increased main-sequence lifetimes.  One might then imagine a scenario whereby stars are created elsewhere and migrate to the Galactic centre, where the presence of dark matter extends their lifetimes.  This might provide an alternative explanation for either the S stars or the outer stellar discs of OB-type stars.  However, such an explanation is incomplete.  The problem with models where stars are created elsewhere and migrate to the centre of the Galaxy is that the inspiralling timescale is typically very large compared to their main sequence lifetime.  One would therefore expect that stars should have left the main sequence by the time they arrive at the central region.  Furthermore, we have shown that more massive stars require higher dark matter densities than low mass ones to experience any structural changes; it is highly unlikely that a star as massive as 10\,M$_\odot$ could capture enough WIMPs to significantly alter its main-sequence evolution on any realistic orbit near the Galactic centre.  One possibility is that such a star could reach the end of its main sequence lifetime during the migration, arrive at the Galactic centre and then begin capturing large numbers of WIMPs.  If burning WIMPs during its post-main-sequence evolution made such a star begin to resemble an OB or Wolf-Rayet star, or revert to looking outwardly like a main sequence star, this could provide an additional explanation for the dense stellar discs or the S stars, respectively.  We will consider the prospect of post-main-sequence dark stars in a later paper.

Whilst such an explanation for the Paradox of Youth must be considered improbable, it is interesting to remember that the S stars are indeed on more elliptical orbits than other stars at the Galactic centre, which would be consistent with them having accreted far more dark matter than others \citep{Zhu:2008qy}.

More promising is the possibility that future observations of the Galactic centre will reveal fainter, lower-mass stars \citep[although there is some suggestion that the initial mass function within the central parsec could be top heavy;][]{Maness:2007py}.  If the binary disruption scenario is indeed the source of the S stars, one would expect that the bursts of star formation which created them in the outer discs would also have produced lower-mass stars.  Some such stars would form in binaries, and could conceivably follow the same path to the Galactic centre as the S stars.  This would produce a population of low-mass stars in the central square arcsecond with randomly-oriented, highly elliptical orbits similar to those of the S stars.  Likewise, most of the other explanations for the origin of the S stars could also involve formation and subsequent migration/disruption of a lower-mass population alongside the S stars, also leading to a population of potential low-mass dark stars.

Finally, the prospect of dark stars forming in binary systems opens a very promising channel through which they might be observed.  Stars within a binary can be compared photometrically if they have similar brightness, allowing their masses and evolutionary states to be determined.  In some cases, binaries might consist of a low-mass star which is significantly affected by WIMP capture and annihilation, and a high-mass partner which is too massive to show any effects whatsoever.  The most striking example of this would be a binary consisting of a low-mass star `frozen' by WIMP burning (resembling a protostar) and a higher-mass companion which had evolved all the way into a white dwarf, though such a system would be difficult to observe at the Galactic centre because of the faintness of the white dwarf.

\section{Conclusions}
\label{conclusions}
When the energy injected due to the annihilation of WIMPs approaches that of nuclear burning, the capture of weakly interacting dark matter will significantly alter the structure and evolution of stars on the main sequence.  Stars on circular orbits in the Milky Way are extremely unlikely to achieve sufficient capture rates for this to occur unless they are present in binaries.  Stars orbiting close to the Galactic centre on elliptical orbits have their capture rates strongly boosted in comparison to those on circular orbits.  The velocity distribution of dark matter near the Galactic centre may be highly non-Gaussian, further boosting capture rates on elliptical orbits by nearly an order of magnitude.  Assuming that the nuclear-scattering cross-sections are equal to their current experimental limits, that dark matter forms a spike around the supermassive black hole at the Galactic centre, and that the dark matter distribution on larger scales has undergone adiabatic contraction, stars of 1\,M$_\odot$ and below will reach break-even between annihilation and fusion energy on orbits with periods of up to 50 years and eccentricities as low as 0.9.  1.5\,M$_\odot$ stars can achieve the same goal with comparable orbital eccentricities if they orbit the central black hole in 10 years or less.  Without adiabatic contraction of the galactic halo, orbits at least as short as this and eccentricities of about 0.99 are required for stars of a solar mass and below to become dark stars.  

These requirements are likely to be significantly relaxed for stars in binary systems.  A binary consisting of a low-mass protostar and a highly-evolved massive star would make the impact of WIMP annihilation very hard to deny.

The observation of one or more stars at the Galactic centre exhibiting the properties we have described would strongly suggest the influence of WIMP dark matter.  Conversely, since we have assumed scattering cross-sections compatible with the current experimental limits, the observation of even a single completely normal star or binary on the orbits we have discussed would allow one to place stringent limits on the properties of dark matter and its density at the Galactic centre.  If one instead assumed a particular halo model for dark matter at the Galactic centre, the dependence of the capture rate upon the WIMP mass and spin-dependent scattering cross-section would allow one to derive limits on these parameters which are highly competitive with current direct-detection sensitivities.  If a star were seen on an orbit where we expect effects even without adiabatic contraction of the Galactic halo on large scales (i.e. $M\lesssim 1\,\mathrm{M}_\odot$, $P\lesssim 10\,\mathrm{yr}$, $e\gtrsim 0.99$), then the derived limits could be made mostly independent of the halo model.

\section*{Acknowledgments}

We thank the referee David Dearborn for strengthening the conclusions of this paper by pointing out the capacity of binaries to further boost capture rates.  We are very grateful to J\"urg Diemand for making the raw results of the Via Lactea simulation available, and Ross Church, Melvyn Davies, Fabio Iocco and Paolo Gondolo for helpful discussions on this work.  P.S. thanks the European Network of Theoretical Astroparticle Physics ILIAS/N6 under contract number RII3-CT-2004-506222 for enabling a visit to CERN, where some of this work was performed, and J.E. thanks the Swedish Research Council (VR) for funding support.  This work has made use of NASA's Astrophysics Data System and SPIRES.

\bibliography{DMbiblio,AbuGen}

\bsp .. and brought to you by the letter W.

\label{lastpage}

\end{document}